\documentclass[preprint,12pt]{elsarticle}
\usepackage[utf8]{inputenc}
\usepackage{color,dsfont,bm}
\pdfoutput=1
\usepackage{float}
\usepackage{amsmath,amssymb,amstext,dsfont,tikz,graphicx}
\usetikzlibrary{arrows}
\usepackage{hyperref}

\DeclareMathOperator{\Tr}{Tr}
\DeclareMathOperator{\diag}{diag}
\def\calA{\mathcal{A}_c}
\def\calB{\mathcal{B}^{\mu}_v}
\def\Ca{\mathsf{C}_{a}}
\def\Cb{\mathsf{C}_{b}}
\def\Sa{\mathsf{S}_{a}}
\def\Sb{\mathsf{S}_{b}}
\def\Ta{\mathsf{T}_{a}}
\def\Tb{\mathsf{T}_{b}}

\def\x{{\bm{x}}}
\def\y{{\bm{y}}}

\def\M{\cal{M}}
\def\N{\cal{N}}

\newcommand{\ket}[1]{| #1 \rangle}

\journal{Annals of Physics}

\begin{document}

\begin{frontmatter}

\title{Absence of Finite Temperature Phase Transitions \\ in the X-Cube Model and its $\mathbb{Z}_{p}$ Generalization
}

\author[wustl]{Zack Weinstein}

\author[sunypoly,dartmouth]{Emilio Cobanera}

\author[iub]{Gerardo Ortiz}

\author[wustl]{Zohar Nussinov}
\ead{zohar@wustl.edu}

\address[wustl]{Department of Physics, Washington University, St.\ Louis, MO 63130, USA}
\address[sunypoly]{Department of Mathematics and Physics, SUNY Polytechnic Institute, Utica, NY 13502, USA}
\address[dartmouth]{Department of Physics and Astronomy, Dartmouth College, Hanover, NH 03755, USA}
\address[iub]{Department of Physics, Indiana University, Bloomington, IN 47405, USA}

\date{\today}

\begin{abstract} 
We investigate thermal properties of the X-Cube model and its $\mathbb{Z}_{p}$ ``clock-type'' 
($p$X-Cube) extension. In the latter, the elementary spin-1/2 operators of the X-Cube model are replaced by elements of the Weyl algebra. We study different boundary condition realizations of these models and analyze their finite temperature dynamics and thermodynamics. We find that (i) no finite temperature phase transitions occur in these systems. In tandem, employing bond-algebraic dualities, we show that for Glauber type solvable baths, (ii) thermal fluctuations might not enable system size dependent time autocorrelations at all positive temperatures (i.e., they are thermally fragile). Qualitatively, our results demonstrate
that similar to Kitaev's toric code model, the X-Cube model (and its $p$-state clock-type descendants) may be mapped to  simple classical Ising ($p$-state clock) chains in which neither phase transitions nor anomalously slow glassy dynamics might appear.  
\end{abstract}


\end{frontmatter}
\tableofcontents

\section{Introduction} 

Notwithstanding the triumphs of the Landau symmetry breaking para\-digm \cite{Landau, Nishimori-Ortiz11}, it is not powerful enough to describe numerous physical systems.  A detailed pedagogical discussion of these non-Landau symmetry breaking aspects appears in \cite{Nishimori-Ortiz11}. Let us briefly mention several of these. In gauge theories, including those describing the fundamental interactions, symmetry breaking is prohibited by Elitzur's theorem \cite{Elitzur,Batista05}. Consequently, in the absence of matter fields, only expectation values associated with closed loops (so-called ``Wilson loops'') may acquire finite expectation values and differentiate between the various phases \cite{Wegner,Wilson,Kogut}. Another important theory in which symmetry breaking cannot occur is the classical two-dimensional XY model \cite{Brezinskii,Kosterlitz1973}. Herein, the binding and unbinding of topological excitations (vortices) characterizes the different phases of the system. 

Along very different lines, the remarkably accurate quantization of the conductance plateau in quantum Hall systems can stem from topological invariance \cite{TKNN}. Motivated by these and related considerations and the prospect of spin-liquids \cite{Anderson73,Kalmeyer}, the notion of ``topological order'' has been introduced, e.g., \cite{Wen03,Wen04}. This endeavor has been bolstered by the prospect of employing topological matter for quantum computation and by the construction of elegant soluble spin models in which basic notions of topological quantum information come to life \cite{Kitaev03,Kitaev06}. Indeed, these and other early
investigations, e.g., \cite{Kitaev-wire,Poulin,Preskill} discussed or were partially motivated
by the quest of achieving  ``fault-tolerant'' topological quantum hardware. The last decade has also witnessed a flurry of experimental findings of materials that exhibit topological effects \cite{top-ins}. 

More recently, various quantum spin models believed to exhibit ``fracton topological order'' \cite{PhysRevB.95.155133,PhysRevB.94.235157,Nan1,Halasz,Shirley1,Shirley2,Shirley3,Shirley4,Shirley5,Slagle1,elastic-fracton,devakul,You,Sid2,Slagle17} were studied. Fracton topological order is a proposed state of matter in which fundamental excitations, termed ``fractons'',
exhibit rich behaviors. In particular, fractons may exhibit confinement along certain spatial directions and further display hindered dynamics at low energies \cite{Nan1}. Individual fractons may be somewhat immobile. However, collectively, several fractons may more readily move together in a constrained correlated fashion at low energies. As a consequence of these constraints on their motion, fractons have been believed to exhibit slow, glassy dynamics, ideal for finite temperature quantum memories \cite{PhysRevB.95.155133,Nan1,Haah,Bravyi}. For a more general current perspective of related issues in localization, symmetry, and topology, see \cite{Sid}.

Prototypical fracton models display symmetries originally known as ``$d$-dimensional Gauge Like symmetries''. These symmetries include, what have been later termed, ``Generalized Global Symmetries'', ``subsystem symmetries'', ``$p$-form symmetries'', or ``higher symmmetries'' \cite{Batista05,Long-TQO,PNAS,holography,gaiotto,lake,Dominic2018,Devakul2,Grozdanov,Glorioso,xgwen-higher}. The relation between $d$-dimensional Gauge Like symmetries and topological quantum order has been established in Refs. \cite{Long-TQO,PNAS,holography}. A key ingredient is the generalization of Elitzur's theorem from conventional gauge (i.e., local ($d=0$)) symmetries to $d>0$ Gauge Like symmetries. These are symmetries that are neither local nor global. For a system residing in $D$ spatial dimensions, the latter symmetries act on a region of intermediate spatial dimensionality $d$ such that $0 <d <D$. These symmetries often allow for low-energy excitations that are typically common to $d$ spatial dimensions to appear in the full $D$-dimensional system. Just as excitations in low-dimensional systems eradicate long lived structures, as a consequence of these symmetries, certain topological memories (such as Kitaev's toric code model \cite{Kitaev03}) may similarly become susceptible to thermal fluctuations - a phenomenon known as ``thermal fragility'' \cite{Long-TQO,PNAS,holography,fragility,fragile2,fragile3,fragile4,fragile5}. This susceptibility is not to be confused with the existence of topological order at finite temperatures \cite{Long-TQO,PNAS}. Imprints of thermal fragility also appear in measures such as the entanglement entropy \cite{Castelnovo}. Different three-dimensional variants of Kitaev's original (two-dimensional) toric code model \cite{fragility} can further exhibit finite temperature robust correlations coexisting with thermally fragile properties \cite{fragility,Castelnovo}. Compass models \cite{compass,compass_rev} similarly exhibit topological order and finite temperature transitions and were also suggested as candidate systems for topological memories \cite{Bacon}.  A broad perspective on quantum memories is given in \cite{Review_memory}. The current paper extends these earlier studies of topological order at finite temperatures to fracton topological order. 

Various conservation laws for effective charges and constrained mobility are often associated with pure gauge symmetries and/or $d > 0$ Gauge like symmetries \cite{Long-TQO, PNAS, holography, hermele_pyrochlore_2004, nussinov_high-dimensional_2007, castelnovo_magnetic_2008,elastic1,elastic2,elastic3}. Restricted dynamics are also known to arise in tensorial gauge field theoretic formulations of elasticity. In these theories \cite{elastic1,elastic2,elastic3}, conservation laws that are captured by the tensorial gauge fields allow for motion of elastic defects only along certain directions (in particular, the well known ``glide'' of dislocations). Herein, dynamics were included by examining elasticity of a medium in space-time (with the kinetic energy corresponding to the energy of elastic deformations along the temporal direction). In this approach, Noether currents may be computed and conservation laws can be reformulated in terms of a gauge invariance \cite{elastic2}. Such a gauge invariance leads to constraints on the motion of dislocations in solids and allows the derivation of the lower dimensional restricted ``glide'' motion. Illuminating works \cite{elastic-fracton,Pai} related these constraints to those pertaining to fractons. 

One particularly popular fracton lattice system is the X-Cube model \cite{PhysRevB.94.235157}, often regarded as a quintessential example of ``type-I fracton topological order'' \cite{Nan1}. The model is given by an exactly solvable commuting Pauli Hamiltonian, traditionally defined on a cubic lattice, but the model can alternatively be defined on many different lattices \cite{Slagle1}. Compared to other fracton models, such as Haah's code and the Chamon model \cite{Haah,Bravyi,holography,Chamon}, the X-Cube model is arguably more intuitive in its structure and excitations, likely leading to its widespread popularity. 

The primary focus of this paper is to investigate the properties of the X-Cube model at finite temperatures by exactly solving its partition function for both open and cylindrical boundary conditions. Additionally, we will use these results to investigate corresponding correlation functions and the finite temperature dynamics of the model. We will employ two different approaches towards this end: first, we will solve each partition function utilizing a brute-force trace calculation, making liberal use of the binomial theorem; second, we will apply a bond-algebraic mapping, in which each Pauli operator contained in the Hamiltonian is mapped to a classical Ising-like spin variable in such a way as to preserve the algebra of the model. We will find that these two approaches yield exactly identical solutions. We will also exactly compute the partition function of the $p$X-Cube model, the natural $\mathbb{Z}_p$ generalization of the ($\mathbb{Z}_2$) X-Cube model, under the same boundary conditions. 

Bond-algebraic mappings akin to those contained in this paper have frequently been used to investigate the finite temperature properties of quantum spin models \cite{Long-TQO,PNAS,holography,fragility,Nussinov08b,solns-bond,bond-PRL,ADP,clock,Maj-dual,Hol-symm,haah_lattice_2013}. Using this strategy, the properties of many complex quantum spin models may be determined from the familiar properties of classical Ising and Ising-like models. In particular, the application of bond algebras first demonstrated that the finite temperature partition function of a prototypical example of topological order - Kitaev's toric code model \cite{Kitaev03} - a model that exhibits $d=1$ dimensional Gauge Like Symmetries, is identical to that of classical one-dimensional Ising chains \cite{Long-TQO}. Similarly, it was shown in Ref. \cite{holography} that the $XXYYZZ$, or Chamon, model maps onto four decoupled Ising chains and also displays the phenomenon of 
``dimensional reduction''. 

More explicitly, a bond-algebraic mapping utilizes a (generally non-local) unitary transformation $X \rightarrow U^{\dagger}X U$ to map each operator in the Hamiltonian to a simple ``classical'' product of $z$-Pauli spins, from which properties such as the partition function, correlation functions, and more, can be evaluated. In particular, this approach will readily enable us to study the finite temperature properties of the X-Cube model.  

We note that the X-Cube model is most commonly assumed to possess periodic boundary conditions, and that although we will be able to exactly solve the partition function for the case of open and cylindrical boundaries, we will be unable to completely solve (in closed form) for the partition function of the model under fully periodic boundary conditions. However, because the corrections to each partition function due to its boundary conditions will appear at high orders of system size, the corresponding corrections to the free energy will be negligible in the thermodynamic limit.
Indeed, in section \ref{periodic}, we will prove that in the thermodynamic limit, the free energy of the X-Cube model with periodic boundary conditions is identically the same as that with open or cylindrical boundaries. 

\section{Main Results of This Work}

A central result of the current work is that: 

{\bf{(1)}} Much as for Kitaev's toric code model which may be mapped onto classical Ising spin chains \cite{Long-TQO,PNAS,fragility}, the X-Cube model does not exhibit a finite temperature phase transition, because of the phenomenon of dimensional reduction.

Along similar lines, we find that:

{\bf{(2)}} There are, at least, some thermal bath realizations which do not lead to super-Arrhenius glassy dynamics of fractons but rather to those of the conventional activated form found in classical Ising spin chains \cite{fragility}. 

Both of these behaviors also appear in the $\mathbb{Z}_p$ generalizations of the X-Cube model, i.e., 
the $p$X-Cube model.

\section{Outline}

The remainder of this paper is organized as follows.

In section \ref{setup:x}, we review the X-Cube model, write its formal partition function, and set up the framework for the high temperature series expansion that we will employ. In section \ref{sec:open}, we compute its partition function (section \ref{sec:openpart}, Eq. (\ref{opensoln})) in the presence of open boundary conditions. This is achieved using both the high temperature series expansion and a bond-algebraic technique that yield identical results.
We further compute (section \ref{corr:open}) general finite temperature correlation functions. In section \ref{sec:cyl}, we perform similar calculations for the X-Cube model endowed with cylindrical boundary conditions (Eq. (\ref{ZTrCylindrical2})). These partition functions lead to free energy densities that exhibit no non-analyticities at any finite temperature. In section \ref{periodic},
we turn to the X-Cube model with periodic boundary conditions. Here, the partition function does not admit a simple closed form expression. However, as we demonstrate (and as is generally anticipated for bulk thermodynamic properties), the free energy density of the periodic system is identical to that when open or cylindrical boundary conditions are present. Thus, also in the presence of periodic boundary conditions, the system exhibits no finite temperature transitions. In section \ref{sec:dynamics}, using simple duality transformations, we arrive at closed form expressions for the autocorrelations. These calculations demonstrate that for a heat bath generated by dualizing a general Glauber heat bath, the X-Cube model does not exhibit long time correlations. That is, here, notwithstanding the topological character of the theory and the constraints for low energy motion, memory of the initial state may be lost after a finite (system size independent) autocorrelation time. We caution that our results do not exclude the possibility of long lived glassy memory when other heat baths are assumed for the system. In section \ref{sec:p}, we discuss the $p$-state clock type generalization of the X-Cube model (the original X-Cube is the $p=2$ realization of this more general model). We compute the partition functions of this $p$X-Cube model under open and cylindrical boundary conditions.
In the $p \to \infty$ limit (section \ref{sec:p>>1}), the $\mathbb{Z}_{p}$ symmetry of the discrete $p$X-Cube models becomes a continuous $U(1)$ symmetry. Given our exact results concerning the lack of finite temperature phase transitions and possible glassy dynamics, we step back, and discuss in section \ref{sec:qualitative} qualitative aspects of low energy motion. We conclude in section \ref{sec:conclude} with general remarks. We discuss various qualitative features of low energy excitations in the Appendix.

Apart from deriving results for the X-Cube models, our analysis highlights the utility of the bond-algebraic duality techniques. Indeed, although we derived the partitions functions for these models using both traditional high temperature series expansions and bond-algebraic duality mappings, the careful reader may readily appreciate the ease with which bond-algebras may enable us to derive results that may be far more cumbersome to arrive at by the more conventional high temperature series expansions. 

\section{General Elements of the X-Cube Model and its Partition Function}
\label{setup:x}

In this Section, we first briefly review the Hamiltonian of the X-Cube model as introduced by Vijay, Haah, and Fu \cite{PhysRevB.94.235157,Nan1}. We will then turn to our main objective of analyzing the system at finite temperatures. Towards that end, we will then formally write down its partition function invoking the well known high temperature series expansion.
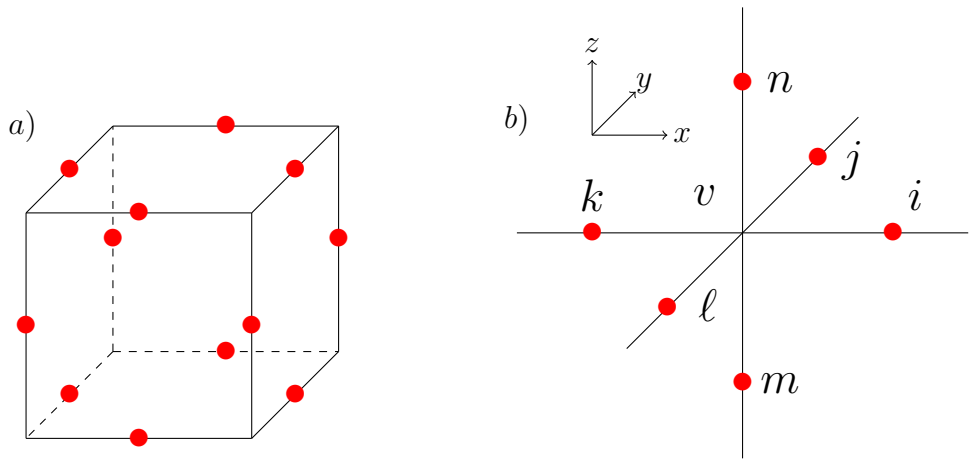
\begin{figure} [h]
\hspace*{0.5cm}
\begin{tikzpicture}
\node at (-1.2,3,0) {$a)$};
\draw (0, 3, 0) -- (0, 3, 3);
\draw (0, 0, 3) -- (0, 3, 3);
\draw (0, 0, 3) -- (3, 0, 3);
\draw (3, 0, 0) -- (3, 0, 3);
\draw (0, 3, 0) -- (3, 3, 0);
\draw (3, 0, 0) -- (3, 3, 0);
\draw (0, 3, 3) -- (3, 3, 3);
\draw (3, 0, 3) -- (3, 3, 3);
\draw (3, 3, 0) -- (3, 3, 3);
\draw [dashed] (0, 0, 0) -- (0, 0, 3);
\draw [dashed] (0, 0, 0) -- (0, 3, 0);
\draw [dashed] (0, 0, 0) -- (3, 0, 0);

\foreach \x in {0,3} {
	\foreach \y in {0,3} {
		\node [red] at (1.5, \x, \y) {\Large \textbullet};
		\node [red] at (\x, 1.5, \y) {\Large \textbullet};
		\node [red] at (\x, \y, 1.5) {\Large \textbullet};
	}
}
\end{tikzpicture}
\hspace*{1.5cm}
\begin{tikzpicture}
\node at (-3,1.5,0) {$b)$};
\draw (3, 0, 0) -- (-3, 0, 0);
\draw (0,3,0) --(0,-3,0);
\draw (0,0,4) -- (0,0,-4);
\foreach \x in {-2,2} {
	\node [red] at (\x, 0, 0) {\Large \textbullet};
	\node [red] at (0, \x, 0) {\Large \textbullet};
	\node [red] at (0, 0, 1.3*\x ) {\Large \textbullet};
}
\node at (2.3, .5, 0) {\Large $i$};
\node at (.5, 0, -2.5) {\Large $j$};
\node at (-2, .5, 0) {\Large $k$};
\node at (.5, 0, 2.5) {\Large $\ell$};
\node at (.5, -2, 0) {\Large $m$};
\node at (.5, 2, 0) {\Large $n$};
\node at (-0.5,0.5,0) {\Large $v$};

\draw [->] (-2,1.3,0) -- (-1,1.3,0);
\node at (-0.8,1.3,0) {$x$};
\draw [->] (-2,1.3,0) -- (-2,2.3,0);
\node at (-2,2.5,0) {$z$};
\draw [->] (-2,1.3,0) -- (-2,1.3,-1.5);
\node at (-2, 1.3, -1.8) {$y$};
\end{tikzpicture}
\caption{Left: A simple $1 \times 1 \times 1$ cube $c$. The qubits associated with its $A_c$ operator are marked as red bullets. Right: A vertex $v$ and its surrounding qubits, labeled to designate the qubits associated with each $B^{\mu}_v$ operator in equation (\ref{B}).}
\label{fig_AB}
\end{figure}

{\it {\rm X}-Cube Model ---}
Consider an $L \times L \times L$ cubic lattice, with  qubits (or spin-$\frac{1}{2}$'s) located at each edge $n$ of the lattice (see figure \ref{fig_AB}). The total number of qubits $N$ in this lattice depends on the choice of boundary conditions, as will be discussed in their respective sections. Each qubit is associated with a two-dimensional Hilbert space $\mathcal{H}_n=\mathbb{C}^2$. The total state space of the system is then given by $\bigotimes_{n = 1}^N \mathcal{H}_n$ with dimension $2^N$. 

For each elementary cube $c$ of the lattice, we define the operator $A_c$ by:
\begin{equation}
\label{A}
A_c \equiv \prod_{n \in \partial c} \sigma^x_n ,
\end{equation}
where $\sigma^x_n$ is the $x$-Pauli operator acting on $\mathcal{H}_n$. $A_c$ is therefore a product of twelve $x$-Pauli operators, each associated with a qubit on one of the twelve edges of the simple cube shown in figure \ref{fig_AB}a.

In addition, for each vertex $v$ of the lattice, label the six surrounding qubits as $i$, $j$, $k$, $\ell$, $m$, and $n$, as in figure \ref{fig_AB}b. The operators $B^{\mu}_v$, $\mu \in \{x,y,z\}$,  are then the four link ``stars" defined by:
\begin{equation}
\label{B}
B^x_v \equiv \sigma^z_j \sigma^z_n \sigma^z_{\ell} \sigma^z_m , \quad B^y_v \equiv \sigma^z_i \sigma^z_n \sigma^z_k \sigma^z_m, \quad B^z_v \equiv \sigma^z_i \sigma^z_j \sigma^z_k \sigma^z_{\ell} .
\end{equation}
Each $B^{\mu}_v$ is the product of the four $z$-Pauli operators surrounding $v$ forming a plane perpendicular to the direction $\mu$. From (\ref{B}), it is quickly seen that:
\begin{equation}
\label{xyz=1}
B^x_v B^y_v B^z_v = \mathds{1} .
\end{equation}
Alternatively, using $(\sigma^z_n)^2 = \mathds{1}$, this can be written as:
\begin{equation}
\label{xy=z}
B^x_v B^y_v = B^z_v, \quad B^x_v B^z_v = B^y_v , \quad B^y_v B^z_v = B^x_v .
\end{equation}

First note that each of the operators $A_c$ and $B^{\mu}_v$ commute. If the vertex $v$ is not a vertex of the simple cube $c$, then $A_c$ and $B^{\mu}_v$ act on no common qubits and therefore must commute. If $v$ is a vertex of $c$, then $A_c$ and $B^{\mu}_v$ act on two common qubits. Since $\sigma^x_n$ and $\sigma^z_n$ anticommute, $\sigma^x_n \sigma^x_m$ and $\sigma^z_n \sigma^z_m$ will commute, and therefore:
\begin{equation}
\label{ABcommutator}
[A_c , B^{\mu}_v] = 0, \ \ \ \forall \ c,v,\mu .
\end{equation}
It is also trivially verified that $A_c$ and $B^{\mu}_v$ are Hermitian operators with eigenvalues $\pm 1$, and that $(A_c)^2 = (B^{\mu}_v)^2 = \mathds{1}$.

The X-Cube model is defined by the stabilizer Hamiltonian \cite{PhysRevB.94.235157,Nan1}:
\begin{equation}
\label{hamiltonian}
H = -a\sum_c A_c - b\sum_{\mu, v} B^{\mu}_v ,
\end{equation}
where $a>0$ and $b>0$ are constant parameters. The first sum is performed over all $L^3$ simple cubes of the lattice, while the second sum is performed over each vertex $v$ and each of the three cardinal directions $\mu$. The particular vertices to be included in the sum will depend on the choice of boundary conditions. 

\begin{figure}
\hspace*{0.5cm}
    \begin{tikzpicture}
    \node at (-1.4,3.7,0) {$a)$};
    \foreach \x in {0,4} {
    \foreach \y in {0,2,4} {
    \draw [line width = 2.0pt, green] (\x,\y,0) -- (\x,\y,4);
    \draw [line width = 2.0pt, green] (\x,0,\y) -- (\x,4,\y);
    \draw [line width = 2.0pt, green] (0,\x,\y) -- (4,\x,\y); 
    \draw [line width = 2.0pt, red, dashed] (2,\y,0) -- (2,\y,4);
    \draw [line width = 2.0pt, red, dashed] (2,0,\y) -- (2,4,\y);
    \draw [line width = 2.0pt, green] (0,2,\y) -- (4,2,\y);
    }
    }
    \node at (1.8,4.5,0) {\Large $\bar{P}^x_1$};
     \end{tikzpicture}
\hspace*{1cm}
    \begin{tikzpicture}
     \node at (-1.4,3.7,0) {$b)$};
    \foreach \x in {0,2,4} {
    \foreach \y in {0,2,4} {
    \draw [line width = 2.0pt, green] (\x,\y,0) -- (\x,\y,4);
    \draw [line width = 2.0pt, green] (0,\x,\y) -- (4,\x,\y);
    }
    \draw [line width = 2.0pt, green] (\x,0,0) -- (\x,4,0);
    \draw [line width = 2.0pt, green] (\x,0,4) -- (\x,4,4);
    \draw [line width = 2.0pt, red, dashed] (\x,0,2) -- (\x,2,2);
    \draw [line width = 2.0pt, green] (\x,2,2) -- (\x,4,2);
    }
    \end{tikzpicture}
\caption{Left: By taking the product of each $\sigma^y_n$ operator corresponding to each edge of the red-dashed plane, we obtain one form of the symmetry operator $\mathcal{U}^{\mu}_i$. Here, $\mu = x$, $i=1$, and $L=2$. Right: By taking the product of each $\sigma^z_n$ operator corresponding to the red-dashed edges, we obtain one form of the symmetry operator $\mathcal{V}$. Here, $L=2$.}
\label{fig_UV}
\end{figure}
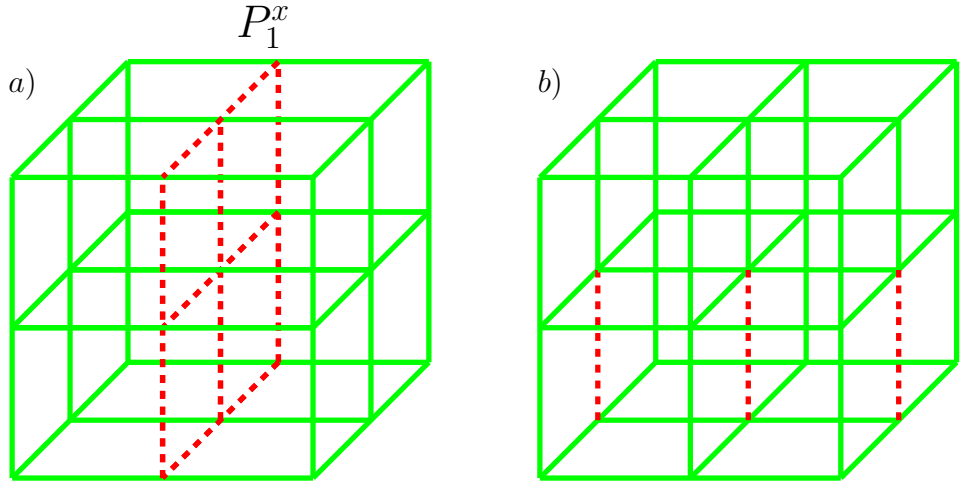

{\it Symmetries ---}
Hamiltonian (\ref{hamiltonian}) is host to several symmetries. $H$ is considered invariant under the symmetry transformation $\ket{\psi} \rightarrow U \ket{\psi}$ if $U^{\dagger} H U = H$. For instance, because each $A_c$ and $B^{\mu}_v$ commutes, $H$ is invariant under the local (gauge) symmetries defined by $U = A_c$ or $U = B^{\mu}_v$ for any $A_c$ or $B^{\mu}_v$. In addition, let $\mathcal{U}^{\mu}_i=\prod_{n \in \bar{P}^{\mu}_i}\sigma^y_n$ be the product of $\sigma^y_n$ for each qubit in the $i$th plane perpendicular to the direction $\mu$, indicated by $\bar{P}^{\mu}_i$ (see figure \ref{fig_UV}a). $1 \leq i \leq L$ for each periodic dimension $\mu$, and $0 \leq i \leq L$ for each open dimension --- see Sections \ref{sec:open} and \ref{sec:cyl} for the precise meaning of open and periodic dimensions. For any individual $n$th qubit in the plane, $\sigma^x_n \rightarrow \sigma^y_n \sigma^x_n \sigma^y_n = -\sigma^x_n$ and $\sigma^z_n \rightarrow \sigma^y_n \sigma^z_n \sigma^y_n = -\sigma^z_n$, so $\mathcal{U}^{\mu}_i$ flips the sign of $\langle \sigma^x_n \rangle$ and $\langle \sigma^z_n \rangle$. However, since any $A_c$ is composed of either zero or four spins in any $i$th plane, and any $B^{\mu}_v$ is composed of zero, two, or four spins in any $i$th plane, $H$ is left invariant under the transformations defined by $\mathcal{U}^{\mu}_i$; these are $d=2$-dimensional Gauge like symmetries. Finally, consider the product $\mathcal{V}$ of each $z$-Pauli operator forming a line parallel to the direction $\mu$, lying on links perpendicular to $\mu$ (see figure \ref{fig_UV}b). For instance, labeling one corner of the lattice as the origin, these may be the $z$-Pauli operators lying at locations $(i, J, K + \frac{1}{2})$ for all $1 \leq i \leq L$ and for some particular $J$ and $K$. Because $\mathcal{V}$ is composed of $z$-Pauli operators, it automatically commutes with each $B^{\mu}_v$, and because $\mathcal{V}$ shares zero or two common spin sites with each $A_c$, $\mathcal{V}$ commutes with each $A_c$ as well, and $H$ is therefore invariant under the transformations defined by $\mathcal{V}$; these are $d=1$-dimensional Gauge like symmetries.

{\it Partition function ---}
The X-Cube model partition function is:
\begin{equation}
\label{partition}
\mathcal{Z} = \Tr \left[ \exp \left( \beta a \sum_c A_c + \beta b \sum_{\mu, v} B^{\mu}_v \right) \right] ,
\end{equation}
where $\beta \equiv 1/k_B T$ is the inverse temperature. Since each operator in (\ref{hamiltonian}) commutes, we may rewrite (\ref{partition}) as:
\begin{equation}
\label{partition2}
\mathcal{Z} = \Tr \left[ \prod_c \left( \exp (\beta a A_c) \right) \prod_{\mu, v} \left( \exp (\beta b B^{\mu}_v) \right) \right] .
\end{equation}
Throughout the following sections,
when performing a high temperature (small $\beta$) series expansion, we will utilize the following properties for simplifying (\ref{partition2}): Using $(A_c)^2 = \mathds{1}$, we can rewrite each exponential as
\begin{equation}
\label{exptocoshsinh}
\begin{split}
\exp (\beta a A_c) &= \mathds{1} + \beta a A_c + \frac{1}{2}(\beta a)^2 \mathds{1} + \frac{1}{6}(\beta a)^3 A_c + \ldots \\
&= \mathds{1}\cosh(\beta a) + A_c \sinh (\beta a) .
\end{split}
\end{equation}
The first product in (\ref{partition2}) can then be written as:
\begin{equation}
\label{Aproduct}
\prod_c \exp (\beta a A_c) = \Ca^{L^3} \prod_c \left[ \mathds{1} + A_c \Ta \right] ,
\end{equation}
where we have defined $\Ca \equiv \cosh(\beta a)$ and $\Ta \equiv \tanh(\beta a)$ (and later $\Sa \equiv \sinh(\beta a)$) for the sake of brevity. The righthand product in (\ref{Aproduct}) will contain one linear term for every possible combination of $A_c$ operators. For clarity, the first few terms of the product are:
\begin{equation}
\label{Aproductexplicit}
\prod_c \left[ \mathds{1} + A_c \Ta \right] = \mathds{1} + \sum_c A_c \Ta + \sum_{c < d} A_c A_d \Ta^2  + \sum_{c<d<e} A_c A_d A_e \Ta^3 + \ldots .
\end{equation}
We thus obtain a series in powers of $\Ta$ realizing a high temperature series expansion.
Using the linearity of the trace, the following feature will be used to discard significant portions of (\ref{partition}) which contribute no trace. Because each Pauli operator is traceless, and each product of Pauli operators is also a Pauli operator, the only terms in (\ref{partition2}) contributing a trace will be those proportional to the identity. These terms will depend on the choice of boundary conditions.

{\it Free energy ---}
Once each partition function is found, the corresponding free energy density is given by:
\begin{equation}
    \label{eq:freeenergy}
    f(\beta) = -\frac{1}{\beta L^3} \log \mathcal{Z}
\end{equation}
For our purposes, the free energy density is important for two reasons: first, we will show that the free energy density of the X-Cube model is independent of our choice of boundary conditions in the thermodynamic limit; second, we will show that this thermodynamic free energy density is completely analytic, indicating the absence of finite temperature phase transitions.

\begin{figure} [h]
\begin{center}
\begin{tikzpicture}
 \node at (-1.5,4.1,0) {$a)$};
\foreach \x in {0,...,3} {
\foreach \y in {0,...,3} {
	\draw [line width = 2.0pt, green] (0,1.5*\x,1.5*\y) -- (4.5,1.5*\x,1.5*\y);
	\draw [line width = 2.0pt, green] (1.5*\x,1.5*\y,0) -- (1.5*\x,1.5*\y,4.5);
	\draw [line width = 2.0pt, red, dashed] (1.5*\x,0,1.5*\y) -- (1.5*\x,4.5,1.5*\y);
}}
 \node at (6,4.1,0) {$b)$};
\foreach \x in {0,...,3} {
\draw [line width = 2.0pt, red, dashed] (7, 1.5*\x, 2.25) -- (11.5, 1.5*\x, 2.25);
\draw [line width = 2.0pt, green] (7+1.5*\x, 0, 2.25) -- (7+1.5*\x, 4.5, 2.25);
}
\end{tikzpicture}
\end{center}
\caption{Left: A cubic lattice constructed of $L+1$ square $L \times L$ lattices (marked in green) and $(L+1)^2$ rungs of $L$ edges each (marked in red-dashed). Right: A square lattice constructed of $L+1$ rungs of $L$ edges horizontally (marked in red-dashed) and $L+1$ rungs of $L$ edges vertically (marked in green). Here, $L = 3$.}
\label{cubiclattice}
\end{figure}
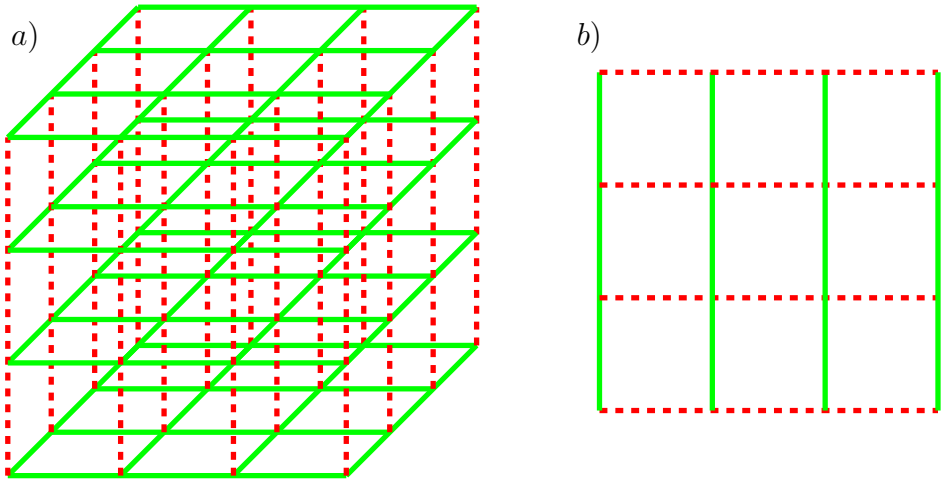

\section{Open Boundary Conditions}
\label{sec:open}
We first assume our $L \times L \times L$ lattice has open boundary conditions. By this, we mean that our lattice is fully non-periodic and has boundaries. This choice of boundary conditions will lead to the simplest solution for (\ref{partition}), but the result is no less meaningful: we will soon see that in the large system size (thermodynamic) limit, any corrections (arising from our choice of boundary conditions) to an extensive physical quantity calculated via derivatives of the partition function will only appear to order $L^2$ and higher.

The $L \times L \times L$ lattice is constructed using $L+1$ horizontal square $L \times L$ lattices, with $(L+1)^2$ vertical rungs of $L$ edges each connecting the horizontal lattices, as seen in figure \ref{cubiclattice}a. Each square lattice consists of $2L(L+1)$ edges: $L+1$ rungs of $L$ edges horizontally, and $L+1$ rungs of $L$ edges vertically, as seen in figure \ref{cubiclattice}b. Then, the total number  of qubits $N$ is given by $3L^3 + 6L^2 + 3L$. 

Using open boundary conditions, we must be careful in how we define the second sum in (\ref{hamiltonian}): for vertices on the boundary of the lattice, $B^{\mu}_v$ will only be properly defined for $\mu$ perpendicular to the boundary. Therefore, we include only the $(L-1)^3$ interior vertices in the sum. We emphasize that this choice will not significantly affect the resulting partition function, and will have no effect whatsoever on the free energy in the thermodynamic limit. We will remark on how the partition function trivially changes for different choices after obtaining our solution.

\subsection{Partition Function}
\label{sec:openpart}

\subsubsection{High Temperature Series Expansion}
To evaluate (\ref{partition2}), we begin by expanding the first set of exponentials using (\ref{exptocoshsinh}). We then note that, since each $\sigma^x_n$ and $\sigma^z_n$ is traceless, and $\sigma^x_n \sigma^z_n = -i\sigma^y_n$ is traceless, the only terms contributing to the trace will be those which are proportional to $\mathds{1}$. However, no terms of (\ref{Aproductexplicit}) containing $A_c$ operators can yield the identity -- for any connected section of elementary cubes, the operators $\sigma^x_n$ corresponding to the boundaries of the section will appear in the product only once. Furthermore, no product of non-identity $A_c$ and $B^{\mu}_v$ operators will yield the identity: because $(\sigma^x_n)^2 = (\sigma^z_n)^2 = \mathds{1}$, any non-identity product of $A_c$ operators will simply be a product of $\sigma^x_n$ operators to the first power, each of which cannot be canceled by a $\sigma^z_n$ of power zero or one. Equation (\ref{partition}) therefore reduces to:
\begin{equation}
\label{Zopen}
\begin{split}
\mathcal{Z}_{\text{Open}} &= \Ca^{L^3} \Tr \left[ \prod_c \left[ \mathds{1} + A_c \Ta \right] \left( \prod_{\mu, v} \exp (\beta b B^{\mu}_v) \right) \right] \\
 &= \Ca^{L^3} \Tr \left[ \prod_{\mu, v} \exp (\beta b B^{\mu}_v) \right] ,
 \end{split}
\end{equation}
To further simplify (\ref{Zopen}), we apply the same procedure, but we must be careful of the constraint (\ref{xyz=1}). We start by expanding the exponentials in the same manner as (\ref{exptocoshsinh}) and (\ref{Aproductexplicit}):
\begin{equation}
\label{Bproduct}
\begin{split}
\prod_{\mu, v} \exp (\beta b B^{\mu}_v) &= \Cb^{3(L-1)^3} \prod_{\mu, v} \left[ \mathds{1} + B^{\mu}_v \Tb \right] \\
 &= \Cb^{3(L-1)^3} \left[\mathds{1} + \sum_{\mu, v} B^{\mu}_v \Tb + \ldots \right] ,
\end{split}
\end{equation}
where $\Cb \equiv \cosh(\beta b)$ and $\Tb \equiv \tanh(\beta b)$ (and later $\Sb \equiv \sinh(\beta b)$). The trace-contributing terms are those proportional to the identity. In particular, each term of (\ref{Bproduct}) will be proportional to the identity if and only if all three $B^{\mu}_v$ are included or excluded at once for each vertex. Each term containing all three $B^{\mu}_v$ for a given set of $n$ vertices will carry a factor of $\Tb^{3n}$, and there are $(L-1)^3 \choose n$ configurations of $n$ vertices out of $(L-1)$ total vertices. Equation (\ref{Bproduct}) can therefore be written as:
\begin{equation}
\label{Bproduct2}
\prod_{\mu, v} [\mathds{1} + B^{\mu}_v \Tb] = \left[ \sum_{n=0}^{(L-1)^3} { (L-1)^3 \choose n} \Tb^{3n} \right] \mathds{1} + \text{t.t.} ,
\end{equation}
where ``t.t." stands for ``traceless terms". 
Using the binomial theorem, and including the factor of $\Cb^{3(L-1)^3}$, we can evaluate the trace in (\ref{Zopen}):
\begin{equation}
\Tr \left[ \prod_{\mu, v} \exp(\beta b B^{\mu}_v) \right] = \Cb^{3(L-1)^3} \left[ 1 + \Tb^3 \right]^{(L-1)^3} \Tr[\mathds{1}] .
\end{equation}
Note that the second product after the equality can be expanded by choosing, for each of $(L-1)^3$ factors of $[1 + \Tb^3]$, either a factor of 1 or of $\Tb^3$. This term altogether can therefore be interpreted to represent the sum of all possible choices for including or excluding all three $B^{\mu}_v$ at each of $(L-1)^3$ vertices. We will therefore sometimes skip the binomial theorem altogether, when the meaning of such a term is clear.

The trace of the identity is $2^N=2^{3L^3 + 6L^2 + 3L}$, the dimension of the total state space. By combining the $\cosh$ and $\tanh$ terms, (\ref{Zopen}) is finally given by:
\begin{equation}
\label{opensoln}
\mathcal{Z}_{\text{Open}} = 2^{3L^3 + 6L^2 + 3L} \Ca^{L^3} \left[ \Cb^3 + \Sb^3 \right]^{(L-1)^3} .
\end{equation}

Alternatively, we may use the following to rewrite the latter product in terms of exponentials:
\begin{equation}
\label{cosh3sinh3}
\Cb^3 + \Sb^3 = \frac{1}{4}(e^{3\beta b} + 3e^{-\beta b}) .
\end{equation}
Rewriting in terms of exponentials, and combining powers of two, (\ref{opensoln}) is given in the form:
\begin{equation}
\label{final-open}
\mathcal{Z}_{\text{Open}} = 2^{L^3 + 12L^2 - 3L + 2} \Ca^{L^3} (e^{3\beta b} + 3e^{-\beta b})^{(L-1)^3} .
\end{equation}
The partition function found by a high temperature series expansion is a regular function. Thus, for all finite $L$ and $\beta$, the high temperature series expansion that we invoked leads to a convergent answer. 

Additionally, we may now consider alternative definitions of the vertex sum in (\ref{hamiltonian}). In addition to the aforementioned treatment of considering only the $(L-1)^3$ interior vertices, there are two other sensible methods of performing this sum. First, we may suppose that for each boundary vertex, we only include $B^{\mu}_v$ operators which are already properly defined -- for instance, a vertex on a boundary perpendicular to the $x$ direction only has a properly defined $B^x_v$ operator, and no vertex on any edge or corner of a boundary has any properly defined $B^{\mu}_v$ operators. In this case, (\ref{opensoln}) is exactly identical to its current form: these newly introduced boundary $B^{\mu}_v$ operators are traceless, cannot be fully canceled by any product of operators, and have no such constraint (\ref{xyz=1}). Therefore, any term in (\ref{Bproduct}) containing a boundary $B^{\mu}_v$ operator has trace zero. 

Alternatively, we could additionally include partial $B^{\mu}_v$ operators at the boundary -- that is, if not all four qubits necessary to define a given $B^{\mu}_v$ are present, we simply include in every $B^{\mu}_v$ only the $z$-Pauli operators which are defined. For instance, in the extreme case, the $B^{\mu}_v$ operators corresponding to each of the eight corner vertices consist of only two $z$-Pauli operators each. In this case, each boundary vertex has three $B^{\mu}_v$ operators satisfying (\ref{xyz=1}). This choice makes (\ref{Zopen}) particularly difficult to solve, as it introduces $3(L+1)$ additional planar constraints. Such planar constraints will be discussed in following sections, and we will see in section \ref{periodic} that such a partition function is very difficult to solve. However, we will still find that such a choice leads to exactly the same free energy in the thermodynamic limit.

Given (\ref{opensoln}), the corresponding free energy density is:
\begin{equation}
    \label{fopen}
    f_{\text{Open}} = -\frac{1}{\beta} \left[ \frac{3L^3\! +\! 6L^2\! +\! 3L}{L^3}\log 2 + \log \Ca + \frac{(L-1)^3}{L^3} \log (\Cb^3+\Sb^3) \right] .
\end{equation}
Note that in the thermodynamic, $L \rightarrow \infty$, limit, our particular choice of definition of the 
model at the boundaries of the system does not affect the free energy density, indicating the bulk nature 
of this physical quantity. Also note that (\ref{fopen}) is a regular function for all finite $\beta$.

\subsubsection{Bond Algebraic Duality}
In addition to the brute-force high temperature series expansion, (\ref{opensoln}) can be obtained using a far simpler approach, the bond-algebraic method. Since each operator in (\ref{hamiltonian}) commutes, we may map each operator to the classical spins $r_m$ and $s^n_j$, $j=1,2$,  as follows (see figures \ref{dualityIsingA} and \ref{dualityIsingB}):
\begin{figure}
    \begin{center}
        \begin{tikzpicture}
        \filldraw[green, fill=red!20] (0,0,4) -- (4,0,4)
                                              -- (4,4,4)
                                              -- (0,4,4)
                                              -- cycle;
        \filldraw[green, fill=red!20] (0,4,4) -- (4,4,4)
                                              -- (4,4,0)
                                              -- (0,4,0)
                                              -- cycle;
        \filldraw[green, fill=red!20] (4,0,4) -- (4,4,4)
                                              -- (4,4,0)
                                              -- (4,0,0)
                                              -- cycle;
        \foreach \x in {0,2,4} {
            \foreach \y in {0,2,4} {
                \foreach \z in {0,2,4} {
                    \draw [line width = 2.0pt, green] (\x,\y,0) -- (\x,\y,4);
                    \draw [line width = 2.0pt, green] (\x,0,\y) -- (\x,4,\y);
                    \draw [line width = 2.0pt, green] (0,\x,\y) -- (4,\x,\y);
                }
            }
        }
        \draw[implies-implies, double equal sign distance] (5.25,2,2) -- (6.75,2,2);
        \foreach \x in {8,10} {
            \foreach \y in {1,3} {
                \foreach \z in {1,3} {
                    \node[red] at (\x,\y,\z) {\Huge \textbullet};
                }
            }
        }
        \draw[-implies, double equal sign distance] (10.5,0.5) -- (10.5,2.5);
        \node at (10.5, 0) {\Large $h = a$};
        
        \draw[implies-implies, double equal sign distance] (9,1,4) -- (8,-0.5,4);
        \draw[line width = 2.0pt, black] (1,-1,4) -- (11,-1,4);
        \foreach \x in {1,2.25,3.5,4.75,6,7.25,8.5,9.75,11} {
            \node[red] at (\x,-1,4) {\Huge \textbullet};
        }
        \node at (12.2,-1,4) {\Large $J=a$};
        \end{tikzpicture}
    \end{center}
    \caption{Under open boundary conditions, the $L^3$ $A_c$ operators of the X-Cube model are dual to $L^3$ isolated spins in a magnetic field, which are in turn dual to an open Ising chain of length $L^3 + 1$. Here, $L=2$.}
    \label{dualityIsingA}
\end{figure}
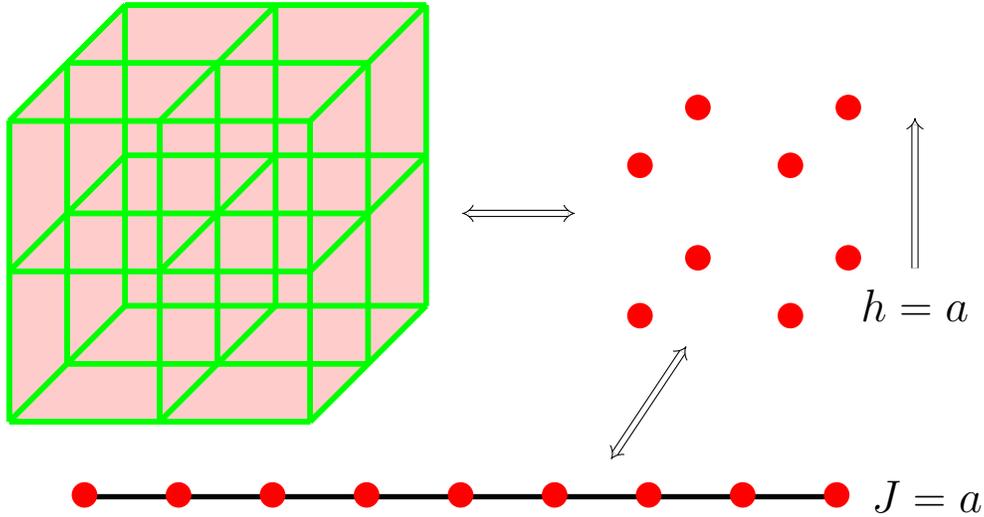
\begin{figure}
\begin{center}
\begin{tikzpicture}[scale=0.72]
\filldraw[fill = red!20, opacity = 0.5]
		(0,-2,-2)
	-- (0,-2,2)
	--(0,2,2)
	--(0,2,-2)
	-- cycle;
	\filldraw[fill = blue!20, opacity = 0.5]
		(-2,0,-2)
	-- (-2,0,2)
	--(2,0,2)
	--(2,0,-2)
	-- cycle;
\filldraw[fill = green!20, opacity = 0.5]
		(-2,-2,0)
	-- (-2,2,0)
	--(2,2,0)
	--(2,-2,0)
	-- cycle;
\draw (2, 0, 0) -- (-2, 0, 0);
\draw (0,2.5,0) --(0,-2.5,0);
\draw (0,0,2) -- (0,0,-2);
\foreach \x in {-2,2} {
	\node [red] at (\x, 0, 0) {\textbullet};
	\node [red] at (0, \x, 0) {\textbullet};
	\node [red] at (0, 0, \x ) {\textbullet};
}
\draw[implies-implies, double equal sign distance] (2.5, 0) -- (4, 0);
\filldraw[fill = blue!20]
	(5,-.1)
	--(6.5,-.1)
	--(6.5,.1)
	--(5,.1)
	-- cycle;
\draw[red, fill = red!20] (5,0) circle (3ex);
\draw[green, fill = green!20] (7,0) circle (3ex);
\node at (6, 1) {\large $J = h = b$};
\draw[implies-implies, double equal sign distance] (0.5, -2.75) -- (1, -4);
\draw[implies-implies, double equal sign distance] (4, -3.75) -- (5, -1.75);
\draw (1,-5) -- (4, -5);
\draw (1, -5) -- (2.5, -7);
\draw (2.5, -7) -- (4, -5);
\draw[red, fill = red!20] (1,-5) circle (3ex);
\draw [green, fill = green!20] (4, -5) circle (3ex);
\draw [blue, fill = blue!20] (2.5, -7) circle (3ex);
\node at (5, -6) {\large $J = b$};
\end{tikzpicture}
\end{center}
\caption{Under open boundary conditions, each set of three $B^{\mu}_v$ operators at a vertex $v$ are dual to two coupled Ising spins under a magnetic field, which are in turn dual to a periodic Ising chain of three Ising spins.}
\label{dualityIsingB}
\end{figure}
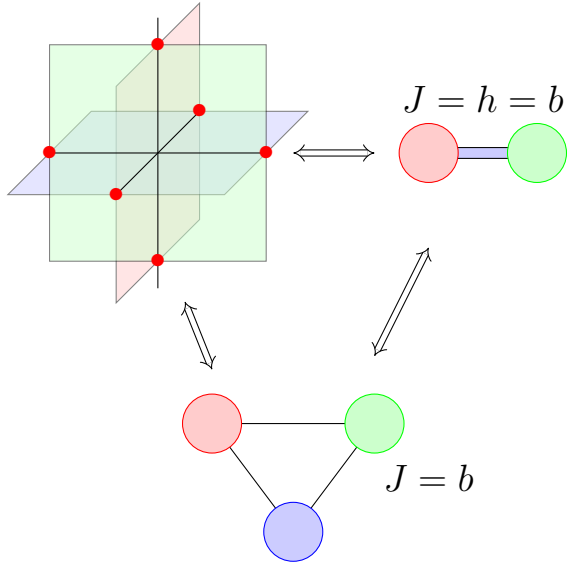\begin{equation}
\label{openclassical}
\begin{split}
A_c \rightarrow r_m, \quad & 1 \leq m \leq L^3 , \\
B^x_v \rightarrow s^n_1, \quad B^y_v \rightarrow s^n_2, \quad B^z_v & \rightarrow s^n_1 s^n_2, \quad 1 \leq n \leq (L-1)^3 .
\end{split}
\end{equation}
The mapping (\ref{openclassical}) preserves the bond algebra of the system -- namely, it respects the conditions (\ref{xy=z}). We therefore expect this classical mapping to preserve the spectrum of (\ref{hamiltonian}), with degeneracies that can only differ by a global power of two. By extension, (\ref{openclassical}) will maintain the form of (\ref{partition}) up to a power of two. By explicitly writing the sum over $\mu$ in (\ref{partition}) and applying the mapping (\ref{openclassical}), we obtain the partition function:
\begin{equation}
\mathcal{Z}_{\text{Open}} = 2^{\bar{N}} \sum_{ \{r_m, s^n_j \} } \prod_{m = 1}^{L^3} \prod_{n = 1}^{(L-1)^3} \exp[\beta a r_m + \beta b (s^n_1 + s^n_2 + s^n_1 s^n_2)] .
\end{equation}

This expression is easily summed directly to yield
\begin{equation}
\label{ZopenClassical}
\mathcal{Z}_{\text{Open}} = 2^{\bar N} (2\Ca)^{L^3}(e^{3\beta b} + 3e^{-\beta b} )^{(L-1)^3} .
\end{equation}
The value of ${\bar N}$ is then determined to be $12L^2-3L+2$ by taking the infinite temperature ($\beta \rightarrow 0$) limit and demanding that $\mathcal{Z}_{\text{Open}}$ is equal to the dimension of the total state space in this limit. Equation (\ref{ZopenClassical}) is then quickly verified to agree with (\ref{opensoln}) and therefore its free energy with (\ref{fopen}). This factor of $2^{\bar{N}}$ indicates that the two dual models have the same spectra, but degeneracies at each energy level that differ by a global factor of $2^{\bar{N}}$. 

We also note that the mapping (\ref{openclassical}) could also have been accomplished by mapping each $B^{\mu}_v$ to the classical spins (see figure \ref{dualityIsingB}):
\begin{equation}
\label{openclassical2}
B^x_v \rightarrow s^n_1 s^n_2, \quad B^y_v \rightarrow s^n_2 s^n_3 , \quad B^z_v \rightarrow s^n_3 s^n_1 .
\end{equation}
These mappings and the corresponding partition function calculations show that (\ref{hamiltonian}) under open boundary conditions is dual to the classical Hamiltonian of a single open Ising chain of length $L^3 + 1$ with bond variables $r_m$ and $(L-1)^3$ two-site Ising chains under a magnetic field (or equivalently, $(L-1)^3$ periodic three-site Ising chains).

\subsection{Cube-Star Correlation Functions}
\label{corr:open}
Correlation functions can also be easily determined under open boundary conditions, using either brute-force trace calculations or bond-algebraic mappings. For $\M$ cubic operators $A_{c_1}, A_{c_2}, \ldots A_{c_{\M}}$ and $\N$ vertex operators $B^{\mu_1}_{v_1}, B^{\mu_2}_{v_2}, \ldots B^{\mu_{\N}}_{v_{\N}}$, we wish to calculate:
\begin{equation}
\label{correlation}
\langle A_{c_1} \ldots A_{c_{\M}} B^{\mu_1}_{v_1} \ldots B^{\mu_{\N}}_{v_{\N}} \rangle = \frac{\Tr [ e^{-\beta H} A_{c_1} \ldots A_{c_{\M}} B^{\mu_1}_{v_1} \ldots B^{\mu_{\N}}_{v_{\N}}]}{\Tr[e^{-\beta H}]} ,
\end{equation}
which is the most general nonvanishing correlation function one can compute: any spin product not expressible as a product of $A_c$ and $B^{\mu}_v$ operators will have a spectrum of $\pm 1$ with equal Boltzmann weights on each eigenvalue, and will therefore have an expectation value of zero.

In this calculation, we will assume without loss of generality that each of the $\M$ operators $A_{c_m}$ and each of the $\N$ operators $B^{\mu_n}_{v_n}$ are distinct. Because $(A_{c_m})^2 = (B^{\mu_n}_{v_n})^2 = \mathds{1}$, and because all operators commute, any instances of indistinct operators can be discarded. In addition, we may assume that each of the $\N$ $B^{\mu_n}_{v_n}$ operators correspond to distinct vertices. If $v_n = v_{n'}$ for $n \neq n'$, then we may use (\ref{xy=z}) to reduce $B^{\mu_n}_{v_n} B^{\mu_{n'}}_{v_{n'}}$ to a single operator. If $v_n = v_{n'} = v_{n''}$ for $n \neq n' \neq n''$, then we may use (\ref{xyz=1}) to eliminate the three operators altogether. Finally, for the sake of brevity, we define $\mathbb{A}\equiv A_{c_1} \ldots A_{c_{\M}}$ and $\mathbb{B}\equiv B^{\mu_1}_{v_1} \ldots B^{\mu_{\N}}_{v_{\N}}$.

In order to calculate (\ref{correlation}), we start by using (\ref{Aproduct}) and (\ref{Bproduct}):
\begin{equation}
\label{correlation2}
\langle \mathbb{A}\mathbb{B} \rangle = \frac{\Tr \left[ \prod_c [\mathds{1} + A_c\Ta] \prod_{\mu, v}[\mathds{1} + B^{\mu}_v\Tb] \mathbb{A}\mathbb{B} \right]}{\Tr \left[ \prod_c [\mathds{1} + A_c\Ta] \prod_{\mu, v}[\mathds{1} + B^{\mu}_v\Tb] \right]} ,
\end{equation} 
where we have preemptively canceled the $\Ca$ and $\Cb$ powers in the numerator and denominator. As before, the only terms inside each trace are those proportional to the identity. Starting with the $A_c$ operators, because no nontrivial product of $A_c$ operators can yield the identity, the only term of (\ref{Aproductexplicit}) which can yield the identity when multiplied by $\mathbb{A}$ is the term containing $\mathbb{A}$ itself, which carries with it a factor of $\Ta^{\M}$. In the denominator, the only term of (\ref{Aproductexplicit}) proportional to the identity is $\mathds{1}$ itself. Equation (\ref{correlation2}) therefore reduces to:
\begin{equation}
\label{correlation3}
\langle \mathbb{A}\mathbb{B} \rangle = \frac{\Ta^{\M} \Tr \left[\prod_{\mu, v}[\mathds{1} + B^{\mu}_v\Tb] \mathbb{B} \right] }{\Tr \left[ \prod_{\mu, v}[\mathds{1} + B^{\mu}_v\Tb]  \right]} .
\end{equation}
For each $B^{\mu_n}_{v_n}$, let $\mu_n'$ and $\mu_n''$ denote the two cardinal directions of $\{x,y,z\}$ not equal to $\mu_n$. Each term in the numerator of (\ref{correlation3}) with nonzero trace must cancel each $B^{\mu_n}_{v_n}$, which can only be canceled by $B^{\mu_n}_{v_n}$ itself or by $B^{\mu_n'}_{v_n} B^{\mu_n''}_{v_n}$. Each term with nonvanishing trace in the numerator of (\ref{correlation3}) equates to picking either $B^{\mu_n}_{v_n}$ with a factor of $\Tb$ or $B^{\mu_n'}_{v_n} B^{\mu_n''}_{v_n}$ with a factor of $\Tb^2$ for each $1 \leq n \leq \N$. In addition, just as in calculating (\ref{Bproduct2}), we may pick any number of vertices $v$ out of the $(L-1)^3 - \N$ vertices not included in $\{v_1, \ldots v_{\N} \}$ and include $B^x_v B^y_v B^z_v$. Therefore, (\ref{correlation3}) is written as:
\begin{equation}
\langle \mathbb{A}\mathbb{B} \rangle = \Ta^{\M} \frac{\left( \Tb + \Tb^2 \right)^{\N} \left( \sum_{n = 0}^{(L-1)^3 - \N} {(L-1)^3 - {\N} \choose n} \Tb^{3n} \right) \Tr[\mathds{1}]}{\left( \sum_{n = 0}^{(L-1)^3} {(L-1)^3 \choose n} \Tb^{3n} \right) \Tr[\mathds{1}]} .
\end{equation} 
Evaluating each summation using the binomial theorem and simplifying yields the solution:
\begin{equation}
\label{corrsoln}
\langle \mathbb{AB} \rangle = \Ta^{\M} \left( \frac{\Tb + \Tb^2}{1 + \Tb^3} \right)^{\N} .
\end{equation}
From (\ref{corrsoln}), we immediately see that each $A_c$ and each $B^{\mu}_v$  on distinct vertices are entirely uncorrelated. That is, the expectation value (\ref{correlation}) is simply given by the product of individual expectation values:
\begin{equation}
    \langle A_{c_1} \ldots A_{c_{\cal{M}}} B^{\mu_1}_{v_1} \ldots B^{\mu_{\cal{N}}}_{v_{\cal{N}}} \rangle =  \langle A_{c_1} \rangle \ldots \langle A_{c_{\cal{M}}} \rangle \langle B^{\mu_1}_{v_1} \rangle \ldots \langle B^{\mu_{\cal{N}}}_{v_{\cal{N}}} \rangle
\end{equation}
This is independent of both the size of the system and the relative placements of cubes and vertices. The expectation values of $A_c$ operators for two adjacent cubes are just as uncorrelated as those of cubes very far apart, and the same is true of vertex operators $B^{\mu}_v$ for distinct vertices. Under open boundary conditions, neither the quantum mechanics nor the thermodynamics of (\ref{hamiltonian}) ``know" about the geometry of the lattice: each cube and each vertex form their own isolated system.

The same solution is easily obtained using the mapping (\ref{openclassical}). First, we map each $B^{\mu}_v$ to some $s^n_1$: although we previously mapped $B^x_v$ in particular to each $s^n_1$, we may instead choose for each vertex individually how we map bijectively from $\{ B^x_v,B^y_v,B^z_v\}$ to $ \{s^n_1, s^n_2, s^n_1 s^n_2 \}$. We obtain the same result with any such mapping, but for ease of calculations, we choose our mapping such that each $B^{\mu}_v$ included in (\ref{correlation}) is mapped to an $s^n_1$. In addition, we will without loss of generality enumerate our mapping such that $A_{c_m} \rightarrow r_m$ for each $1 \leq m \leq \M$ and $B^{\mu_n}_{v_n} \rightarrow s^n_1$ for each $1 \leq n \leq \N$. Defining $\mathbb{R}\equiv r_1 \ldots r_{\M}$ and $\mathbb{S} \equiv s^1_1 \ldots s^{\N}_1$ for brevity,  (\ref{correlation}) is then mapped to:
\begin{equation}
\label{classicalcorr}
\langle \mathbb{R}\mathbb{S} \rangle = \frac{\sum_{ \{r_m , s^n_j \}} \mathbb{RS} \prod_{m=1}^{L^3} \prod_{n = 1}^{(L-1)^3} \exp[\beta a r_m + \beta b (s^n_1 + s^n_2 + s^n_1 s^n_2)]}{\sum_{ \{r_m , s^n_j \}} \prod_{m=1}^{L^3} \prod_{n = 1}^{(L-1)^3} \exp[\beta a r_m + \beta b (s^n_1 + s^n_2 + s^n_1 s^n_2)]} .
\end{equation} 
This expression is easily evaluated by directly summing the first $\M$ variables $r_m$ and the first $\N$ variables $s^n_1$ and $s^n_2$. In the numerator, each factor of $r_m e^{\beta a r_m}$, when summed over $r_m = \pm 1$, yields a factor of $e^{\beta a} - e^{-\beta a}$. Similarly, each factor of $s^n_1 \exp[\beta b(s^n_1 + s^n_2 + s^n_1 s^n_2)]$, when summed over $s^n_j = \pm 1$, yields a factor of $e^{3\beta b} - e^{-\beta b}$. In the denominator, each factor of $e^{\beta a r_m}$ sums to a factor of $e^{\beta a} + e^{-\beta a}$, and each factor of $\exp[\beta b(s^n_1 + s^n_2 + s^n_1 s^n_2)]$ sums to a factor of $e^{3\beta b} + 3e^{-\beta b}$. Canceling the remaining sums over $\{ r_m , s^n_j\}$ for $m > \M$ and $n > \N$, which are identical in the numerator and denominator, (\ref{classicalcorr}) therefore evaluates to:
\begin{equation}
\langle \mathbb{R}\mathbb{S} \rangle = \frac{(e^{\beta a} - e^{-\beta a})^{\M} (e^{3\beta b} - e^{-\beta b})^{\N}}{(e^{\beta a} + e^{-\beta a})^{\M} (e^{3\beta b} + 3e^{-\beta b})^{\N}} 
= \Ta^{\M} \left( \frac{\Tb + \Tb^2}{1 + \Tb^3} \right)^{\N} .
\end{equation}

\section{Cylindrical Boundary Conditions}
\label{sec:cyl}
\begin{figure} [h]
\hspace*{0.5cm}
    \begin{tikzpicture}
    \node at (-1.4,3.7,0) {$a)$};
    \foreach \x in {0,2,4} {
    \foreach \y in {0,2} {
    \filldraw[green, fill=red!20] (\y,0,0) -- (\y,0,4)
                                          -- (\y,4,4)
                                          -- (\y,4,0)
                                          --cycle;
    \filldraw[green, fill=red!20] (0,\x,0) -- (0,\x,4)
                                           -- (2,\x,4)
                                           -- (2,\x,0)
                                           --cycle;
    \filldraw[green, fill=red!20] (0,0,\x) -- (0,4,\x)
                                           -- (2,4,\x)
                                           -- (2,0,\x)
                                           --cycle;
    }
    }
    \foreach \x in {0,2,4} {
    \foreach \y in {0,2,4} {
    \draw [line width = 2.0pt, green] (\x,\y,0) -- (\x,\y,4);
    \draw [line width = 2.0pt, green] (0,\x,\y) -- (4,\x,\y);
    \draw [line width = 2.0pt, green] (\x,0,\y) -- (\x,4,\y);
    }
    }
    
    \node at (1,4.8,0) {\Large $P^x_i$};
     \end{tikzpicture}
\hspace*{1cm}
    \begin{tikzpicture}
    \node at (-1.4,3.7,0) {$b)$};
    \foreach \x in {0,4} {
    \foreach \y in {0,2,4} {
    \draw [line width = 2.0pt, green] (\x,\y,0) -- (\x,\y,4);
    \draw [line width = 2.0pt, green] (\x,0,\y) -- (\x,4,\y);
    \draw [line width = 2.0pt, green] (0,\x,\y) -- (4,\x,\y); 
    \draw [line width = 2.0pt, red, dashed] (2,\y,0) -- (2,\y,4);
    \draw [line width = 2.0pt, red, dashed] (2,0,\y) -- (2,4,\y);
    \draw [line width = 2.0pt, green] (0,2,\y) -- (4,2,\y);
    }
    }
    \node at (1.8,4.8,0) {\Large $\bar{P}^x_i$};
    \end{tikzpicture}
\caption{Left: an $x$-plane of simple cubes, denoted by $P^x_i$. Right: an $x$-plane of vertices, denoted by $\bar{P}^x_i$. In both figures, $i=1$ and $L=2$.}
\label{fig:planes}
\end{figure}
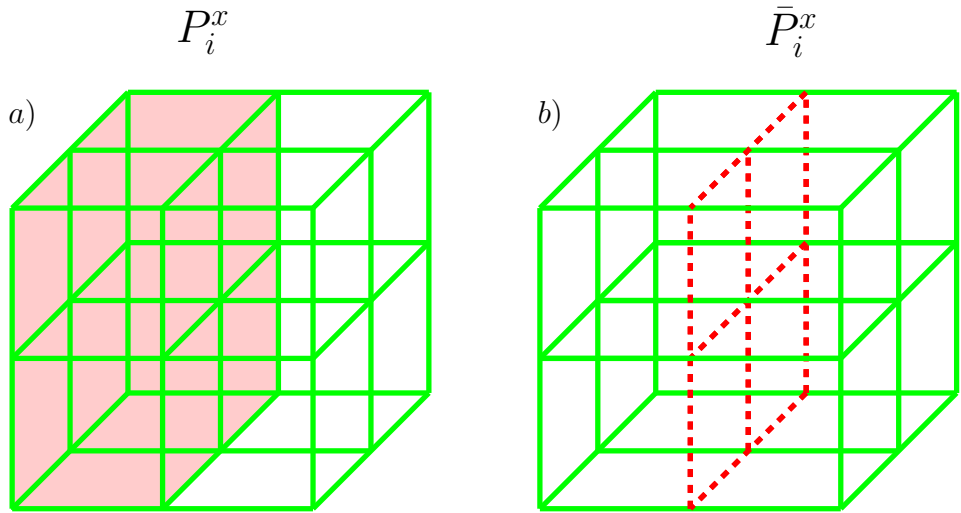

Next, we consider a cubic lattice periodic along its $y$ and $z$ axes, but open along its $x$ axis. Under this cylindrical topology, a plane of simple cubes extending in the $y$ and $z$ directions (that is, perpendicular to the $x$ direction) will contain cubes sharing common edges across the periodic boundaries. In this plane of cubes, all edges parallel to the $x$ direction will be shared by four common cubes, while all edges parallel to the $y$ and $z$ directions will be shared by two common cubes --- see figure \ref{fig:planes}a. The $A_c$ operators corresponding to these cubes will therefore cancel along a plane $P^x_i$ of simple cubes perpendicular to the $x$ axis, 
forming $L$ additional constraints of the form:
\begin{equation}
\label{AplanarCylindrical}
\prod_{c \in P^x_i} A_c = \mathds{1}, \quad 1 \leq i \leq L .
\end{equation}

Similarly, any set of $B^x_v$ operators corresponding to a plane of vertices perpendicular to the $x$ axis will share each of its four qubits with exactly one other $B^x_v$ due to the periodicity in the $y$ and $z$ directions. We therefore have an additional $L-1$ constraints given by:
\begin{equation}
\label{BplanarCylindrical}
\prod_{v \in \bar{P}^x_i} B^x_v = \mathds{1}, \quad 1 \leq i \leq L-1 ,
\end{equation}
where $\bar{P}^x_i$ here refers to a plane of vertices perpendicular to the $x$ axis --- see figure \ref{fig:planes}b.

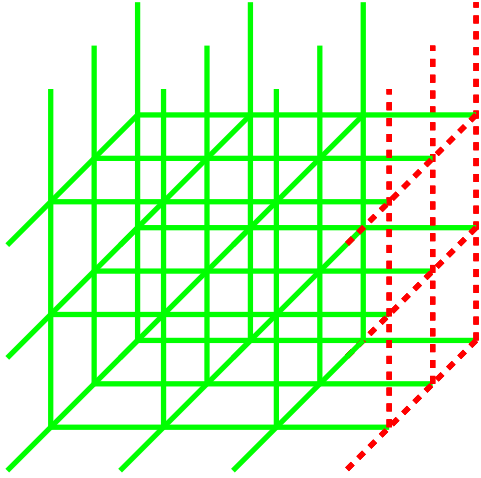
\begin{figure} [h]
\begin{center}
\begin{tikzpicture}
\foreach \x in {0,1.5,3} {
\foreach \y in {0,1.5,3} {
	\draw [line width=2.0pt, green] (0,\x,\y) -- (4.5,\x,\y);
	\draw [line width=2.0pt, green] (\x,\y,0) -- (\x,\y,4.5);
	\draw [line width=2.0pt, green] (\x,0,\y) -- (\x,4.5,\y);
	\draw [line width=2.0pt, red, dashed] (4.5, \x, 0) -- (4.5, \x, 4.5);
	\draw [line width=2.0pt, red, dashed] (4.5, 0, \x) -- (4.5, 4.5, \x);
}}

\end{tikzpicture}
\end{center}
\caption{The periodic $L \times L \times L$ lattice (shown in green) is closed along the $x$ axis with the addition of $2L^2$ additional edges (shown in red-dashed). Here, $L = 3$.}
\label{fig_cylindricalqubits}
\end{figure}

Under this topology, the number $N$ of qubits on the lattice is given by $3L^3 + 2L^2$. This is obtained by starting with the $3L^3$ qubits in the periodic lattice, and closing the lattice in the $x$ direction with an additional $2L^2$ edges (see figure \ref{fig_cylindricalqubits}). In addition, we again must be careful in defining the second sum in (\ref{hamiltonian}). Whereas each of the $L^2$ vertices on an $x$-plane interior to the lattice have all three $B^{\mu}_v$ operators properly defined, vertices on the two exterior $x$-planes have no properly defined $B^y_v$ or $B^z_v$ operators. Therefore, we include only the $L^2(L-1)$ interior vertices in the sum. Other sensible definitions, as well as their implications, are the same as those discussed in the end of section \ref{sec:openpart}.

\subsection{Partition Function} \label{sec:cylpart}

\subsubsection{High Temperature Series Expansion}

First, we evaluate (\ref{partition}) directly, using (\ref{Aproduct}) and (\ref{Bproduct}). Starting with the $A_c$ operators, the terms proportional to the identity in (\ref{Aproductexplicit}) are those satisfying (\ref{AplanarCylindrical}) for some number $n$ of $x$-planes with $0 \leq n \leq L$. Out of $L$ total $x$-planes, there are $L \choose n$ configurations of $n$ planes, with each plane contributing a factor of $\Ta^{L^2}$. We may therefore rewrite (\ref{Aproductexplicit}) under cylindrical boundary conditions as:
\begin{equation}
\label{cylindricalAproduct}
\prod_c [\mathds{1} + A_c \Ta] = \left( \sum_{n = 0}^L {L \choose n} \Ta^{nL^2} \right) \mathds{1} + 
\text{t.t.} .
\end{equation}
Using the binomial theorem and combining with the factor of $\Ca^{L^3}$, (\ref{partition2}) is reduced to:
\begin{equation}
\label{ZTrCylindrical}
\mathcal{Z}_{\text{Cylindrical}} = \left[ \Ca^{L^2} + \Sa^{L^2} \right]^L \Tr \left[ \mathds{1} \prod_{\mu, v} \exp(\beta b B^{\mu}_v ) \right] .
\end{equation}
The terms contributing to a nonvanishing trace in (\ref{ZTrCylindrical}) are those satisfying (\ref{xyz=1}) and (\ref{BplanarCylindrical}). However, we must be careful when using (\ref{BplanarCylindrical}): due to (\ref{xy=z}), we may replace $B^x_v$ in any product forming an $x$-plane with $B^y_v B^z_v$ and still yield the identity. These terms will have an additional factor of $\Tb$, since two operators are included for the vertex $v$ instead of one. We therefore sum (\ref{Bproduct}) for cylindrical boundary conditions as follows: given $n$ $x$-planes of vertices, with $0 \leq n \leq L - 1$, there are $L - 1 \choose n$ configurations of these planes. A priori, each of these planes is composed of $B^x_v$ operators alone, and contributes a factor of $\Tb^{L^2}$. However, for each of the $nL^2$ vertices on the $n$ planes, we may replace $m$ of these $B^x_v$ operators with $B^y_v B^z_v$ in $nL^2 \choose m$ different configurations, with each vertex replacement contributing an additional factor of $\Tb$. Finally, out of the $L^2(L-1) - nL^2$ vertices not lying on any $x$-plane, we may include any $\ell$ products of the form (\ref{xyz=1}) in $L^2(L-1) - nL^2 \choose \ell$ different configurations, each contributing a factor of $\Tb^3$. (\ref{Bproduct}) can therefore be written under cylindrical boundary conditions as:
\begin{equation}
\label{cylindricalBproduct}
\prod_{\mu, v} [\mathds{1} + B^{\mu}_v\Tb] = \left( \sum_{n = 0}^{L-1} {L-1 \choose n} \Tb^{nL^2} \sum_{m = 0}^{nL^2} {nL^2 \choose m} \Tb^m \sum_{\ell = 0}^{Q} {Q \choose \ell} \Tb^{3\ell} \right) \mathds{1} + \text{t.t.} ,
\end{equation} 
where $Q \equiv L^2(L-1) - nL^2$. Equation (\ref{cylindricalBproduct}) is easily summed using three applications of the binomial theorem. First, we evaluate the sums over $m$ and $\ell$ at once, since they share no dependencies. Equation (\ref{ZTrCylindrical}) becomes:
\begin{equation}
\begin{split}
\mathcal{Z}_{\text{Cylindrical}} &= \left[ \Ca^{L^2} + \Sa^{L^2} \right]^L \Cb^{3L^2(L-1)}  \\
& \quad \times \sum_{n = 0}^{L-1} {L-1 \choose n} \Tb^{nL^2} [ 1 + \Tb]^{nL^2} \left[ 1 + \Tb^3 \right]^{L^2(L-1) - nL^2} \Tr[\mathds{1}] .
\end{split}
\end{equation} 
Isolating the $n$ dependencies, applying the binomial theorem once again, and combining $\Cb$ and $\Tb$ terms, we obtain our solution:
\begin{equation}
\label{ZTrCylindrical2}
\begin{split}
\mathcal{Z}_{\text{Cylindrical}} &= \left[ \Ca^{L^2} + \Sa^{L^2} \right]^L \left[ \left[ \Cb^3 + \Sb^3 \right]^{L^2} + \left[ \Cb^2 \Sb + \Cb \Sb^2 \right]^{L^2} \right]^{L-1} \Tr[\mathds{1}] \\
&= 2^{3L^3+2L^2} \left[ \Ca^{L^2} + \Sa^{L^2} \right]^L \left[ \left[ \Cb^3 + \Sb^3 \right]^{L^2} + \left[ \Cb^2 \Sb + \Cb \Sb^2 \right]^{L^2} \right]^{L-1} ,
\end{split}
\end{equation}
where the righthand trace is $2^N=2^{3L^3 + 2L^2}$, the dimension of the total state space. In addition, the $\Cb$ and $\Sb$ terms may be rewritten as exponentials using (\ref{cosh3sinh3}) and:
\begin{equation}
\Cb^2  \Sb + \Cb\Sb^2 = \frac{1}{4}(e^{3\beta b} - e^{-\beta b}) .
\end{equation}
Combining powers of two, the partition function is finally given by:
\begin{equation}
\begin{split}
\label{cylindricalsoln}
\mathcal{Z}_{\text{Cylindrical}} = {} & 2^{L^3 + 4L^2} \left[ \Ca^{L^2} + \Sa^{L^2} \right]^L \\
{} & \times \left[ [e^{3\beta b} + 3e^{-\beta b}]^{L^2} + [e^{3\beta b} - e^{-\beta b}]^{L^2} \right]^{L-1} .
\end{split}
\end{equation}

Note that this expression appears very similar to that of (\ref{opensoln}) in the high-temperature ($\beta \rightarrow 0$) limit. In fact, by dividing (\ref{cylindricalsoln}) by (\ref{opensoln}), we obtain:
\begin{equation}
\begin{split}
\frac{\mathcal{Z}_{\text{Cylindrical}}}{\mathcal{Z}_{\text{Open}}} = {} & 2^{-8L^2 + 3L - 2} \left[ 1 + \Ta^{L^2} \right]^L \\
{} & \times \left[ \left( e^{3\beta b} + 3e^{-\beta b} \right)^{2L - 1} \left[ 1 + \left( \frac{e^{3\beta b} - e^{- \beta b}}{e^{3\beta b} + 3e^{- \beta b}} \right)^{L^2} \right] \right]^{L-1} .
\end{split}
\end{equation}
This function is perfectly regular for all $\beta \geq 0$.
This illustrates, that similar to the case of open boundary conditions, our high temperature series expansion converges for all $\beta$. 

The free energy density corresponding to (\ref{ZTrCylindrical2}) is: 
\begin{equation}
\label{eq:fcyl}
    \begin{split}
    f_{\text{Cylindrical}} = -\frac{1}{\beta} &\left[ \frac{3L^3+2L^2}{L^3} \log 2 + \frac{1}{L^2} \log \left[ \Ca^{L^2} + \Sa^{L^2} \right] \right. \\
    &\quad \left. + \frac{L-1}{L^3} \log \left[ [\Cb^3 + \Sb^3]^{L^2} + [\Cb^2 \Sb + \Cb \Sb^2]^{L^2} \right] \right] ,
    \end{split}
\end{equation}
which could alternatively be written as:
\begin{equation}
\label{eq:fcyl2}
    \begin{split}
    f_{\text{Cylindrical}} = -\frac{1}{\beta} &\left[ \frac{3L^3+2L^2}{L^3}\log 2 + \log \Ca + \frac{L-1}{L}\log (\Cb^3 + \Sb^3) \right. \\
    & \left. + \frac{1}{L^2}\log [1+\Ta^{L^2}] + \frac{L-1}{L^3}\log\left[ 1+ \left( \frac{\Tb + \Tb^2}{1+\Tb^3} \right)^{L^2} \right] \right] .
    \end{split}
\end{equation}
Written this way, we see that (\ref{eq:fcyl}) rapidly converges with (\ref{fopen}) as $L$ increases, as the final two terms of (\ref{eq:fcyl2}) quickly approach zero. This is yet another indication that our choice of boundary conditions should not affect the thermodynamics of the model at large system sizes. 

\subsubsection{Bond Algebraic Duality}
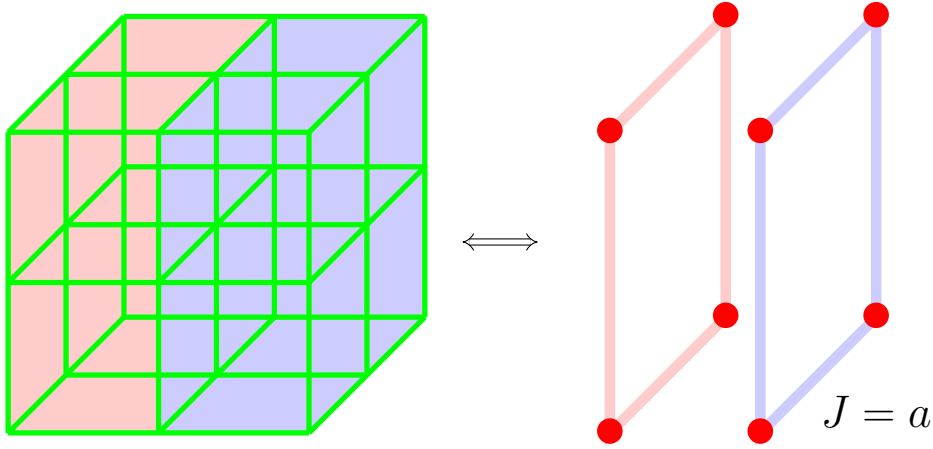
\begin{figure}
\begin{center}
\begin{tikzpicture}
\foreach \x in {0,2} {
\filldraw[green, fill=red!20] (\x,0,0) -- (\x,0,4)
                                       -- (\x,4,4)
                                       -- (\x,4,0)
                                       -- cycle;
\filldraw[green, fill=blue!20] (\x+2,0,0) -- (\x+2,0,4)
                                          -- (\x+2,4,4)
                                          -- (\x+2,4,0)
                                          -- cycle;
}
\foreach \x in {0,2,4} {
\filldraw[green, fill=red!20] (0,\x,0) -- (0,\x,4)
                                       -- (2,\x,4)
                                       -- (2,\x,0)
                                       -- cycle;
\filldraw[green, fill=red!20] (0,0,\x) -- (0,4,\x)
                                       -- (2,4,\x)
                                       -- (2,0,\x)
                                       -- cycle;
\filldraw[green, fill=blue!20] (2,\x,0) -- (2,\x,4)
                                       -- (4,\x,4)
                                       -- (4,\x,0)
                                       -- cycle;
\filldraw[green, fill=blue!20] (2,0,\x) -- (2,4,\x)
                                       -- (4,4,\x)
                                       -- (4,0,\x)
                                       -- cycle;
}
\foreach \x in {0,2,4} {
\foreach \y in {0,2,4} {
\draw[line width = 2.0pt, green] (\x, \y, 0) -- (\x, \y, 4);
\draw[line width = 2.0pt, green] (\x, 0, \y) -- (\x, 4, \y);
\draw[line width = 2.0pt, green] (0, \x, \y) -- (4, \x, \y);
}
}
\draw[implies-implies,double equal sign distance] (4.5,1) -- (5.5,1);
\draw[line width = 4.0pt, red!20] (8,0,0) -- (8,0,4)
                                          -- (8,4,4)
                                          -- (8,4,0)
                                          -- cycle;
\draw[line width = 4.0pt, blue!20] (10,0,0) -- (10,0,4)
                                          -- (10,4,4)
                                          -- (10,4,0)
                                          -- cycle;
\foreach \x in {8,10} {
\foreach \y in {0,4} {
\foreach \z in {0,4} {
\node [red] at (\x, \y, \z) {\Huge \textbullet};
}
}
}
\node at (10,-1.25) {\Large $J = a$};
\end{tikzpicture}
\end{center}
\caption{Under cylindrical boundary conditions, the $L^3$ $A_c$ operators of the X-Cube model are dual to $L$ independent periodic Ising chains of length $L^2$ each. Here, $L = 2$.}
\label{dualitycyl}
\end{figure}

\begin{figure}
\begin{center}
\begin{tikzpicture}
    \draw[black] (0,-1,-3) -- (0,-1,3);
    \draw[black] (0,1,-3) -- (0,1,3);
    \draw[black] (0,-3,-1) -- (0,3,-1);
    \draw[black] (0,-3,1) -- (0,3,1);
    \foreach \x in {-1,1} {
    	\foreach \y in {-1,1} {
    		\draw[black] (-1,\x,\y) -- (1,\x,\y);
    		\filldraw[fill=green!20, opacity=0.5]
    			(-0.9,\x-0.9,\y) -- (-0.9,\x+0.9,\y)
    							 -- (0.9,\x+0.9,\y)
    							 -- (0.9,\x-0.9,\y)
    							 -- cycle;
    		\filldraw[fill=red!20, opacity=0.5] 
    			(0,\x-0.9,\y-0.9) -- (0,\x+0.9,\y-0.9)
    							  -- (0,\x+0.9,\y+0.9)
    							  -- (0,\x-0.9,\y+0.9)
    							  -- cycle;
    		\filldraw[fill=blue!20, opacity=0.5] 
    			(-0.9,\x,\y-0.9) -- (0.9,\x,\y-0.9)
    							 -- (0.9,\x,\y+0.9)
    							 -- (-0.9,\x,\y+0.9)
    							 -- cycle;
    	}
    }
    \draw[implies-implies, double equal sign distance] (2,0,0) -- (3.5,0,0);
    \foreach \x in {-1.5,1.5} {
        \draw[line width = 4.0pt, red!20] (5,\x,-1.5) -- (5,\x,1.5);
        \draw[line width = 4.0pt, red!20] (5,-1.5,\x) -- (5,1.5,\x);
    }
    \foreach \x in {-1.5,1.5} {
        \node[blue!70] at (5,\x,0) {\Huge $\times$};
        \node[blue!70] at (5,0,\x) {\Huge $\times$};
        \foreach \y in {-1.5,1.5} {
            \node[red] at (5,\x,\y) {\Huge \textbullet};
        }
    }
\end{tikzpicture}
\end{center}
\caption{Under cylindrical boundary conditions, the $3L^2(L-1)$ $B^{\mu}_v$ operators of the X-Cube model are dual to $L-1$ independent Ising-gauge chains of length $L^2$ each.}
\label{fig:cylBdual}
\end{figure}
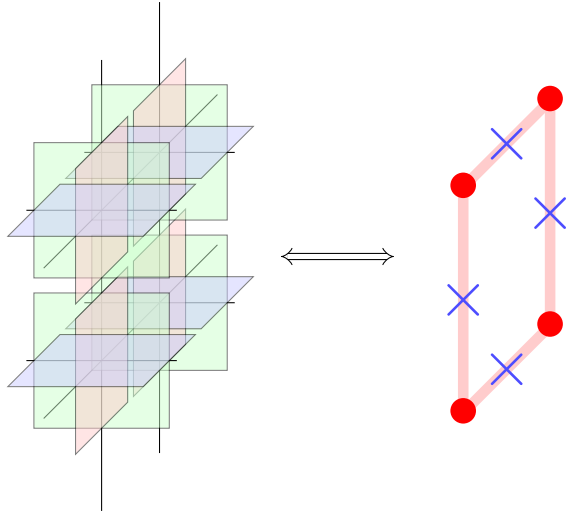

Once again, the most economical way to evaluate (\ref{partition}) is via a duality mapping each $A_c$ and $B^{\mu}_v$ to classical spins preserving the original bond algebra. This is achieved through the following mapping:
\begin{equation}
\label{cylindricalmapping}
\begin{split}
A_c \rightarrow r^i_m r^i_{m+1} , \quad B^x_v \rightarrow s^j_n s^j_{n+1}, \quad B^y_v \rightarrow t^j_n, \quad B^z_v \rightarrow s^j_n s^j_{n+1} t^j_n, \\
1 \leq i \leq L, \quad 1 \leq m \leq L^2, \quad 1 \leq j \leq L-1, \quad 1 \leq n \leq L^2 ,
\end{split}
\end{equation}
where the upper $i$ index in $r^i_m$ is constant along each of the $L$ cubic $x$-planes, and the upper $j$ index in $s^j_n$ and $t^j_n$ is constant along each of the $L-1$ vertex $x$-planes. Effectively, each $x$-plane of $A_c$ operators is mapped to a periodic Ising chain of length $L^2$ (see figure \ref{dualitycyl}), while each $x$-plane of $B^{\mu}_v$ operators is mapped to what might be thought of as an Ising-gauge chain (see figure \ref{fig:cylBdual}). Under this mapping, (\ref{AplanarCylindrical}) and (\ref{BplanarCylindrical}) are mapped to:
\begin{equation}
\prod_{c \in P^x_i} A_c \rightarrow \prod_{m = 1}^{L^2} r^i_m r^i_{m+1} = 1, \quad \prod_{v \in \bar{P}^x_i} B^x_v \rightarrow \prod_{n = 1}^{L^2} s^j_n s^j_{n+1} = 1 ,
\end{equation}
where $L^2 + 1 \equiv 1$ in the above products. Similarly, the constraint (\ref{xyz=1}) is mapped to:
\begin{equation}
B^x_v B^y_v B^z_v \rightarrow s^j_n s^j_{n+1} t^j_n s^j_n s^j_{n+1} t^j_n = 1 .
\end{equation}
We therefore see that this mapping indeed preserves the bond algebra of the operators in (\ref{hamiltonian}). 

The partition function is then given by:
\begin{equation}
\mathcal{Z}_{\text{Cylindrical}} = 2^{\tilde N} \! \! \! \! \! \sum_{ \{r^i_m, s^j_n, t^j_n \}} \prod_{i = 1}^L \prod_{j = 1}^{L-1} \prod_{m,n = 1}^{L^2} \!\! \exp [\beta a r^i_m r^i_{m+1} + \beta b (s^j_n s^j_{n+1} + t^j_n + s^j_n s^j_{n+1} t^j_n)] ,
\end{equation}
where the first and third terms inside the parenthesis represent the Ising-gauge chain (see figure \ref{fig:cylBdual}). This expression is most easily summed by first summing the $r^i_m$ spins, then the $t^j_n$ spins, and finally the $s^j_n$ spins. The $r^i_m$ spins simply form $L$ independent periodic Ising chains each of length $L^2$:
\begin{equation}
\label{Zcylindrical}
\begin{split}
\mathcal{Z}_{\text{Cylindrical}} = {} & 2^{\tilde N} \left[ (2\Ca)^{L^2} + (2\Sa)^{L^2} \right]^L \\
{} & \times \sum_{ \{s^j_n, t^j_n \}} \prod_{j = 1}^{L-1} \prod_{n = 1}^{L^2} \exp[\beta b(s^j_n s^j_{n+1} + t^j_n + s^j_n s^j_{n+1} t^j_n)] .
\end{split}
\end{equation}
Next, because each $t^j_n$ is uncoupled, they can be summed directly:
\begin{equation}
\label{Zcylindrical2}
\begin{split}
\mathcal{Z}_{\text{Cylindrical}} = {} & 2^{\tilde N} \left[ (2\Ca)^{L^2} + (2\Sa)^{L^2} \right]^L \\
{} & \times \sum_{ \{s^j_n \}} \prod_{j = 1}^{L-1} \prod_{n = 1}^{L^2} \left( \exp[\beta b (2s^j_n s^j_{n+1} + 1)] + \exp[-\beta b] \right) .
\end{split}
\end{equation}
The remaining sum is evaluated using the transfer matrix method: let $T$ be a $2 \times 2$ matrix given by:
\begin{equation}
\begin{split}
\langle s^j_n | T | s^j_{n+1} \rangle = \exp[\beta b (2s^j_n s^j_{n+1} + 1)] + \exp[-\beta b] \\
\implies T = \begin{bmatrix}
e^{3\beta b} + e^{-\beta b} & 2e^{-\beta b} \\ 
2e^{-\beta b} & e^{3\beta b} + e^{-\beta b}
\end{bmatrix} .
\end{split}
\end{equation}
Then (\ref{Zcylindrical2}) is reduced to the form:
\begin{equation}
\label{Zcylyndrical3}
\mathcal{Z}_{\text{Cylindrical}} = 2^{\tilde N} \left[ (2\Ca)^{L^2} + (2\Sa)^{L^2} \right]^L \left[ \Tr T^{L^2} \right]^{L-1} .
\end{equation}
The eigenvalues of $T$ are given by $e^{3\beta b} + 3e^{-\beta b}$ and $e^{3\beta b} - e^{-\beta b}$, reducing (\ref{Zcylyndrical3}) to:
\begin{equation}
\label{Zcylyndrical4}
\begin{split}
\mathcal{Z}_{\text{Cylindrical}} = {} & 2^{\tilde N} \left[ (2\Ca)^{L^2} + (2\Sa)^{L^2} \right]^L \\
{} & \times \left[ [e^{3\beta b} + 3e^{-\beta b}]^{L^2} + [e^{3\beta b} - e^{-\beta b}]^{L^2} \right]^{L-1} .
\end{split}
\end{equation}
Once again, the value of $\tilde N$ is determined as $4L^2$ by taking the infinite temperature ($\beta \rightarrow 0$) limit and demanding that $\mathcal{Z}_{\text{Cylindrical}}$ is given by $2^N$ in this limit, where $N = 3L^3 + 2L^2$. Combining powers of two, we see that the solution given by the mapping (\ref{cylindricalmapping}) exactly matches that of (\ref{cylindricalsoln}), showing that (\ref{hamiltonian}) under cylindrical boundary conditions is dual to the classical Hamiltonian of $L$ periodic Ising chains of length $L^2$ and $L-1$ periodic Ising-gauge chains. The factor of $2^{\tilde{N}}$ again indicates that the dual models have the same energy levels, with degeneracies that differ only by a global factor of $2^{\tilde{N}}$.

\subsection{Cube-Star Correlation Functions}
To compute (\ref{correlation}) under this topology, we must be particularly careful of $A_c$ operators lying on the same cubic $x$-plane and $B^{\mu}_v$ operators lying on the same vertex $x$-plane. Labeling the $L$ cubic $x$-planes 1 through $L$ and the $L-1$ vertex $x$-planes 1 through $L-1$, let ${\M}_i$ for $1 \leq i \leq L$ denote the number of $A_c$ operators lying in the $i$th plane that are included in the expectation value (\ref{correlation}), and let ${\N}_j$ for $1 \leq j \leq L-1$ denote the number of $B^{\mu}_v$ operators lying in the $j$th plane that are included in (\ref{correlation}). Furthermore, let $A_{i;m}$ denote the $m$th included operator in the $i$th cubic plane, and let $B^{\mu_{j;n}}_{j;n}$ denote the $n$th included operator in the $j$th vertex plane. Equation (\ref{correlation}) is then given by:
\begin{equation}
\label{cylindricalcorrelation}
\left\langle \prod_{i = 1}^L \prod_{m = 1}^{{\M}_i} A_{i;m} \prod_{j = 1}^{L-1} \prod_{n = 1}^{{\N}_j} B^{\mu_{j;n}}_{j;n} \right\rangle = \frac{\Tr \left[ e^{-\beta H} \prod_{i = 1}^L \prod_{m = 1}^{{\M}_i} A_{i;m} \prod_{j = 1}^{L-1} \prod_{n = 1}^{{\N}_j} B^{\mu_{j;n}}_{j;n} \right]}{\Tr[e^{-\beta H}]} .
\end{equation}
Once again, (\ref{cylindricalcorrelation}) is the most general nonvanishing correlation function: any spin product not composed of the $A_c$ and $B^{\mu}_v$ operators composing (\ref{hamiltonian}) will necessarily have an expectation value of zero. In addition, we again assume in our calculation of (\ref{cylindricalcorrelation}) that each $A_{i;m}$ and $B^{\mu_{j;n}}_{j;n}$ is distinct, and that each $B^{\mu_{j;n}}_{j;n}$ lies on distinct vertices: any product of $A_c$ and $B^{\mu}_v$ operators can be reduced to this form using $(A_c)^2 = (B^{\mu}_v)^2 = \mathds{1}$ and (\ref{xy=z}). Just as in the case of open boundary conditions, we calculate (\ref{cylindricalcorrelation}) using an expansion of the form (\ref{correlation2}). Under cylindrical boundary conditions, each cubic planar product $\prod_{m = 1}^{{\M}_i} A_{i;m}$ can be canceled by one of two products: either $\prod_{m = 1}^{{\M}_i} A_{i;m}$ itself, or the remaining $L^2 - {\M}_i$ cubic operators in the $i$th plane not included in $\prod_{m = 1}^{{\M}_i} A_{i;m}$. Therefore, each term of (\ref{Aproductexplicit}) contributing nonvanishing trace is given by choosing for each plane either the product $\prod_{m = 1}^{{\M}_i} A_{i;m}$ with a factor of $\Ta^{{\M}_i}$ or the entire plane except $\prod_{m = 1}^{{\M}_i} A_{i;m}$ with a factor of $\Ta^{L^2 - {\M}_i}$. The $A_c$ operators in the denominator are handled as in (\ref{ZTrCylindrical}), reducing (\ref{cylindricalcorrelation}) to:
\begin{equation}
\label{cylindricalcorrelation2}
\begin{split}
\left\langle \prod_{i = 1}^L \prod_{m = 1}^{{\M}_i} A_{i;m} \prod_{j = 1}^{L-1} \prod_{n = 1}^{{\N}_j} B^{\mu_{j;n}}_{j;n} \right\rangle & = \prod_{i = 1}^L \left( \frac{\Ta^{{\M}_i} + \Ta^{L^2 - {\M}_i}}{1 + \Ta^{L^2} } \right) \\
& \times \frac{\Tr \left[ \prod_{\mu,v} [\mathds{1} + B^{\mu}_v \Tb] \prod_{j = 1}^{L-1} \prod_{n = 1}^{{\N}_j} B^{\mu_{j;n}}_{j;n} \right]}{\Tr[\prod_{\mu,v}[\mathds{1} + B^{\mu}_v \Tb] ]} .
\end{split}
\end{equation}
Each product $\prod_{n = 1}^{{\N}_j} B^{\mu_{j;n}}_{j;n}$ can be canceled in one of two ways. One possibility is that they are canceled pointwise, where each $B^{\mu}_v$ is matched by an identical $B^{\mu}_v$ or by (\ref{xy=z}). A priori, this incurs a factor of $\Tb^{{\N}_j}$, with an additional factor of $\Tb^n$ for each choice of $n$ operators canceled using (\ref{xy=z}). We may also include $m$ factors of $\Tb^3$ for $0 \leq m \leq L^2 - {\N}_j$ by applying (\ref{xyz=1}) to $m$ of the $L^2 - {\N}_j$ vertices not included in $\prod_{n = 1}^{{\N}_j} B^{\mu_{j;n}}_{j;n}$. The other method of canceling $\prod_{n = 1}^{{\N}_j} B^{\mu_{j;n}}_{j;n}$ is to include the remaining factors necessary to satisfy (\ref{BplanarCylindrical}). To do this, we must account for the particular directions $\mu_{j;n}$ involved: let ${\cal{X}}_j$ be the number of $B^x_v$ included in the $j$th $x$-plane. For each such operator, we may either simply include each such $B^x_{j;n}$ in the product, or we may include (\ref{xyz=1}) with a factor of $\Tb^3$. For each additional ${\N}_j - {\cal{X}}_j$ operators included in $\prod_{n = 1}^{{\N}_j} B^{\mu_{j;n}}_{j;n}$, we may either include $B^{\mu_{j;n}'}_{j;n}$ for $\mu_{j;n}' \neq x$, or we may include $B^{\mu_{j;n}}_{j;n} B^x_{j;n}$, each converting $B^{\mu_{j;n}}_{j;n}$ to the required $B^x_{j;n}$ to satisfy (\ref{BplanarCylindrical}). For the remaining $L^2 - {\N}_j$ vertices in the $j$th plane, we may either include $B^x_v$ or $B^y_v B^z_v$ to satisfy (\ref{BplanarCylindrical}), contributing factors of $\Tb$ and $\Tb^2$ respectively. The numerator trace of (\ref{cylindricalcorrelation2}) is therefore given by:
\begin{equation}
\begin{split}
 \Tr \! \left[ \! \prod_{\mu,v} [\mathds{1} + B^{\mu}_v \Tb] \! \prod_{j = 1}^{L-1} \prod_{n = 1}^{{\N}_j} B^{\mu_{j;n}}_{j;n} \! \right] \! = \!\left( \prod_{j = 1}^{L-1} \! \left( \sum_{n = 0}^{{\N}_j} \! \sum_{m = 0}^{L^2 - {\N}_j} \! \! {L^2 - {\N}_j \choose m} \! {{\N}_j \choose n}  \Tb^{{\N}_j+n+3m} \right. \right.  \\ 
 + \left. \left. \sum_{\ell = 0}^{{\cal{X}}_j}  \sum_{q = 0}^{{\N}_j - {\cal{X}}_j} \sum_{r = 0}^{L^2 - {\N}_j} \! {{\cal{X}}_j \choose \ell}   {{\N}_j - {\cal{X}}_j \choose q}    {L^2 - {\N}_j \choose r} \Tb^{3\ell -{\cal{X}}_j+q+L^2 + r} \right) \right) \Tr[\mathds{1}] .
\end{split}
\end{equation} 
Using the binomial theorem and combining with the results of (\ref{ZTrCylindrical2}), (\ref{cylindricalcorrelation2}) can be simplified to:
\begin{equation}
\label{cylcorrsoln}
\begin{split}
&\left\langle \prod_{i = 1}^L \prod_{m = 1}^{{\M}_i} A_{i;m} \prod_{j = 1}^{L-1} \prod_{n = 1}^{{\N}_j} B^{\mu_{j;n}}_{j;n} \right\rangle = \prod_{i = 1}^L \left( \frac{\Ta^{{\M}_i} + \Ta^{L^2 - {\M}_i}}{1 + \Ta^{L^2} } \right) \\
&  \times \prod_{j = 1}^{L-1} \left( \frac{\left( \Tb + \Tb^2 \right)^{{\N}_j} \left( 1 + \Tb^3 \right)^{L^2 - {\N}_j} + \left( \Tb + \Tb^2 \right)^{L^2 - {\cal{X}}_j} \left( 1 + \Tb^3 \right)^{{\cal{X}}_j}}{\left( \Tb + \Tb^2 \right)^{L^2} + \left( 1 + \Tb^3 \right)^{L^2}} \right) .
\end{split}
\end{equation}
First, note that cubes and stars are once again uncorrelated: $\langle A_c B^{\mu}_v \rangle = \langle A_c \rangle \langle B^{\mu}_v \rangle$ for each $c,\mu,v$. This is a general feature of the X-Cube model, independent of boundary conditions, as the partition function will always factor into the product of a partition function for the $A_c$ operators and a partition function for the $B^{\mu}_v$ operators. Additionally, note that each cubic $x$-plane and each vertex $x$-plane are entirely uncorrelated: for any cubes $c$ and $d$ on distinct cubic $x$-planes, $\langle A_c A_d \rangle = \langle A_c \rangle \langle A_d \rangle$, and the same holds for two $B^{\mu}_v$ operators on different vertex $x$-planes. Each $x$-plane of cubes or vertices forms its own isolated system, as suggested by the duality (\ref{cylindricalmapping}). Finally, note that cubes and stars within the same $x$-plane are very weakly correlated in large system sizes: (\ref{cylcorrsoln}) is trivially verified to agree with (\ref{corrsoln}) in the thermodynamic limit.

\section{Periodic Boundary Conditions} \label{periodic}
Care must be exercised in evaluating (\ref{partition}) when using fully periodic boundary conditions. Whereas the constraints (\ref{cylindricalAproduct}) were entirely independent of each other, the corresponding constraints under boundary conditions periodic in all three cardinal directions are not so simple. 

In particular, let $P^{\mu}_i$ denote a set of simple cubes $c$ forming an $L \times L$ plane perpendicular to the direction $\mu$. Then, we have the $3L$ constraints:
\begin{equation}
    \label{Aplanar}
    \prod_{c \in P^{\mu}_i} A_c = \mathds{1}, \quad 1 \leq i \leq L, \quad \mu \in \{x,y,z\} .
\end{equation}
Note that, unlike the constraints (\ref{AplanarCylindrical}), these constraints do not all function independently from one another: for instance, two perpendicular planes of simple cubes share a common line of $L$ cubes at their intersection, and three perpendicular planes additionally share one common cube at their shared intersection. We will find as a consequence that properly counting the possible products in (\ref{Aproductexplicit}) with nonzero trace becomes far less trivial than in the previous two cases, and that the binomial theorem alone will prove insufficient for obtaining a closed-form partition function under this topology.

Similarly, let $\bar{P}^{\mu}_i$ denote any set of vertices $v$ forming a plane perpendicular to the direction $\mu$. We then also have the $3L$ constraints:
\begin{equation}
\label{Bplanar}
    \prod_{v \in \bar{P}^{\mu}_i} B^{\mu}_v = \mathds{1}, \quad 1 \leq i \leq L, \quad \mu \in \{x,y,z\} .
\end{equation}
These constraints will pose different difficulties than the cubic constraints (\ref{Aplanar}): while two perpendicular planes of $B^{\mu}_v$ operators will not share any common operators, the additional constraints (\ref{xyz=1}) will lead to many more possible trace-containing factors, further complicating the trace calculation. These difficulties, along with those from the constraints (\ref{Aplanar}), will render our previous strategies insufficient in finding a closed form solution to (\ref{partition}). 

Finally, we note that fully periodic boundary conditions leads to a very natural choice of the second sum in (\ref{hamiltonian}): we simply sum over all $L^3$ vertices in the lattice, as the elimination of hard boundaries allow us to properly define each $B^{\mu}_v$ for all vertices. Additionally, the counting of spins is trivial: each of the $L^3$ vertices is simply associated with the three edges located directly in the positive $x$, $y$, and $z$ directions from the vertex, giving the total number of qubits $N$ as simply $3L^3$.

\subsection{Partition Function} \label{sec:perpart}
We might attempt to simplify (\ref{partition2}) using (\ref{Aproduct}) as follows: the terms in (\ref{Aproductexplicit}) proportional to the identity are those for which (\ref{Aplanar}) can be used to eliminate products of $A_c$ operators -- that is, each product corresponding to some set of planes of cubes will be proportional to the identity. A priori, if a given linear term in (\ref{Aproductexplicit}) contains the products of $n$ $x$-planes, $m$ $y$-planes, and $\ell$ $z$-planes, it will carry a factor of $\Ta^{(n+m+\ell)L^2}$ corresponding to the $L^2$ cubes in each of the $n+m+\ell$ planes. However, we must be cautious when dealing with perpendicular planes: because each linear term in (\ref{Aproductexplicit}) may contain at most one of each $A_c$ operator, and two perpendicular planes share a common intersection of $L$ cubes, two perpendicular planes cannot each individually contribute $L^2$ factors of $\Ta$. Instead, using $A_c^2 = \mathds{1}$, we include no operators corresponding to the linear intersection of two given planes. Because each of the $nm +n\ell + m\ell$ intersections of planes are double-counted initially, we remove $2L$ factors of $\Ta$ at each intersection. Finally, at each of the $nm\ell$ intersections of three perpendicular planes, we must add back in the $A_c$ operator and its corresponding factor of $\Ta$: using $A_c = A_c^3$, only one $A_c$ is necessary to satisfy all three constraints (\ref{Aplanar}) for three perpendicular planes. Since these intersections are initially counted three times and then removed six times, we must add four factors of $\Ta$ back to the product for these $nm\ell$ planes. Therefore, each product corresponding to $n$ $x$-planes, $m$ $y$-planes, and $\ell$ $z$-planes carries a factor of:
\begin{equation}
    \Ta^{(n+m+\ell)L^2 - 2(nm +n\ell + m\ell)L + 4nm\ell} .
\end{equation}
Unfortunately, more care is needed than simply summing each of $n$, $m$, and $\ell$ from $0$ to $L$ with the appropriate binomial coefficients: because the constraints (\ref{Aplanar}) are not all independent of each other, a given trace-containing product of $A_c$ operators can be interpreted as one of several possible planar configurations. As a simple example, the product of all $L^3$ cubes can be interpreted as $L$ $x$-planes, or $L$ $y$-planes, or $L$ $z$-planes, or the intersection of all $3L$ planes. Indeed, while this particular example is trivial to see, there exist many more nontrivial examples of a given operator product having multiple interpretations in terms of the constraints (\ref{Aplanar}). As a result, the simple counting argument employed in Sections \ref{sec:openpart} and \ref{sec:cylpart} will lead to dramatic overcounting of the terms proportional to the identity. 

These same sort of difficulties are not present immediately in counting the trace-containing terms in (\ref{Bproduct}), but the resulting computation is no less difficult. We might attempt to proceed as follows: a given product corresponding to $n$ $x$-planes, $m$ $y$-planes, and $\ell$ $z$-planes of vertices will a priori contribute a factor of $\Tb^{(n+m+\ell)L^2}$. Then, each vertex in the lattice lies on exactly zero, one, two, or three planes. Each vertex lying on zero planes could contribute either zero or three additional factors of $\Tb$ by choosing to include factors of the form (\ref{xyz=1}). Each vertex lying on exactly one plane may contribute an additional factor of $\Tb$ by substituting $B^{\mu}_v$ contained in (\ref{Bplanar}) with $B^{\mu'}_v B^{\mu''}_v$ using (\ref{xy=z}). Each vertex lying on exactly two perpendicular planes may contribute one less factor of $\Tb$ using (\ref{xy=z}) in the same manner. Finally, each vertex lying on three perpendicular planes may contribute three fewer factors of $\Tb$ using (\ref{xyz=1}). The problem of counting the trace-containing terms of (\ref{Bproduct}) is then simply reduced to properly counting the number of vertices lying on each of zero, one, two, or three planes given $n$ $x$-planes, $m$ $y$-planes, and $\ell$ $z$-planes. Unfortunately, the constraints (\ref{xyz=1}) introduce the same overcounting problems as before: as a trivial example, the product of all $3L^3$ $B^{\mu}_v$ operators can be thought of as either the product of all $3L$ planes and zero vertices or of zero planes and of all $L^3$ vertices.

\subsection{Absence of Phase Transitions}
In place of a closed-form partition function, we will instead give a more general argument against the presence of finite temperature phase transitions in the model. Starting from (\ref{Aproductexplicit}), we gather all terms proportional to the identity and denote their prefactors by the function $\mathcal{T}_a$:
\begin{equation}
    \begin{split}
    \prod_c [\mathds{1} + A_c \Ta] &= \left[ 1 + a_1 \Ta + a_2 \Ta^2 + \ldots + a_{L^3} \Ta^{L^3} \right] \mathds{1} + \text{t.t.} \\
    &= \mathcal{T}_a \mathds{1} + \text{t.t.} \ .
    \end{split}
\end{equation}
$\mathcal{T}_a$ is a finite-order power series with nonnegative integer coefficients. Each coefficient $a_n$ represents the number of unique products of $n$ $A_c$ operators yielding the identity. We do the same for (\ref{Bproduct}), denoting the prefactors of the trace-containing terms by $\mathcal{T}_b$:
\begin{equation}
\label{calTb}
    \begin{split}
        \prod_{\mu,v} [\mathds{1} + B^{\mu}_v \Tb] &= \left[ 1 + b_1 \Tb + b_2 \Tb^2 + \ldots + b_{3L^3} \Tb^{3L^3} \right] \mathds{1} + \text{t.t.} \\
        &= \mathcal{T}_b \mathds{1} + \text{t.t.} \ .
    \end{split}
\end{equation}
As usual, the traceless terms may not combine to yield any terms containing a trace. Using the functions $\mathcal{T}_a$ and $\mathcal{T}_b$, the fully periodic partition function is given by:
\begin{equation}
    \begin{split}
    \mathcal{Z}_{\text{Periodic}} &= \Ca^{L^3} \Cb^{3L^3} \mathcal{T}_a \mathcal{T}_b \Tr [\mathds{1}] 
    = 2^{3L^3} \Ca^{L^3} \Cb^{3L^3} \mathcal{T}_a \mathcal{T}_b .
    \end{split}
\end{equation}
In order to investigate the possibility of a phase transition, we wish to understand how $\mathcal{T}_a$ and $\mathcal{T}_b$ behave in the thermodynamic ($L \rightarrow \infty$) limit. We start with $\mathcal{T}_a$ by noting that each $a_n$ must be bounded above by $2^{3L}$. Because each trace-containing term of (\ref{Aproductexplicit}) is reduced to the identity using the constraints (\ref{Aplanar}), each trace-containing term may be found using the $2^{3L}$ possible products of these $3L$ constraints. Although we know that each of these $2^{3L}$ products will not yield a unique trace-containing term, there can certainly be no greater than $2^{3L}$ trace-containing terms in (\ref{Aproductexplicit}), and therefore no greater than $2^{3L}$ trace-containing terms for each power of $\Ta^n$. Additionally, we note that $a_n = 0$ for $1 \leq n < L^2$, as there are no nontrivial trace-containing terms outside the order $L^2$ terms arising from (\ref{Aplanar}). We therefore find:
\begin{equation}
\label{Tabound}
    \begin{split}
    \mathcal{T}_a - 1 &= \sum_{n = 1}^{L^3} a_n \Ta^n 
    = \Ta^{L^2} \sum_{n = 0}^{L^3 - L^2} a_{n + L^2} \Ta^n 
    \leq \Ta^{L^2} 2^{3L} \sum_{n = 0}^{L^3 - L^2} \Ta^n .
    \end{split}
\end{equation}
We may now safely take the limit of $L \rightarrow \infty$: as $\Ta < 1$ for all nonzero temperatures, the above sum converges to a finite value while $\Ta^{L^2} 2^{3L}$ goes to zero. We therefore find that $\mathcal{T}_a = 1$ in the thermodynamic limit. This result is exactly as expected: it is simply a restatement of the idea that the partition function of (\ref{hamiltonian}) should be asymptotically the same in the thermodynamic limit for any choice of boundary conditions. 

Unfortunately, the same strategy does not follow as easily for $\mathcal{T}_b$. Because we expect the free energy to be asymptotically the same as (\ref{fopen}), we expect $\mathcal{T}_b$ to go asymptotically as:
\begin{equation} 
\label{Tblim}
    \mathcal{T}_b \sim [1 + \Tb^3]^{L^3} .
\end{equation}
Note the exponent of $L^3$ in place of $(L-1)^3$, arising from our definition of the second sum in (\ref{hamiltonian}) under periodic boundary conditions. Of course, this change has no effect on the free energy density in the thermodynamic limit. From this, we do not expect the coefficients $b_n$ in $\mathcal{T}_b$ to be bounded so easily as in (\ref{Tabound}). 

Instead, we examine the differences in the free energy density to that obtained by presupposing a large $L$ $\mathcal{T}_b$ limit of the form (\ref{Tblim}). That is, we wish to examine:
\begin{equation}
\label{fdiff}
    \begin{split}
    f_{\text{Periodic}}^{(0)} - f_{\text{Periodic}} &= \frac{1}{\beta L^3} \log \frac{\mathcal{Z}_{\text{Periodic}}}{\mathcal{Z}_{\text{Periodic}}^{(0)}} 
    = \frac{1}{\beta L^3} \log \frac{2^{3L^3} \Ca^{L^3} \Cb^{3L^3} \mathcal{T}_a \mathcal{T}_b}{2^{3L^3} \Ca^{L^3} \Cb^{3L^3} \mathcal{T}_a [1+\Tb^3]^{L^3}} \\
    &= \frac{1}{\beta L^3} \log \frac{\mathcal{T}_b}{[1+\Tb^3]^{L^3}} .
    \end{split}
\end{equation}
In order to bound $\mathcal{T}_b$, we perform the calculation described in section \ref{sec:perpart} without regard for possible overcounting. That is, we count every possible combination of planar and vertex products proportional to the identity without worrying about potentially multi-counting a particular configuration in order to obtain an upper bound. This is done as follows: given a product of $n$ $x$-planes, $m$ $y$-planes, and $\ell$ $z$-planes, there exist $nm\ell$ vertices lying on all three planes, $(nm + n\ell + m\ell)L - 3nm\ell$ vertices lying on exactly two planes, $(n+m+\ell)L^2 - 2(nm+n\ell+m\ell)L+3nm\ell$ vertices lying on exactly one plane, and $L^3-(n+m+\ell)L^2+(nm+n\ell+m\ell)L-nm\ell$ vertices lying on no planes. Each vertex lying on zero or three planes may contribute either a factor of $1$ or $\Tb^3$, and each vertex lying on exactly one or two planes may contribute either a factor of $\Tb$ or $\Tb^2$. We therefore find that $\mathcal{T}_b$ is bounded above by:
\begin{equation}
    \begin{split}
    \mathcal{T}_b &\leq \sum_{n = 0}^L \sum_{m = 0}^L \sum_{\ell = 0}^L {L \choose n} {L \choose m} {L \choose \ell} [1+\Tb^3]^{nm\ell} [\Tb + \Tb^2]^{(nm + n\ell + m\ell)L - 3nm\ell} \\
    \times& [\Tb\! +\! \Tb^2]^{(n+m+\ell)L^2 - 2(nm+n\ell+m\ell)L+3nm\ell} [1\! +\! \Tb^3]^{L^3-(n+m+\ell)L^2+(nm+n\ell+m\ell)L-nm\ell} \\
    &= [1\! +\! \Tb^3]^{L^3} \sum_{n = 0}^L \sum_{m = 0}^L \sum_{\ell = 0}^L {L \choose n} {L \choose m} {L \choose \ell} \left( \frac{\Tb+\Tb^2}{1+\Tb^3} \right)^{(n+m+\ell)L^2-(nm+n\ell+m\ell)L} .
    \end{split}
\end{equation}
The term in parenthesis is less than or equal to one for all temperatures, and its exponent is non-negative. Additionally, each binomial is strictly less than $2^L$. We therefore have:
\begin{equation}
    \begin{split}
    \mathcal{T}_b &\leq [1+\Tb^3]^{L^3} 2^{3L} \sum_{n = 0}^L \sum_{m = 0}^L \sum_{\ell = 0}^L 1 
    = [1+\Tb^3]^{L^3} 2^{3L}(L+1)^3 .
    \end{split}
\end{equation}
We also have that $\mathcal{T}_b$ is strictly larger than $[1+\Tb^3]^{L^3}$, as $\mathcal{T}_b$ contains all zero plane trace-containing terms. Dividing by $[1+\Tb^3]^{L^3}$ and raising to the power of $1/L^3$, we find:
\begin{equation}
    1 \leq \left( \frac{\mathcal{T}_b}{[1+\Tb^3]^{L^3}} \right)^{1/L^3} \leq 2^{3/L^2}(L+1)^{3/L^3} .
\end{equation}
In the thermodynamic limit, we see that the difference in free energies (\ref{fdiff}) goes to zero. Then, in that limit, the free energy density is simply given by:
\begin{equation}
    \begin{split}
    f_{\text{Periodic}} &= \lim_{L \rightarrow \infty} -\frac{1}{\beta L^3} \log (2^{3L^3}\Ca^{L^3}\Cb^{3L^3} \mathcal{T}_a \mathcal{T}_b) \\
    &= -\frac{1}{\beta} \left[ \log 8 +\log \Ca + \log (\Cb^3 + \Sb^3) \right] .
    \end{split}
\end{equation}
As expected, this is the same thermodynamic limit as in (\ref{fopen}) and (\ref{eq:fcyl2}), proving that the order $L^2$ constraints (\ref{Aplanar}) and (\ref{Bplanar}) do not matter in the thermodynamic limit, and that (\ref{hamiltonian}) under periodic boundary conditions cannot display a finite temperature phase transition.

\section{Dynamics of the X-Cube Model at Finite Temperature}
\label{sec:dynamics}
In addition to equilibrium correlation functions, we also wish to estimate the time dependent autocorrelation function of each $A_c$ and $B^{\mu}_v$. The autocorrelation function of a generic operator $X$ as a function of time $t$ is given by:
\begin{equation}
\label{autocorr}
\langle X(0) X(t) \rangle = \Tr \left[ \rho X(0) X(t) \right] = \Tr \left[ \rho X\, U^{\dagger}(t) X U(t)  \right] ,
\end{equation}
where $\rho$ is the density matrix of the  system, and $U$ is a unitary time-evolution operator. A priori, since both $A_c$ and $B^{\mu}_v$ commute with $H$, neither will evolve in time under the Heisenberg picture. To allow for thermalization, we therefore imagine perturbing (\ref{hamiltonian}) into a thermal Hamiltonian with an infinitesimal applied field given by:
\begin{equation}
H^{\text{therm}} = H - \lambda \sum_{ n} \sigma^z_{n} - \gamma \sum_{ n} \sigma^x_{n} ,
\end{equation}
where $\lambda > 0$ and $\gamma > 0$ determine field strengths, and the sum over $n$ includes  all edges.

Equation (\ref{autocorr}) is most easily estimated using the duality map defined by (\ref{openclassical}): explicitly, it defines an isometry realized by a unitary transformation $\mathcal{U}$, mapping each 
operator $X$ corresponding to the X-Cube model to the operator $X_{\sf d}$ in the dual representation via:
\begin{equation}
X \rightarrow X_{\sf d} = \mathcal{U}^{\dagger} X \mathcal{U} .
\end{equation}
While the dual representations given by (\ref{openclassical}) and (\ref{openclassical2}) are written in terms of classical variables, the same bond algebra is achieved if each classical variable is thought of as a $\sigma^z$ operator. This representation of the duality allows us to estimate (\ref{autocorr}) by calculating the corresponding autocorrelations of simple Ising models using Glauber dynamics \cite{doi:10.1063/1.1703954}:
\begin{equation}
\begin{split}
\langle X_{\sf d}(0) X_{\sf d}(t) \rangle &= \Tr \left[ {\rho}_{\sf d} X_{\sf d} \, {U}_{\sf d}^{\dagger}(t) 
X_{\sf d} {U}_{\sf d}(t) \right] \\
&= \Tr \left[ (\mathcal{U}^{\dagger} \rho \mathcal{U}) (\mathcal{U}^{\dagger} X \mathcal{U}) (\mathcal{U}^{\dagger} U(t) \mathcal{U}) (\mathcal{U}^{\dagger} X \mathcal{U}) (\mathcal{U}^{\dagger} U(t) \mathcal{U}) \right]  \\
&= \langle X(0) X(t) \rangle ,
\end{split}
\end{equation}
where ${\rho}_{\sf d}$ and ${U}_{\sf d}$ are the corresponding Ising density matrix and time-evolution operator. We may therefore find the time evolution of each $A_c$ and $B^{\mu}_v$ from the time evolution of their classical Ising duals. Although the following results are derived explicitly assuming open boundary conditions, corrections to these results due to boundary conditions such as (\ref{Aplanar}) or (\ref{Bplanar}) are of order $L^2$. We therefore expect that these results will accurately describe the bulk material of the system to very high order in the thermodynamic limit of very large $L$. 

The dynamics of Ising-type models are investigated thoroughly in \cite{doi:10.1063/1.1703954}, with important results cited here. The most important assumption of Glauber dynamics is that of detailed balance: given a system of classical spins $\{s_i\}$ and a corresponding Hamiltonian $H$, the transition rate $w_i(s)$ of each $i$th spin $s_i$ in the state $s$ is related to the probability $P(s_i=s)$ of finding the spin in that state by:
\begin{equation}
\label{detailedbalance}
\frac{P(s_i = s)}{P(s_i = -s)} = \frac{w_i(-s)}{w_i(s)} .
\end{equation}
That is, the ratio of the rates at which each spin transitions out of and into the state $s$, and correspondingly into and out of state $-s$, is equal to the ratio of probabilities of finding that spin in the state $s$ or $-s$ in the first place. 

Each probability in (\ref{detailedbalance}) is given by the corresponding Boltzmann factor. Utilizing the mappings (\ref{openclassical}) and (\ref{openclassical2}), we write the classical dual Hamiltonian as:

\begin{equation}
{H}_{\sf d} = \sum_{m = 1}^{L^3} H^A_m + \sum_{n = 1}^{(L-1)^3} H^B_n ,
\end{equation}
where $H^A_m$ and $H^B_n$ are simply given by:
\begin{equation}
H^A_m = -a r_m, \quad H^B_n = -b(s^n_1 s^n_2 + s^n_2 s^n_3 + s^n_3 s^n_1) .
\end{equation}
From (\ref{detailedbalance}), we see that the Glauber dynamics of each $r_m$ are governed solely by the effective Hamiltonian $H^A_m$, and those of each $s^n_j$ are governed solely by the effective Hamiltonian $H^B_n$, as each Boltzmann factor contribution from uncorrelated spins divides out on the right-hand side. This is simply a restatement of the result that (\ref{hamiltonian}) under open boundary conditions is dual to $L^3$ single spins in a magnetic field and $(L-1)^3$ three-site Ising chains. 

First, to find the autocorrelation function of each $A_c$, the expectation value of each $r_m$ is given by that of a single spin in a constant magnetic field. This problem is investigated in \cite{doi:10.1063/1.1703954}, with the following result:
\begin{equation}
\langle r_m(t) \rangle = \langle r_m(0) \rangle e^{-\alpha t} + (1-e^{-\alpha t})\Ta ,
\end{equation}
where $\alpha>0$ is a parameter corresponding to the transition probability between states. A priori, this parameter could be constant or may have a nontrival temperature dependence. Note that $\langle r_m(t) \rangle$ at long times corresponds with the result of (\ref{corrsoln}). We may find (\ref{autocorr}) using the law of total expectation:
\begin{equation}
\begin{split}
\langle r_m(0) r_m(t) \rangle &= \langle r_m(0) r_m(t) | r_m(0) = +1 \rangle P(r_m(0) = +1) \\
& \quad + \langle r_m(0) r_m(t) | r_m(0) = -1 \rangle P(r_m(0) = -1) \\
&= \left[ (+1) \left( (+1)e^{-\alpha t} + (1-e^{-\alpha t})\Ta \right) \right]\frac{e^{\beta a}}{e^{\beta a} + e^{-\beta a}} \\
& \quad + \left[ (-1) \left( (-1)e^{-\alpha t} + (1-e^{-\alpha t})\Ta \right) \right] \frac{e^{-\beta a}}{e^{\beta a} + e^{-\beta a}} ,
\end{split}
\end{equation}
where $\langle \cdot | \cdot \rangle$ here denotes conditional expectation, and the equilibrium probability of finding $r_m$ at $\pm 1$ is proportional to the corresponding Boltzmann factor. We therefore find our $A_c$ autocorrelation function:
\begin{equation}
\label{Aautocorr}
\langle A_c(0) A_c(t) \rangle = \langle r_m(0) r_m(t) \rangle = e^{-\alpha t} + (1-e^{-\alpha t})\Ta^2 .
\end{equation}

In the standard case of a constant, temperature independent, $\alpha$, the autocorrelation function (\ref{Aautocorr}) exhibits trivial high and low temperature limits. Specifically, for all $\beta$, the autocorrelation function (\ref{Aautocorr}) has an asymptotic long time decay set by $1/\alpha$. By contrast, the spin-spin correlations in conventional, infinite size, Ising chains and other systems become increasingly longer (in both spatial and temporal separation) as the temperature is lowered.
In finite size systems (and especially for effective single site systems such as those associated with the decoupled terms $H^A_m$ on a lattice with open boundary conditions), by construction, long range spatial correlations are impossible and long time autocorrelations are more readily destroyed by thermal fluctuations. Effectively, in the equations of \cite{doi:10.1063/1.1703954} determining the autocorrelations, the absence of coupling to spins that would appear in the infinite chain Ising model and are absent in the trivial finite size system can be emulated by setting $\beta=0$. The infinite chain case (typically examined by Glauber dynamics) leads to coupled differential equations whose solution exhibits a divergent correlation length (and time) as the temperature veers to zero. In our case, because the ``chain" is finite (an effective single site in our calculation for the $A_c$ autocorrelations), the correlation length is trivially bounded by the effective length of the system (a single spin $r_m$) so that the infinite set of recursive coupled equations that appear in the standard textbook Markov chain analysis for Ising chains is truncated and no divergent correlations appear (in space and thus also weaker correlations in time). In finite size systems, autocorrelations are weaker than those on infinite size lattices. Given that even infinite length Ising chains do not exhibit divergent memory times at any non-zero temperature, it is no surprise, then, that in (\ref{Aautocorr}) we find that the correlations do not persist on arbitrarily long time scales. Even if the basic elementary ``clock move'' time $1/\alpha$ of Glauber dynamics is adjusted to be of an activated Arrhenius form of an exponential in the inverse temperature, (\ref{Aautocorr}) will not display long time correlations. Exactly at zero temperature, the righthand side of (\ref{Aautocorr}) is identical unity; at all other positive temperatures, no long memory times appear. 

We find qualitatively similar behaviors for the $B^{\mu}_v$ operators. That is, similar to the $A_c$ autocorrelations, at finite temperatures, no long memory times appear in the autocorrelation function of $B^{\mu}_v$ operators. To explicitly compute the Glauber dynamics of the $B^{\mu}_v$ terms, we investigate the dynamics of the classical $s^n_j$ spins. The thermalization of each $s^n_j$ is not investigated explicitly in \cite{doi:10.1063/1.1703954}, but we may reproduce Glauber's arguments easily by applying his techniques to the effective Hamiltonian $H^B_n$. Because each $B^{\mu}_v$ is mapped to some $s^n_i s^n_j$ for $i \neq j$, we seek expectation values of the form $\langle s_i (t) s_j(t) \rangle$, which are given by the functions $r_{i,j}$ in \cite{doi:10.1063/1.1703954}. For the particular case of an ordinary Ising chain, each $r_{i,j}$ satisfies the differential equation in equation (31) of Glauber's paper:

\begin{equation}\hspace*{-0.2cm}
\frac{dr_{i,j}}{dt} = \alpha \left[ -2r_{i,j}(t) + \frac{1}{2}\mathsf{T}_{2b} \left( r_{i-1, j}(t) + r_{i+1,j}(t) + r_{i,j-1}(t) + r_{i,j+1}(t) \right) \right] ,
\end{equation}
where $\mathsf{T}_{2b} = \tanh(2\beta b)$. This differential equation holds a particularly simple form for the case of a periodic three-spin system: using $r_{i,i}(t) = \langle [s_i(t)]^2 \rangle = 1$ and $r_{i,j} = r_{j,i}$, we have for $r_{1,2}$:
\begin{equation}
\frac{dr_{1,2}}{dt} = \alpha \left[ -2r_{1,2}(t) + \frac{1}{2}\mathsf{T}_{2b}\left( r_{1,3}(t) + r_{2,3}(t) + 2 \right) \right] .
\end{equation}
Similar differential equations are found for $r_{1,3}$ and $r_{2,3}$ by exchanging $r_{1,2}$ with $r_{1,3}$ and $r_{2,3}$ respectively. These three equations yield a system of three linear inhomogeneous differential equations, most easily solved in matrix form:

\begin{equation}
\label{matrixODE} 
\frac{d}{dt} \begin{bmatrix}
r_{1,2}(t) \\ 
r_{2,3}(t) \\ 
r_{1,3}(t)
\end{bmatrix} = \begin{bmatrix}
-2\alpha & \frac{1}{2}\alpha \mathsf{T}_{2b} & \frac{1}{2}\alpha \mathsf{T}_{2b} \\ 
\frac{1}{2}\alpha \mathsf{T}_{2b}  & -2\alpha & \frac{1}{2}\alpha \mathsf{T}_{2b} \\ 
\frac{1}{2}\alpha \mathsf{T}_{2b} & \frac{1}{2}\alpha \mathsf{T}_{2b} & -2\alpha
\end{bmatrix} \begin{bmatrix}
r_{1,2}(t) \\ 
r_{2,3}(t) \\ 
r_{1,3}(t)
\end{bmatrix}  + \begin{bmatrix}
\alpha \mathsf{T}_{2b} \\ 
\alpha \mathsf{T}_{2b} \\ 
\alpha \mathsf{T}_{2b}
\end{bmatrix} .
\end{equation}
We solve this system by diagonalizing the matrix. Its eigenvalues and eigenvectors are given by:
\begin{equation}
\begin{split}
\lambda_1 &= -\alpha(2+\frac{1}{2}\mathsf{T}_{2b}), \quad \lambda_2 = -\alpha(2+\frac{1}{2}\mathsf{T}_{2b}), \quad \lambda_3 = -\alpha(2-\mathsf{T}_{2b}) ,\\
v_1 &= \begin{bmatrix}
1 \\ 
-1 \\ 
0
\end{bmatrix}, \quad \quad \quad \quad \ v_2 = \begin{bmatrix}
1 \\
0 \\
-1
\end{bmatrix}, \quad \quad \quad \quad \ v_3 = \begin{bmatrix}
1 \\
1 \\
1
\end{bmatrix} .
\end{split}
\end{equation}
In this eigenbasis, (\ref{matrixODE}) is given by:
\begin{equation}
\label{matrixODEUncoupled}
\frac{d}{dt}\begin{bmatrix}
X(t) \\
Y(t) \\
Z(t)
\end{bmatrix} = \begin{bmatrix}
\lambda_1 & 0 & 0 \\
0 & \lambda_2 & 0 \\
0 & 0 & \lambda_3
\end{bmatrix} \begin{bmatrix}
X(t) \\
Y(t) \\
Z(t)
\end{bmatrix} + \begin{bmatrix}
0 \\
0 \\
\alpha \mathsf{T}_{2b}
\end{bmatrix} ,
\end{equation}
where $X(t), Y(t),$ and $Z(t)$ are given by:
\begin{equation}
\begin{bmatrix}
r_{1,2}(t) \\ 
r_{2,3}(t) \\ 
r_{1,3}(t)
\end{bmatrix} = \begin{bmatrix}
v_1 & v_2 & v_3
\end{bmatrix} \begin{bmatrix}
X(t) \\
Y(t) \\
Z(t)
\end{bmatrix} .
\end{equation}
Equation (\ref{matrixODEUncoupled}) consists of three uncoupled ordinary differential equations, which are easily solved:
\begin{equation}
\begin{bmatrix}
X(t) \\
Y(t) \\
Z(t)
\end{bmatrix} = \begin{bmatrix}
X_0 e^{-\alpha(2+\frac{1}{2}\mathsf{T}_{2b})t} \\
Y_0 e^{-\alpha(2+\frac{1}{2}\mathsf{T}_{2b})t} \\
Z_0 e^{-\alpha(2-\mathsf{T}_{2b})t} + \frac{\mathsf{T}_{2b}}{2-\mathsf{T}_{2b}}
\end{bmatrix} .
\end{equation}
This yields the following solutions for the expectation values:
\begin{equation}
\begin{bmatrix}
r_{1,2}(t) \\ 
r_{2,3}(t) \\ 
r_{1,3}(t)
\end{bmatrix} = \begin{bmatrix}
(X_0 + Y_0)e^{-\alpha(2+\frac{1}{2}\mathsf{T}_{2b})t} + Z_0e^{-\alpha(2-\mathsf{T}_{2b})t} \\
-X_0 e^{-\alpha(2+\frac{1}{2}\mathsf{T}_{2b})t} + Z_0 e^{-\alpha(2-\mathsf{T}_{2b})t} \\
-Y_0 e^{-\alpha(2+\frac{1}{2}\mathsf{T}_{2b})t} + Z_0 e^{-\alpha(2-\mathsf{T}_{2b})t}
\end{bmatrix} + \frac{\mathsf{T}_{2b}}{2-\mathsf{T}_{2b}} \begin{bmatrix}
1 \\
1 \\
1
\end{bmatrix} .
\end{equation}
To determine the constants $X_0$, $Y_0$, and $Z_0$, we demand that each expectation value is equal to its initial value at $t=0$:
\begin{equation}
\begin{bmatrix}
r_{1,2}(0) \\ 
r_{2,3}(0) \\ 
r_{1,3}(0)
\end{bmatrix} = \begin{bmatrix}
X_0 + Y_0 + Z_0 \\
-X_0 + Z_0 \\
-Y_0 + Z_0
\end{bmatrix} + \frac{\mathsf{T}_{2b}}{2-\mathsf{T}_{2b}} \begin{bmatrix}
1 \\
1 \\
1
\end{bmatrix} .
\end{equation}
Solving this system, we find the solution for $r_{1,2}$ is given by:
\begin{equation}
\label{r12}
\begin{split}
r_{1,2}(t) =& \frac{1}{3}[2r_{1,2}(0) - r_{2,3}(0) - r_{1,3}(0)] e^{-\alpha(2+\frac{1}{2}\mathsf{T}_{2b})t} \\
{}& + \frac{1}{3}[r_{1,2}(0) + r_{2,3}(0) + r_{1,3}(0)] e^{-\alpha(2-\mathsf{T}_{2b})t} \\
{}& + (1-e^{-\alpha(2-\mathsf{T}_{2b})t}) \frac{\mathsf{T}_{2b}}{2-\mathsf{T}_{2b}} .
\end{split}
\end{equation}
Similar expressions are given for $r_{2,3}$ and $r_{1,3}$ by respectively exchanging $r_{1,2}(0)$ and $r_{2,3}(0)$ or $r_{1,2}(0)$ and $r_{1,3}(0)$. Note that we may rewrite the final term of (\ref{r12}) as follows:
\begin{equation}
\label{T2b}
\begin{split}
\frac{\mathsf{T}_{2b}}{2-\mathsf{T}_{2b}} &= \left( \frac{e^{2\beta b} - e^{-2\beta b}}{e^{2\beta b} + e^{-2\beta b}} \right) \left( \frac{e^{2\beta b} + e^{-2\beta b}}{2(e^{2\beta b} + e^{-2\beta b}) - e^{2\beta b} + e^{-2\beta b}} \right) \\
&= \frac{e^{2\beta b} - e^{-2\beta b}}{e^{2\beta b} + 3e^{-2\beta b}}
= \frac{e^{3\beta b} - e^{-\beta b}}{e^{3\beta b} + 3e^{-\beta b}} \\
&= \frac{\Tb + \Tb^2}{1+\Tb^3} .
\end{split}
\end{equation}
In this way, we see that (\ref{r12}) also coincides at long times with the result of (\ref{corrsoln}).

Since this expression doesn't assume knowledge of the initial conditions $r_{i,j}(0)$, it can be used to calculate conditional expectation values as well. To finally obtain each autocorrelation function, we again employ the law of total expectation:
\begin{equation}
\label{r12auto}
\begin{split}
&\langle s_1(0) s_2(0) s_1(t) s_2(t) \rangle = \\
& \quad \quad \langle s_1(0)s_2(0)s_1(t)s_2(t) | s_1(0) s_2(0) = +1\rangle P (s_1(0) s_2(0)= +1) \\
& \quad \quad +  \langle s_1(0)s_2(0)s_1(t)s_2(t) | s_1(0) s_2(0) = -1\rangle P (s_1(0) s_2(0)= -1) .
\end{split}
\end{equation}
Each probability is given by the respective Boltzmann constants, as determined by the effective Hamiltonian $H^B_n$. In this Hamiltonian, $H^B_n$ attains a value of $-3b$ for the two configurations corresponding to $s_1 = s_2 = s_3 = \pm 1$, and attains a value of $b$ for all six remaining configurations. The former two states and two of the six latter states correspond to $s_1(0) s_2(0) = +1$, while the remaining four correspond to $s_1(0) s_2(0) = -1$. Therefore the two probabilities are:
\begin{equation}\hspace*{-0.28cm}
P(s_1(0) s_2(0) = +1) = \frac{e^{3\beta b} + e^{-\beta b}}{e^{\beta b} + 3e^{-\beta b}} , \ P(s_1(0) s_2(0) = -1) = \frac{2e^{-\beta b}}{e^{3\beta b} + 3e^{-\beta b}} .
\end{equation}
Since $r_{1,2}(t) = \langle s_1(t)s_2(t) \rangle$, $r_{1,2}(0)$ is determined immediately by the conditionals $s_1(0)s_2(0) = \pm 1$. Additionally, note that $r_{2,3}(0)$ and $r_{1,3}(0)$ will take on different values under these two conditionals: in particular, if $s_1(0)s_2(0) = +1$, then $H^B_n = - 3b$ in the two configurations for which $s_3(0) = s_2(0)$, and $H^B_n = b$ for the two configurations in which $s_3(0) \neq s_2(0)$. On the other hand, $H^B_n = b$ for all four configurations of spins in which $s_1(0) s_2(0) = -1$, giving $s_2(0) s_3(0)$ an expectation value of zero. Letting $r_{i,j}^{\pm}$ denote the conditional expectation of $s_i s_j$ given $s_1 s_2 = \pm 1$, we arrive at the result:
\begin{equation}
r_{2,3}^+(0) = r_{1,3}^+(0)= \frac{e^{3\beta b} - e^{-\beta b}}{e^{3\beta b} + e^{-\beta b}}, \quad r_{2,3}^-(0) = r_{1,3}^-(0) = 0 .
\end{equation}
Utilizing (\ref{T2b}), we therefore evaluate (\ref{r12auto}) as:
\begin{equation}
\label{AutoBBss}
\begin{split}
&\hspace*{-2cm}\langle s_1(0) s_2(0) s_1(t) s_2(t) \rangle = \frac{2}{3} e^{-\alpha(2+\frac{1}{2}\mathsf{T}_{2b})t} + \frac{1}{3}e^{-\alpha(2-\mathsf{T}_{2b})t} \\
& \quad + \frac{2}{3} \left( e^{-\alpha(2-\mathsf{T}_{2b})t} - e^{-\alpha(2+\frac{1}{2}\mathsf{T}_{2b})t} \right) \left( \frac{\Tb + \Tb^2}{1+\Tb^3} \right)  \\
&\quad + (1 - e^{-\alpha(2-\mathsf{T}_{2b})t}) \left( \frac{\Tb + \Tb^2}{1+\Tb^3} \right)^2 .
\end{split}
\end{equation}
Perusing (\ref{AutoBBss}), we indeed see that, similar to (\ref{Aautocorr}), for Glauber dynamics on a lattice with open boundary conditions, the autocorrelation function for the $B^{\mu}_v$ operators indeed does not exhibit divergent autocorrelation times. Even if $1/\alpha$ is made to be exponential in the inverse temperature (as in activated finite temperature dynamics), the correlations of (\ref{AutoBBss}) decay with a finite lifetime at all non-zero temperatures. 

We next analyze a $p$-state generalization of the X-Cube model (the ``$p$X-Cube model'').

\section{Equilibrium Thermodynamics of the $p$X-Cube Model}
\label{sec:p}
We may easily generalize the preceding discussion of the ordinary X-Cube model to an X-Cube model built upon $\mathbb{Z}_p$ clock and shift operators, see, e.g., \cite{Slagle17, vijay_isotropic_2017}. That is, rather than considering qubits at each edge of an $L \times L \times L$ lattice, we consider $p$-qudits with associated $p$-dimensional Hilbert spaces $\mathcal{H}_n$, forming a total state space $\bigotimes_{n=1}^N \mathcal{H}_n$. In place of the ordinary Pauli operators $\sigma^x_n$ and $\sigma^z_n$ acting on each $n$th Hilbert space, following the quantum clock operators of \cite{clock} (see also the subsequent work of \cite{Paul}), one may generalize the qubits of the X-Cube model to the $p$-clock operators $X_n$ and $Z_n$ (which are the $V$ and $U$ operators of Reference 
\cite{clock}, respectively). These are traceless operators with eigenvalues $\omega^m$ for $\omega = e^{2\pi i / p}$ and $0 \leq m \leq p-1$. $X_n$ and $Z_n$ are further defined by their relation:
\begin{equation}
\label{XZcomm}
X_n Z_n = \omega Z_n X_n, \quad X_n Z_m = Z_m X_n \text{ for } n\neq m .
\end{equation}
Unlike the ordinary Pauli operators, $X_n$ and $Z_n$ are not Hermitian for $p > 2$. However, it is easy to verify that they are unitary: in a given $\mathcal{H}_n$, we may diagonalize either $X_n$ or $Z_n$ (but not both simultaneously) as $\diag(1,\omega,\ldots \omega^{p-1})$, in which case $X_n^{\dagger}$ or $Z_n^{\dagger}$ is given by $\diag(1, \overline{\omega},\ldots \overline{\omega}^{p-1})$, where the bar denotes complex conjugation. We therefore quickly see that $X_n^{\dagger} X_n = Z_n^{\dagger} Z_n = \mathds{1}$. Using this result and (\ref{XZcomm}), we may also derive the relations:
\begin{equation}\hspace*{-0.3cm}
X_n X_n^{\dagger} Z_n = Z_n X_n X_n^{\dagger} = \overline{\omega}X_n Z_n X_n^{\dagger}, \quad X_n Z_n^{\dagger} Z_n = Z_n^{\dagger} Z_n X_n = \overline{\omega} Z_n^{\dagger} X_n Z_n .
\end{equation}
From above, and additionally Hermitian conjugating (\ref{XZcomm}), we find:
\begin{equation}
\label{XZcomm2}
X_n^{\dagger} Z_n = \overline{\omega} Z_n X_n^{\dagger}, \quad X_n Z_n^{\dagger} = \overline{\omega} Z_n^{\dagger} X_n, \quad X_n^{\dagger} Z_n^{\dagger} = \omega Z_n^{\dagger} X_n^{\dagger} .
\end{equation}
Additionally, it can easily be verified that $X_n Z_n$ is traceless by taking the trace of both sides of (\ref{XZcomm}), and both $X_n^m$ and $Z_n^m$ can be verified to be traceless for $1 \leq m \leq p-1$ by computing in their respective diagonal bases.

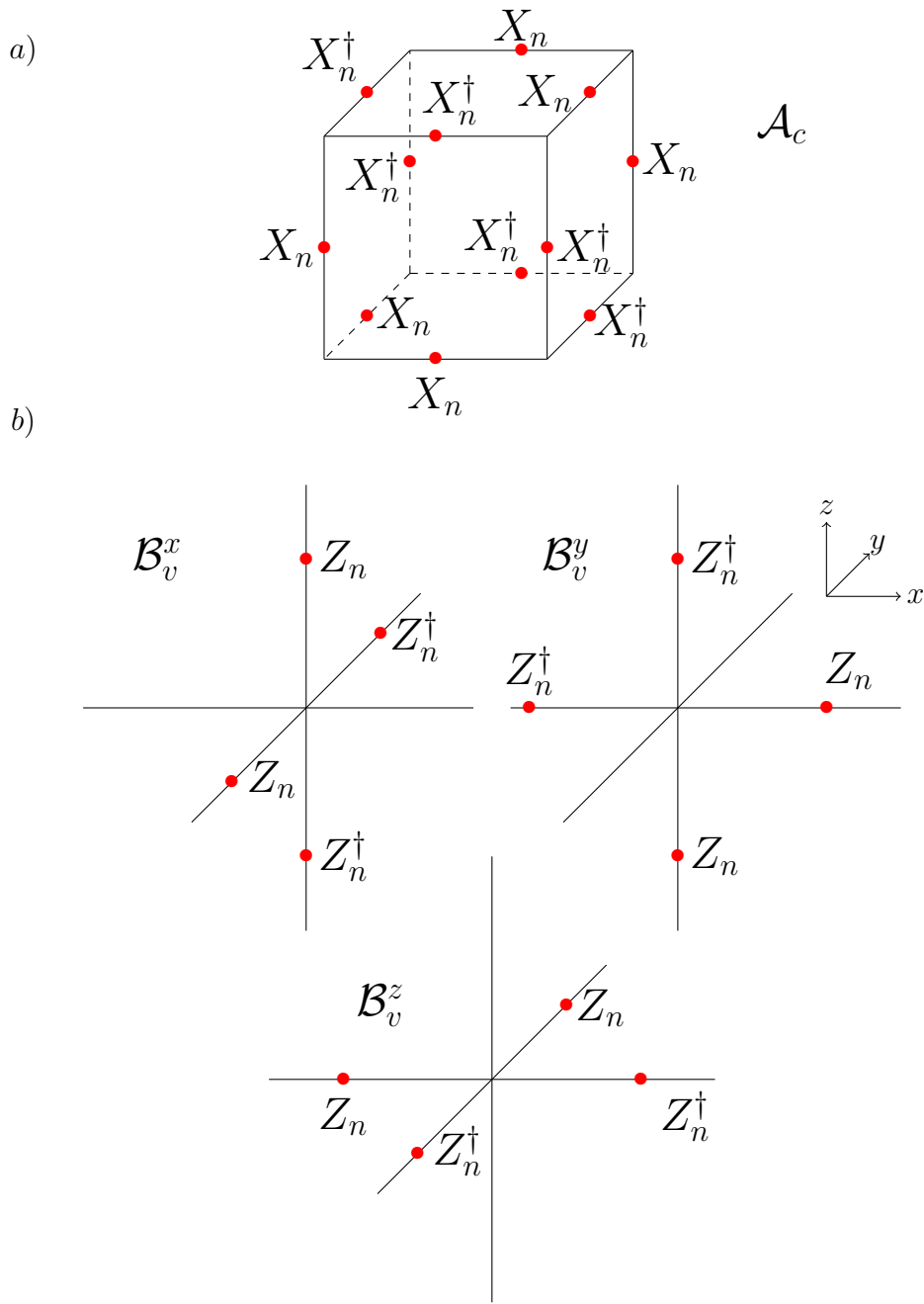
\begin{figure}
\vspace*{-1.5cm}
\begin{tikzpicture}
\node at (-5.2,3,0) {$a)$};
\node at (-5.2,-2,0) {$b)$};
\draw (0, 3, 0) -- (0, 3, 3);
\draw (0, 0, 3) -- (0, 3, 3);
\draw (0, 0, 3) -- (3, 0, 3);
\draw (3, 0, 0) -- (3, 0, 3);
\draw (0, 3, 0) -- (3, 3, 0);
\draw (3, 0, 0) -- (3, 3, 0);
\draw (0, 3, 3) -- (3, 3, 3);
\draw (3, 0, 3) -- (3, 3, 3);
\draw (3, 3, 0) -- (3, 3, 3);
\draw [dashed] (0, 0, 0) -- (0, 0, 3);
\draw [dashed] (0, 0, 0) -- (0, 3, 0);
\draw [dashed] (0, 0, 0) -- (3, 0, 0);

\foreach \x in {0,3} {
	\foreach \y in {0,3} {
		\node [red] at (1.5, \x, \y) {\textbullet};
		\node [red] at (\x, 1.5, \y) {\textbullet};
		\node [red] at (\x, \y, 1.5) {\textbullet};
	}
}
\node at (1.5,-0.5,3) {\Large $X_n$};
\node at (-0.5, 1.5, 3) {\Large $X_n$};
\node at (3.5, 1.5, 3) {\Large $X_n^{\dagger}$};
\node at (3.5, 1.5, 0) {\Large $X_n$};
\node at (-0.5, 1.3, 0) {\Large $X_n^{\dagger}$};
\node at (4, 0.5, 3) {\Large $X_n^{\dagger}$};
\node at (0.5, 0, 1.5) {\Large $X_n$};
\node at (1.1, 0.5, 0) {\Large $X_n^{\dagger}$}; 
\node at (-0.5,3.5,1.5) {\Large $X_n^{\dagger}$};
\node at (1.5, 3.3, 2.5) {\Large $X_n^{\dagger}$};
\node at (1.5, 3.3, 0) {\Large $X_n$};
\node at (2.4, 3, 1.5) {\Large $X_n$};

\node at (5,2,0) {\Large $\mathcal{A}_c$};
\end{tikzpicture}

\begin{center}
\begin{tikzpicture}
\draw (2.25, 0, 0) -- (-3, 0, 0);
\draw (0,3,0) --(0,-3,0);
\draw (0,0,4) -- (0,0,-4);
\foreach \x in {-2,2} {
	\node [red] at (0, \x, 0) {\textbullet};
	\node [red] at (0, 0, 1.3*\x ) {\textbullet};
}
\node at (.5, 0, -2.5) {\Large $Z_n^{\dagger}$};
\node at (.5, 0, 2.5) {\Large $Z_n$};
\node at (.5, -2, 0) {\Large $Z_n^{\dagger}$};
\node at (.5, 2, 0) {\Large $Z_n$};

\draw (8, 0, 0) -- (2.75, 0, 0);
\draw (5,3,0) --(5,-3,0);
\draw (5,0,4) -- (5,0,-4);
\foreach \x in {-2,2} {
	\node [red] at (\x+5, 0, 0) {\textbullet};
	\node [red] at (5, \x, 0) {\textbullet};
}
\node at (7.3, .5, 0) {\Large $Z_n$};
\node at (3, .5, 0) {\Large $Z_n^{\dagger}$};
\node at (5.5, -2, 0) {\Large $Z_n$};
\node at (5.5, 2, 0) {\Large $Z_n^{\dagger}$};

\draw (5.5, -5, 0) -- (-0.5, -5, 0);
\draw (2.5,-2,0) --(2.5,-8,0);
\draw (2.5,-5,4) -- (2.5,-5,-4);
\foreach \x in {-2,2} {
	\node [red] at (\x+2.5, -5, 0) {\textbullet};
	\node [red] at (2.5, -5, 1.3*\x ) {\textbullet};
}
\node at (5.1, -5.5, 0) {\Large $Z_n^{\dagger}$};
\node at (3, -5, -2.5) {\Large $Z_n$};
\node at (0.5, -5.5, 0) {\Large $Z_n$};
\node at (3, -5, 2.5) {\Large $Z_n^{\dagger}$};

\draw [->] (7,1.5,0) -- (8,1.5,0);
\node at (8.2,1.5,0) {$x$};
\draw [->] (7,1.5,0) -- (7,2.5,0);
\node at (7,2.7,0) {$z$};
\draw [->] (7,1.5,0) -- (7,1.5,-1.5);
\node at (7, 1.5, -1.8) {$y$};

\node at (-2,2,0) {\Large $\mathcal{B}^x_v$};
\node at (3.5,2,0) {\Large $\mathcal{B}^y_v$};
\node at (1,-4,0) {\Large $\mathcal{B}^z_v$};
\end{tikzpicture}
\end{center}
\caption{$p$-qudits are marked as red bullets. $a)$ A simple cube representing an $\calA$ operator, with operators $X_n$ used in constructing $\calA$. $b)$ The three $\calB$ operators associated with a given vertex, with  the respective operators $Z_n$ used in constructing each $\calB$.}
\label{calB}
\end{figure}

To generalize our $A_c$ and $B^{\mu}_v$ operators to arbitrary dimension $p$, we construct the operators $\calA$ and $\calB$ as depicted in figure \ref{calB}. Note that each $\calA$ and $\calB$ operator commutes: for any given cube $c$ containing a vertex $v$, $\calA$ and any of $\calB$ share two common qudits. On one qudit, $\calA$ and $\calB$ utilize either $X_n$ and $Z_n$, or $X_n^{\dagger}$ and $Z_n^{\dagger}$. On the other qudit, $\calA$ and $\calB$ utilize either $X_n$ and $Z_n^{\dagger}$, or $X_n^{\dagger}$ and $Z_n$. When $\calA$ and $\calB$ are interchanged, the former will yield a factor of $\omega$ while the latter will yield a factor of $\overline{\omega}$, yielding an overall factor of 1.

Given these $\calA$ and $\calB$ operators, we define the corresponding $p$-clock X-Cube, or $p$X-Cube,  Hamiltonian as:
\begin{equation}
\label{pH}
H_p = - a \sum_c (\calA + \calA^{\dagger}) - b\sum_{\mu,v} (\calB + (\calB)^{\dagger}) .
\end{equation}
As in (\ref{hamiltonian}), the first sum is over each elementary cube $c$, while the second sum is over vertices $v$ and cardinal directions $\mu \in \{x,y,z\}$. Note that $\calA^{\dagger}$ and $(\calB)^{\dagger}$ are included in (\ref{pH}) to ensure its Hermiticity. Since each operator in (\ref{pH}) commutes, we may write the corresponding partition function as:
\begin{equation}
\label{Zp}\hspace*{-0.3cm}
\mathcal{Z}_p = \Tr \left[ \prod_{c} \left( \exp(\beta a \calA) \right) \left( \exp(\beta a \calA^{\dagger} \right) \prod_{\mu,v} \left( \exp(\beta b \calB) \right) \left( \exp(\beta b (\calB)^{\dagger} \right) \right] .
\end{equation}
In the ordinary Pauli case of $p = 2$, we evaluated this sum by noting that $e^{\beta \sigma} = \mathds{1}\cosh(\beta) + \sigma\sinh(\beta)$ for some operator $\sigma$ with $\sigma^2 = \mathds{1}$. Here, the case is not so simple: let $\Sigma$ be an operator with $\Sigma^p = \mathds{1}$. Then, we have:
\begin{align*}
e^{\beta \Sigma} &= \sum_{n=0}^{\infty} \frac{(\beta \Sigma)^n}{n!}
= \sum_{m = 0}^{p-1} \sum_{n=0}^{\infty} \Sigma^m \frac{\beta^{pn+m}}{(pn+m)!}
= \sum_{m = 0}^{p-1} \Sigma^m Q^p_m(\beta) ,
\end{align*}
where we have defined the functions $Q^p_m$ via power series and in closed form:
\begin{equation}
\label{Q}
Q^p_m(x) = \sum_{n=0}^{\infty} \frac{x^{pn+m}}{(pn+m)!} = \frac{1}{p}\sum_{n = 0}^{p-1} \exp \left[ x \, \omega^n - i \frac{2\pi nm}{p}\right], 
\end{equation}
$0 \leq m \leq p-1$. These functions are natural generalizations of $\cosh$ and $\sinh$: for $p = 2$, $Q^2_0$ is simply $\cosh$ and $Q^2_1$ is $\sinh$.

Using these results in (\ref{Zp}), we have:
\begin{equation}
\label{Zp2}
\begin{split}
\mathcal{Z}_p = \Tr & \left[ \prod_c \left( \sum_{m=0}^{p-1} \calA^m Q^p_m(\beta a) \right) \left( \sum_{m=0}^{p-1} (\calA^{\dagger})^m Q^p_m(\beta a) \right) \right. \\
& \times \left. \prod_{\mu,v} \left( \sum_{m=0}^{p-1} (\calB)^m Q^p_m(\beta b) \right) \left( \sum_{m=0}^{p-1} ((\calB)^{\dagger})^m Q^p_m(\beta b) \right) \right] .
\end{split}
\end{equation} 
We may simplify this expression further: since $\calA^p = \calA \calA^{p-1} = \mathds{1}$, we have that $\calA^{\dagger} = \calA^{p-1}$, and we can rewrite the first two summation products (and the latter two similarly) as: 
\begin{equation} \hspace*{-0.38cm} 
\left( \sum_{m = 0}^{p-1} \calA^m Q^p_m(\beta a) \right) \left( \sum_{n = 0}^{p-1} (\calA^{\dagger})^n Q^p_n(\beta a) \right) = \sum_{m = 0}^{p-1} \sum_{n = 0}^{p-1} \calA^{m-n} Q^p_m(\beta a) Q^p_n(\beta a) .
\end{equation}
Because this sum includes negative powers of $\calA$, it contains redundancies: $\calA^{-n}$ is the same as $\calA^{p-n}$. We therefore wish to rewrite this sum in terms of only positive powers of $\calA$. Toward this end, we first split the sum into three components:
\begin{equation} 
\begin{split}\hspace*{-0.5cm} 
\sum_{m = 0}^{p-1} \sum_{n = 0}^{p-1} \calA^{m-n} Q^p_m(\beta a) Q^p_n(\beta a) = \sum_{m = 1}^{p-1} \calA^{-m} \sum_{n = 0}^{p-1-m}Q^p_n(\beta a)Q^p_{m+n}(\beta a)  \\
 + \mathds{1} \sum_{n=0}^{p-1}[Q^p_n(\beta a)]^2 + \sum_{m=1}^{p-1} \calA^m \sum_{n=0}^{p-1-m} Q^p_{n+m}(\beta a)Q^p_n(\beta a) .
\end{split}
\end{equation}
Then, we rewrite the first sum as a sum over positive powers of $\calA$:
\begin{equation}
\sum_{m = 1}^{p-1} \calA^{-m} \sum_{n = 0}^{p-1-m}Q^p_n(\beta a)Q^p_{m+n}(\beta a) = \sum_{m=1}^{p-1} \calA^m \sum_{n=0}^{m-1} Q^p_n(\beta a) Q^p_{p-m+n}(\beta a) .
\end{equation}
Finally, we recombine powers of $\calA$:
\begin{equation}
\begin{split}\hspace*{-0.5cm} 
\sum_{m = 0}^{p-1} \sum_{n = 0}^{p-1} & \calA^{m-n}  Q^p_m(\beta a) Q^p_n(\beta a) = \mathds{1} \sum_{n=0}^{p-1}[Q^p_n(\beta a)]^2 \\
{}& \hspace*{-0.5cm} + \sum_{m=1}^{p-1} \calA^m \left[ \sum_{n=0}^{p-1-m} Q^p_{n+m}(\beta a)Q^p_n(\beta a) + \sum_{n=0}^{m-1} Q^p_n(\beta a) Q^p_{p-m+n}(\beta a) \right] .
\end{split}
\end{equation}
In order to proceed, we once again introduce new functions to manage the algebra. Let $R^p_m$ be defined by:
\begin{equation}
\label{R}\hspace*{-0.5cm} 
R^p_m(x) = \begin{cases}
\displaystyle\sum_{n=0}^{p-1} [Q^p_n(x)]^2 , & m = 0 \\
\displaystyle\sum_{n=0}^{p-1-m} Q^p_{n+m}(x) Q^p_n(x) + \displaystyle\sum_{n=0}^{m-1} Q^p_n(x) Q^p_{p-m+n}(x), & 1 \leq m \leq p-1
\end{cases} .
\end{equation}
We may then rewrite (\ref{Zp2}) as: 
\begin{equation}
\label{Zp3}
\mathcal{Z}_p = \Tr \left[ \prod_{c,\mu,v} \left( \sum_{m = 0}^{p-1} \calA^m R^p_m(\beta a) \right) \left( \sum_{n = 0}^{p-1} (\calB)^n R^p_n(\beta b) \right) \right] .
\end{equation}
In defining the functions $R^p_m$, we have somewhat obscured the meaning of our calculation, but we have made the computation significantly easier by reducing the number of constraints needed to consider. As written in (\ref{Zp3}), each product term for a given $c$ or $(\mu, v)$ contains a sum over all powers zero through $p-1$ of the respective operator $\calA$ or $\calB$. The total product expansion will therefore contain exactly one linear term for every possible combination of operators $\calA$ and $\calB$ with each operator raised to a power zero through $p-1$. All terms not proportional to the identity will be traceless. Had we attempted to expand directly from (\ref{Zp2}), this product would also contain negative powers of $\calA$ and $\calB$, and the number of possible product combinations proportional to the identity would be significantly larger. From here, the calculation will depend on the choice of boundary conditions.

\subsection{Open Boundary Conditions}
Under open boundary conditions, no product of $\calA$ operators can yield the identity. As can be seen in figure \ref{calB}a, each $\calA$ can only be canceled by adjacent $\calA$ operators. But for product of $\calA$ operators forming a connected section of elementary cubes, the $X_n$ operators lying at its boundary will appear in the product only once. We may therefore immediately reduce (\ref{Zp3}) to the form:
\begin{equation}
\label{Zpopen}
\begin{split}
\mathcal{Z}_{p,\text{Open}} &= \Tr \left[ \prod_{c,\mu,v} \left( \mathds{1} R^p_0(\beta a) \right) \left( \sum_{n = 0}^{p-1} (\calB)^n R^p_n(\beta b) \right) \right] \\
&= [R^p_0(\beta a)]^{L^3} \Tr \left[ \prod_{\mu, v} \left( \sum_{n = 0}^{p-1} (\calB)^n R^p_n(\beta b) \right) \right] .
\end{split}
\end{equation}
As in the $p=2$ case, the remaining sum is performed over only the $(L-1)^3$ interior vertices of the system. Additionally, we see from figure \ref{calB} that the constraint (\ref{xyz=1}) carries over to the general case. However, because each $\calB$ may now be raised to nontrivial powers, we have the more general condition:
\begin{equation}
\label{xpypzp=1}
(\mathcal{B}^x_v)^m (\mathcal{B}^y_v)^m (\mathcal{B}^z_v)^m = \mathds{1}, \quad 0 \leq m \leq p-1 .
\end{equation}
Just as in the $p=2$ case, the only products of $\calB$ operators proportional to the identity are those satisfying (\ref{xpypzp=1}). Therefore, we obtain all terms proportional to the identity by choosing, for each of $(L-1)^3$ vertices, the power to which (\ref{xpypzp=1}) is given. Each product of the form (\ref{xpypzp=1}) carries a factor of $[R^p_m(\beta)]^3$. Therefore, we may expand the remaining product as:
\begin{equation}
\prod_{\mu, v} \left( \sum_{n = 0}^{p-1} (\calB)^n R^p_n(\beta b) \right) = \left[ \sum_{n = 0}^{p-1} [R^p_n(\beta b)]^3 \right]^{(L-1)^3} \mathds{1} + \text{t.t.}.
\end{equation}
As before, the number $N$ of $p$-qudits in our system is given by $3L^3 + 6L^2 + 3L$. Because $\Tr[\mathds{1}]$ is given by $p^N$, (\ref{Zpopen}) is finally given by:
\begin{equation}
\label{Zpopen2}
\mathcal{Z}_{p,\text{Open}} = p^{3L^3 + 6L^2 + 3L} [R^p_0(\beta a)]^{L^3} \left[ \sum_{n = 0}^{p-1} [R^p_n(\beta b)]^3 \right]^{(L-1)^3} .
\end{equation}
As before, we remark that the thermodynamic properties of the model captured by this partition function will accurately describe the bulk material in the thermodynamic limit of very large $L$, regardless of our choice of boundary conditions: any corrections due to constraints such as (\ref{Aplanar}) and (\ref{Bplanar}) will appear only at orders $L^2$ and higher.

If $p=2$, we expect this expression to yield the prior solution (\ref{opensoln}). Indeed, $Q^2_0(x)$ is simply $\cosh(x)$ and $Q^2_1(x)$ is $\sinh(x)$. $R^2_0$ and $R^2_1$ are therefore given by:
\begin{equation}
\label{R2}
\begin{split}
R^2_0(x) &= [\cosh(x)]^2 + [\sinh(x)]^2 = \cosh(2x) , \\
R^2_1(x) &= \sinh(x)\cosh(x) + \cosh(x)\sinh(x) = \sinh(2x) .
\end{split}
\end{equation}
By substituting into (\ref{Zpopen2}), we obtain:
\begin{equation}
\mathcal{Z}_{p=2,\text{Open}} = 2^{3L^3 + 6L^2 + 3L} \mathsf{C}_{2a}^{L^3} \left[ \mathsf{C}_{2b}^3 + \mathsf{S}_{2b}^3 \right]^{(L-1)^3} .
\end{equation}
This matches exactly with our prior solution (\ref{opensoln}): for $p = 2$, we have $\calA^{\dagger} = \calA$ and $(\calB)^{\dagger} = \calB$, so (\ref{pH}) is identical to (\ref{hamiltonian}) with $a \rightarrow 2a$ and $b \rightarrow 2b$.

\subsection{Cylindrical Boundary Conditions}
Just as in the $p=2$ case, we gain two additional constraints by imposing cylindrical boundary conditions: letting our system become periodic in the $y$ and $z$ directions, we find:
\begin{equation}
\label{pxplanes}
\prod_{c \in P^x_i} \calA^m = \mathds{1}, \quad \prod_{v \in \bar{P}^x_i} (\mathcal{B}^x_v)^m = \mathds{1}, \quad 0 \leq m \leq p-1 ,
\end{equation}
where $P^x_i$ refers to a plane of elementary cubes perpendicular to the $x$-direction in the first equality, and $\bar{P}^x_i$ refers to a plane of vertices perpendicular to the $x$-direction in the second equality. Once again, the only products of $\calA$ and $\calB$ operators proportional to the identity are those satisfying (\ref{xpypzp=1}) and/or (\ref{pxplanes}). Therefore, the products of $\calA$ operators in (\ref{Zp3}) proportional to the identity are found by choosing, for each of $L$ $x$-planes, the multiplicity $m$ of the entire plane included in the product:
\begin{equation}
\prod_c \left( \sum_{m = 0}^{p-1} \calA^m R^p_m(\beta a) \right) = \left[ \sum_{m = 0}^{p - 1} [R^p_m(\beta a)]^{L^2} \right]^L \mathds{1} + \text{t.t.} .
\end{equation}
Handling the $\calB$ operators is slightly more complex: first, for any $x$-plane of vertices not satisfying (\ref{pxplanes}), we must choose for each vertex to satisfy (\ref{xpypzp=1}) for $0 \leq m \leq p-1$. Second, for any vertex on an $x$-plane satisfying (\ref{pxplanes}), we may choose to replace $(\mathcal{B}^x_v)^m$ with $(\mathcal{B}^y_v \mathcal{B}^z_v)^{p-m}$. We therefore construct each $\calB$ product proportional to the identity as follows: first, for each of $L-1$ $x$-planes, we choose the degree $m$ with $0 \leq m \leq p-1$ with which a given $x$-plane satisfies (\ref{pxplanes}). Then, if $m = 0$ for a given plane, we choose for each vertex on that plane to include a factor of $[R^p_n(\beta b)]^3$ with $0 \leq n \leq p-1$ in order to satisfy (\ref{xpypzp=1}). Next, if $m \geq 1$, we pick for each vertex on the plane to include either a factor of $R^p_m(\beta b)[R^p_0(\beta b)]^2$ corresponding to the use of $(\mathcal{B}^x_v)^m$ for that vertex or a factor of $R^p_0(\beta b) [R^p_{p-m}(\beta b)]^2$ corresponding to the use of $(\mathcal{B}^y_v \mathcal{B}^z_v)^{p-m}$ for the vertex. This gives the following product expansion:
\begin{equation}
\begin{split}
\prod_{\mu, v} & \left( \sum_{n = 0}^{p-1} (\calB)^n R^p_n(\beta b) \right) = \left[ \left[ \sum_{n = 0}^{p-1} [R^p_n(\beta b)]^3 \right]^{L^2} \right. \\
& \left. + \sum_{m = 1}^{p-1} \left[ R^p_m(\beta b) [R^p_0(\beta b)]^2 + R^p_0(\beta b)[R^p_{p-m}(\beta b)]^2 \right]^{L^2} \right]^{L-1} \!\!\!\mathds{1} + \text{t.t.}.
\end{split}
\end{equation}
Finally, since the number $N$ of $p$-qudits in our system under cylindrical boundary conditions is $3L^3 + 2L^2$, the final partition function is given by:
\begin{equation}
\begin{split}
&\mathcal{Z}_{p, \text{Cylindrical}}  = p^{3L^3 + 2L^2} \left[ \sum_{m = 0}^{p - 1} [R^p_m(\beta a)]^{L^2} \right]^L  \\
& \times \left[ \left[ \sum_{m = 0}^{p-1} [R^p_m(\beta b)]^3 \right]^{L^2} \! + \sum_{m = 1}^{p-1} \left[ R^p_m(\beta b) [R^p_0(\beta b)]^2 + R^p_0(\beta b)[R^p_{p-m}(\beta b)]^2 \right]^{L^2} \right]^{L-1} \!\!\! .
\end{split}
\end{equation}
Once again, we check the $p=2$ case to ensure it matches (\ref{ZTrCylindrical2}). Using (\ref{R2}), we have:
\begin{equation}
\begin{split}
\mathcal{Z}_{2,\text{Cylindrical}} ={}& 2^{3L^3 + 2L^2} \left[ \mathsf{C}_{2a}^{L^2} + \mathsf{S}_{2a}^{L^2} \right]^L \\
{}& \times \left[ \left[ \mathsf{C}_{2b}^3 + \mathsf{S}_{2b}^3 \right]^{L^2} + \left[\mathsf{S}_{2b}\mathsf{C}_{2b}^2 + \mathsf{C}_{2b}\mathsf{S}_{2b}^2 \right]^{L^2} \right]^{L-1} .
\end{split}
\end{equation}
This is exactly the solution we previously derived, after rescaling $a \rightarrow 2a$ and $b \rightarrow 2b$.

\subsection{Large-$p$ Limit}
\label{sec:p>>1}
Another limit of particular interest is when $p$ becomes very large. In this case, the $\mathbb{Z}_p$ theory becomes approximated by a $U(1)$ theory
(see, e.g., \cite{clock} for a discussion of clock models). Let $Q^{\infty}_m$ and $R^{\infty}_m$ denote the limits as $p$ goes to infinity of $Q^p_m$ and $R^p_m$, respectively. From its power series representation in (\ref{Q}), it can easily be seen that $Q^{\infty}_m$ is simply given by the $m$th term of the exponential power series:
\begin{equation}
Q^{\infty}_m(x) = \frac{x^m}{m!} .
\end{equation}
Then, each $R^{\infty}_m$ can quickly be verified to be given by modified Bessel functions by comparing Taylor series:
\begin{equation}
R^{\infty}_m(x) = I_m(2x) .
\end{equation}
That $R^{\infty}_0$ becomes the 0th modified Bessel function may not be surprising -- since the eigenvalues of $\calA$ and $\calA^{\dagger}$ are given respectively by $\omega^m$ and $\overline{\omega}^m$, we may evaluate the partition function of a single elementary cube via:
\begin{equation}
\Tr \left[ \exp[\beta a (\calA + \calA^{\dagger})] \right] = \sum_{m = 0}^{p-1} e^{2\beta a \cos(2\pi m/p)} .
\end{equation}
In the large $p$ limit, we may approximate this sum by an integral, in which the above sum is interpreted as a left-handed Riemann sum \cite{clock}:
\begin{equation}
\Tr \left[ \exp[\beta a (\calA + \calA^{\dagger})] \right] \rightarrow p \int_0^1 dx\, e^{2\beta a \cos(2\pi x)} = p I_0(2\beta a) .
\end{equation}
This result appears in particular in (\ref{Zpopen2}), in which the first nontrivial factor is simply the product of $L^3$ such Bessel functions in the large $p$ limit.

\section{Implications of Dualities on the Nature of Fracton Excitations}
\label{sec:qualitative}

It has been well appreciated, for some time by now, that the low energy excitations of the X-Cube model may propagate in a correlated manner in order to avoid further energy penalties.
This intriguing feature raised the possibility of glassy dynamics and associated ``protection'' of quantum information that may be coded in the low energy states of this model \cite{PhysRevB.95.155133,Nan1,Haah,Bravyi}. Given the results of our duality mappings, we now revisit these notions and point to a simple consequence of our dualities. As we demonstrated in section \ref{sec:dynamics}, our dualities imply that with (dual) thermal baths, no excessively long time autocorrelations may persist at positive temperatures.
That is, the finite temperature autocorrelations display, at low energies (or, equivalently, low temperatures), the hallmarks of conventional activated dynamics. Our computed results do not feature any indications of exotic behaviors or particularly slow constrained dynamics. We caution anew that our results, invoking dualities, relate to (generally non-local) duals of Glauber thermal baths. Thus, the Glauber dynamics that we derived in section \ref{sec:dynamics} might differ from those for other baths. However, if the baths are relatively featureless (as typical thermal baths are) then duality transformations might not be expected to alter the system dynamics. Additionally, it is sometimes possible to explicitly construct dualities for which local heat baths in the original model are mapped to local heat baths in the dual model. For a demonstration using the 2D toric code, see the supplementary material of Ref. \cite{weinstein_universality_2019}.

The low energy dynamics are rooted in the character of the corresponding excitations. With that in mind, we wish to stress a simple conceptual point. Our duality mappings establish that the spectra of the X-Cube model and those of Ising chains are {\it identical}. Stated more precisely, the dualities (\ref{openclassical}) and (\ref{cylindricalmapping}) imply that for open and cylindrical boundary conditions, the X-Cube model has a spectrum which is none other than that of Ising chains, with degeneracies that differ only by a global power of two (as is also confirmed by our high temperature series
results of (\ref{final-open}) and (\ref{ZTrCylindrical2})). All exact dualities are unitary maps that preserve the spectrum \cite{bond-PRL,ADP}. Equivalently, the equivalence of the partition function,
\begin{equation}
\label{zge}
\mathcal{Z}(\beta) = \sum_{n = 0}^{N_{\text{max}}} g(E_n) e^{-\beta E_n},
\end{equation}
of two dual models implies that the spectra of the dual models is the same \cite{Nishimori-Ortiz11,bond-PRL,ADP}. 
Here, $g(E_n)$ denotes the degeneracy of each energy $E_n$. Thus, for both cylindrical and open boundary conditions, the spectrum of the X-Cube model including the degeneracy (modulo a global power of two) of all its low energy excitations is {\it precisely} the same as that of Ising chains. (From our results in section \ref{periodic}, the same holds true only in an asymptotic sense for the X-Cube model with periodic boundary conditions.) 

Thus, if the arguments concerning immobility of low energy excitations do not involve the boundaries, one might expect that since the excitations map in a one-to-one manner between the dual models, the {\it energetics of defects in the X-Cube model in the system bulk} (including arguments favoring low energy dynamics of one type or another) {\it will have exact counterparts for the classical Ising chains that are dual to the X-Cube model}. However, because
defects in standard classical Ising chains (i.e., domain walls) do not feature unusual dynamics, the same may be expected for their exact X-Cube duals. Dualities 
are, generally, non-local unitary transformations. Thus, a priori, one might anticipate that a sequence of states in which the energy is progressively altered or remains the same as defects locally move in a given system may involve, in its dual counterpart, a very different sequence of (non-local) moves for the corresponding defects in the dual system. With this in mind, we remark that Lieb-Robinson (LR) bound \cite{Lieb-Robinson} type arguments (that fundamentally restrict the propagation of correlations) are suggestive of local defect motion in two systems that are dual to each other, so long as these systems exhibit local interactions and local operators may be used to define/measure the defects \cite{note}. Thus, any local dynamics of defects in a given system (with this locality also required by the LR bounds) that change the energy in some way mandate the corresponding existence of local dynamics of any locally discernible defects in the dual model. By the unitary character of the duality transformation, the dynamics of the defects in the dual model alter the energy by exactly the same amount. In the Appendix, we very qualitatively discuss in some more detail several aspects of low energy excitations of the X-Cube model. Unlike the calculations in our work thus far, the arguments in that Appendix and in the current Section are by no means rigorous, and are only suggestive.

\section{Conclusion and Outlook}
\label{sec:conclude}

In this paper, we investigated a prototypical fractonic model, the X-Cube model and its
\(p\)-state generalizations, at finite temperature. We computed the partition function
of the models in closed form for open and cylindrical boundary conditions, and we showed
that these partition functions agree in the thermodynamic limit and agree with that
for periodic boundary conditions. These calculations provide compelling evidence that the
X-Cube models have a straightforward thermodynamic limit, insensitive to boundary conditions
as one usually assumes in elementary statistical mechanics. Moreover, we 
find the absence of finite temperature phase transitions and thermal fragility in these models 
 \cite{Long-TQO,PNAS,holography,fragility,fragile2,fragile3,fragile4,fragile5}. From a dynamical point of view, the elementary
excitations of the X-Cube model display highly constrained mobility. Thus,
one may hope that the approach to equilibrium might be extraordinarily slow (``glassy behavior").
To explore this possibility, we leveraged a duality transformation
to set up a simple Glauber model of equilibration for the X-Cube model, and found that conventional activated dynamics may appear, as opposed to glassy dynamics. It is possible that the Glauber dynamics that we study are not generic, but they suffice to demonstrate as a proof of principle that glassy dynamics are not mandatory for fractonic matter. 

Why is fractonic matter susceptible to thermal fragility? At the most basic level, the
problem is that all the efforts in designing fracton models go into engineering the 
energy barrier for low-lying excitations (as in, for example, this type of excitation can
only be created in quartets), but the issue at finite temperature is, of course, the free energy. 
To design matter at finite temperature, one must keep track of both the energy and the entropy,
or, microscopically, the energy levels and the density of states. Now, more concretely,
thermal fragility in numerous models stems from the same specific source - that of {\it dimensional reduction}. Entropic effects prohibit stable finite temperature order in conventional low dimensional systems. In a similar manner, thermal fluctuations eradicate stable orders in models dual to these low dimensional systems. This underscores the importance of entropic effects in this family of models of topological quantum matter. We are currently further investigating
the notion of effective dimensionality and encoding from the point of view of bond algebras; we reserve 
further remarks for a future publication.  

Looking forward, we believe that our results highlight the need for an extended set of
designing principles towards topological quantum memories.

\section{Acknowledgments}
This research was largely supported by NSF Grant 1411229 (CMMT). We also gratefully acknowledge NSF PHY-1607611 for work at the Aspen Center for Physics and grant NSF PHY-1748958 for work at the Kavli Institute for Theoretical Physics (KITP).

\appendix
\section{
\\ Ground States and Low Energy Excitations: \\ A Review and General Remarks}
\label{excite}

In this Appendix, we review and further discuss the ground states of the X-Cube model and their low energy excitations. The aim of this Appendix is to ground the general considerations of section \ref{sec:qualitative}. Unlike most other sections of this paper, the following discussion is largely qualitative.
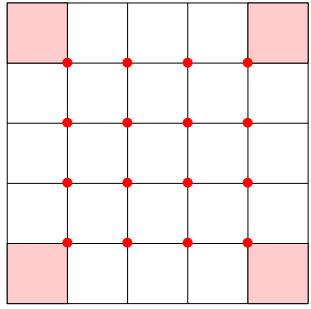
\begin{figure} [h]
\begin{center}
\scalebox{0.8}{
\begin{tikzpicture}
\foreach \x in {0,...,5} {
\draw (\x, 0) -- (\x, 5);
\draw (0, \x) -- (5, \x);
}
\filldraw[fill = red!20]
		(0,0)
	-- (0,1)
	--(1,1)
	--(1,0)
	-- cycle;
\filldraw[fill = red!20]
		(4,0)
	-- (4,1)
	--(5,1)
	--(5,0)
	-- cycle;
\filldraw[fill = red!20]
		(0,4)
	-- (0,5)
	--(1,5)
	--(1,4)
	-- cycle;
\filldraw[fill = red!20]
		(4,4)
	-- (4,5)
	--(5,5)
	--(5,4)
	-- cycle;
\foreach \x in {1,2,3,4} {
\foreach \y in {1,2,3,4} {
	\node [red] at (\x,\y) {\textbullet};
}
}
\end{tikzpicture}
}
\end{center}
\caption{A cross-section displaying a ``membrane" of $\sigma^z$ operators used to construct four cubic excitations at the corners of the membrane  $\mathcal{M}$. The $\sigma^z$ operators are included on the perpendicular outgoing edges at each red dot, and the excitations are shaded red.}
\label{membrane}
\end{figure}

The X-Cube model Hamiltonian of (\ref{hamiltonian}) is a sum of commuting terms (so-called ``stabilizers''). Any ground state $| \psi_0 \rangle$ of the X-Cube model satisfies the ``frustration free'' condition, 
\begin{equation}
\label{ff}
A_c \ket{\psi_0} = B^{\mu}_v \ket{\psi_0} = \ket{\psi_0}, \ \ \  \forall c, \mu, v .
\end{equation}
From (\ref{hamiltonian}), $\ket{\psi_0}$ clearly has the lowest possible energy of the system. A ground state is given by
\begin{equation}
\label{groundstateexplicit}
\ket{\psi_0} = {\cal{N}}_{0} \prod_c \frac{1}{2}(\mathds{1} + A_c) \ket{0} ,
\end{equation}
where ${\cal{N}}_{0}$ is a normalization factor and $\ket{0}$
is a simultaneous ($+1$) $\sigma^z$ eigenstate of each link. The state $\ket{0}$ is a trivial eigenstate of all $B^{\mu}_v$ operators with eigenvalue $+1$ and the commuting projectors
$\frac{1}{2}(\mathds{1} + A_c) $
ensure that $\ket{\psi_0}$ is an eigenstate of $A_c$ (with eigenvalue $+1$) for all cubes $c$. Thus, Eq. (\ref{ff}) is satisfied. The X-Cube model exhibits an exponential (in system length $L$) degeneracy. Notice that such an exponentially large degeneracy may appear in classical models that do not display topological order \cite{compass_rev,Sadegh}.

\begin{figure} [h]
\begin{center}
\scalebox{0.8}{
\begin{tikzpicture}
\foreach \x in {0,...,4} {
\draw (\x, 0) -- (\x, 4);
\draw (0, \x) -- (4, \x);
}
\filldraw[fill = red!20]
		(0,0)
	-- (0,1)
	--(1,1)
	--(1,0)
	-- cycle;

\draw[->] (4.5,2) -- (5.5,2);

\foreach \x in {0,...,4} {
\draw (6+\x, 0) -- (6+\x, 4);
\draw (6, \x) -- (10, \x);
}
\filldraw[fill = red!20]
		(7,0)
	-- (7,1)
	--(8,1)
	--(8,0)
	-- cycle;
\filldraw[fill = red!20]
		(6,1)
	-- (6,2)
	--(7,2)
	--(7,1)
	-- cycle;
\filldraw[fill = red!20]
		(7,1)
	-- (7,2)
	--(8,2)
	--(8,1)
	-- cycle;
\node [red] at (7,1) {\textbullet};

\draw[->] (7,-0.25) -- (5.5,-1.25);

\foreach \x in {0,...,4} {
\draw (3+\x, -5.5) -- (3+\x, -1.5);
\draw (3, \x-5.5) -- (7, \x-5.5);
}
\filldraw[fill = red!20]
		(4,-5.5)
	-- (5,-5.5)
	--(5,-4.5)
	--(4,-4.5)
	-- cycle;
\filldraw[fill = red!20]
		(4,-2.5)
	-- (5,-2.5)
	--(5,-1.5)
	--(4,-1.5)
	-- cycle;
\filldraw[fill = red!20]
		(3,-2.5)
	-- (4,-2.5)
	--(4,-1.5)
	--(3,-1.5)
	-- cycle;
\foreach \x in {-4.5,-3.5,-2.5} {
\node [red] at (4,\x) {\textbullet};
}
\end{tikzpicture}
}
\end{center}
\caption{A cross-section displaying a single excitation moved by the creation of an excitation pair, which can be freely moved ``off to infinity". $\sigma^z$ operators are included on the perpendicular edges at each red dot. The excitations are shaded red.}
\label{fig:fractonmovement}
\end{figure}
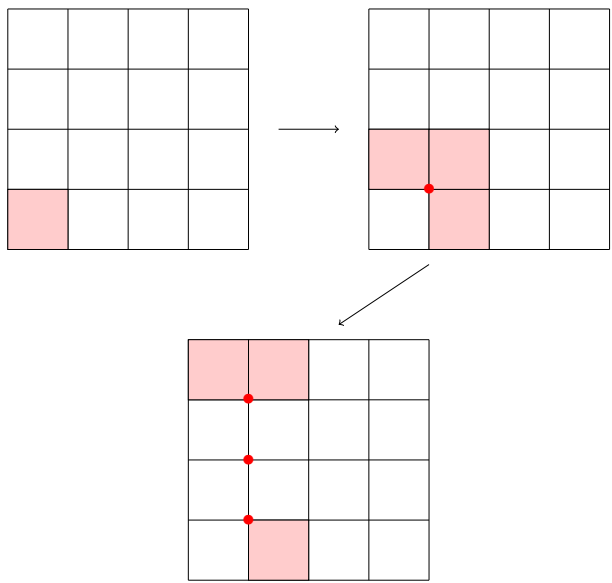

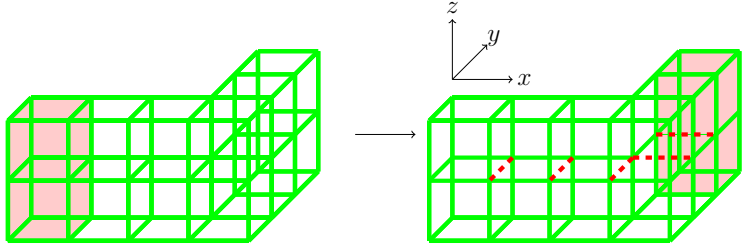
\begin{figure} [t]
    \begin{center}
    \scalebox{0.8}{
        \begin{tikzpicture}
        \foreach \x in {0,1} {
            \foreach \y in {0,1} {
                \filldraw[green, fill = red!20]
                    (0,\x,\y) -- (1,\x,\y)
                             -- (1,\x+1,\y)
                             -- (0,\x+1,\y)
                             -- cycle;
                \filldraw[green, fill=red!20]
                    (\x,\y,0) -- (\x,\y,1)
                              -- (\x,\y+1,1)
                              -- (\x,\y+1,0)
                              -- cycle;
            }
        }
        \foreach \y in {0,1,2} {
            \foreach \z in {0,1} {
                \foreach \x in {0,1,2,3,4} {
                    \draw [line width=2.0pt, green] (0,\y,\z) -- (4,\y,\z);
                    \draw [line width=2.0pt, green] (\x,\y,0) -- (\x,\y,1);
                    \draw [line width=2.0pt, green] (\x,0,\z) -- (\x,2,\z);
                }
                \foreach \x in {0,-1,-2} {
                    \draw [line width=2.0pt, green] (\z+3,\y,0) -- (\z+3,\y,-2);
                    \draw [line width=2.0pt, green] (3,\y,\x) -- (4,\y,\x);
                    \draw [line width=2.0pt, green] (\z+3,0,\x) -- (\z+3,2,\x);
                }
            }
        }
        
        \draw[->] (5,1,-1) -- (6,1,-1);
        
        \foreach \x in {0,1} {
            \foreach \y in {0,1} {
                \filldraw[green, fill = red!20]
                    (10,\x,\y-2) -- (11,\x,\y-2)
                             -- (11,\x+1,\y-2)
                             -- (10,\x+1,\y-2)
                             -- cycle;
                \filldraw[green, fill=red!20]
                    (\x+10,\y,-2) -- (\x+10,\y,-1)
                              -- (\x+10,\y+1,-1)
                              -- (\x+10,\y+1,-2)
                              -- cycle;
            }
        }
        \foreach \y in {0,2} {
            \foreach \z in {0,1} {
                \foreach \x in {0,1,2,3,4} {
                    \draw [line width=2.0pt, green] (7,\y,\z) -- (11,\y,\z);
                    \draw [line width=2.0pt, green] (7+\x,\y,0) -- (7+\x,\y,1);
                    \draw [line width=2.0pt, green] (7+\x,0,\z) -- (7+\x,2,\z);
                }
                \foreach \x in {0,-1,-2} {
                    \draw [line width=2.0pt, green] (\z+10,\y,0) -- (\z+10,\y,-2);
                    \draw [line width=2.0pt, green] (10,\y,\x) -- (11,\y,\x);
                    \draw [line width=2.0pt, green] (\z+10,0,\x) -- (\z+10,2,\x);
                }
            }
        }
        \draw [line width=2.0pt, green] (7,1,0) -- (10,1,0);
        \draw [line width=2.0pt, green] (7,1,1) -- (11,1,1);
        \draw [line width=2.0pt, green] (7,1,0) -- (7,1,1);
        \draw [line width=2.0pt, green] (11,1,1) -- (11,1,-2);
        \draw [line width=2.0pt, green] (10,1,0) -- (10,1,-2);
        \foreach \x in {1,2,3} {
            \draw [line width = 2.0pt, red, dashed] (7+\x,1,0) -- (7+\x,1,1);
        }
        \foreach \z in {0,-1} {
            \draw [line width = 2.0pt, red, dashed] (10,1,\z) -- (11,1,\z);
        }
        \draw [->] (7,2.3,0) -- (8,2.3,0);
        \node at (8.2,2.3,0) {$x$};
        \draw [->] (7,2.3,0) -- (7,3.3,0);
        \node at (7,3.5,0) {$z$};
        \draw [->] (7,2.3,0) -- (7,2.3,-1.5);
\node at (7, 2.3, -1.8) {$y$};
        \end{tikzpicture}
        }
    \end{center}
    \caption{An excitation pair, shaded red, can move (``glide'') freely in the $z$-plane by the application of $\sigma^z$ operators corresponding to the qubits at the dashed red edges. Motion along the $z$-direction (``climb'') costs energy.}
    \label{fig:2dmovement}
\end{figure}

\begin{figure} [h]
    \begin{center}
\scalebox{0.8}{
        \begin{tikzpicture}
            \filldraw[green, fill=red!20]
                (0,0,2) -- (2,0,2)
                        -- (2,2,2)
                        -- (0,2,2)
                        -- cycle;
            \filldraw[green, fill=red!20]
                (2,0,2) -- (2,2,2)
                        -- (2,2,0)
                        -- (2,0,0)
                        -- cycle;
            \filldraw[green, fill=red!20]
                (0,2,2) -- (2,2,2)
                        -- (2,2,0)
                        -- (0,2,0)
                        -- cycle;
             \filldraw[green, fill=yellow!50]
                (2,2,2) -- (4,2,2)
                        -- (4,4,2)
                        -- (2,4,2)
                        -- cycle;
            \filldraw[green, fill=yellow!50]
                (4,2,2) -- (4,4,2)
                        -- (4,4,0)
                        -- (4,2,0)
                        -- cycle;
            \filldraw[green, fill=yellow!50]
                (2,4,2) -- (4,4,2)
                        -- (4,4,0)
                        -- (2,4,0)
                        -- cycle;
             \filldraw[green, fill=blue!20]
                (0,2,4) -- (2,2,4)
                        -- (2,4,4)
                        -- (0,4,4)
                        -- cycle;
            \filldraw[green, fill=blue!20]
                (2,2,4) -- (2,4,4)
                        -- (2,4,2)
                        -- (2,2,2)
                        -- cycle;
            \filldraw[green, fill=blue!20]
                (0,4,4) -- (2,4,4)
                        -- (2,4,2)
                        -- (0,4,2)
                        -- cycle;
            \foreach \x in {0,2,4} {
                \foreach \y in {0,2,4} {
                    \draw[line width = 2.0pt, green] (0,\x,\y) -- (4,\x,\y);
                    \draw[line width = 2.0pt, green] (\x,0,\y) -- (\x,4,\y);
                    \draw[line width = 2.0pt, green] (\x,\y,0) -- (\x,\y,4);
                }
            }
            \draw[implies-implies, double equal sign distance] (2,-1,2) -- (3,-2,2);
            \draw[line width = 2.0pt, black] (-1,-2.5,2) -- (9,-2.5,2);
            \foreach \x in {-1,0.25,1.5,2.75,4,5.25,6.5,7.75,9} {
                \node[red] at (\x,-2.5,2) {\Huge \textbullet};
            }
            \node[red!50] at (0.875,-2.5,2) {\Huge $\cal{X}$};
            \node[blue!50] at (4.625,-2.5,2) {\Huge $\cal{X}$};
            \node[yellow!90] at (5.825,-2.5,2) {\Huge $\cal{X}$};
            \node[black] at (8,2,2) {\Huge $g(E_n) \propto {L^3 \choose n}$};
        \end{tikzpicture}
        }
    \end{center}
    \caption{Under open boundary conditions, the $L^3$ cubic operators of the X-Cube model are dual to the bonds of an open Ising chain of length $L^3 + 1$. In both models, the $n$th energy level above the ground state is given by choosing any $n$ excitations out of $L^3$ possible defects, modulo internal symmetries which do not change the excitations. Each cubic excitation corresponds to a given domain wall in the Ising chain. Here, $L=2$ and $n=3$.}
    \label{fig:fractondwduality}
\end{figure}
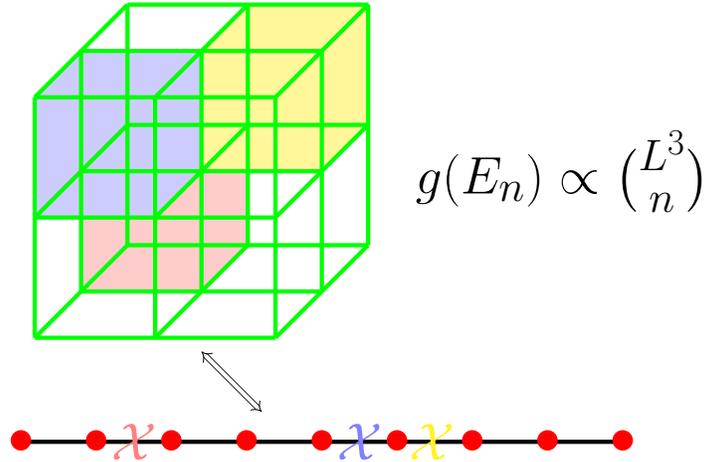

\begin{figure} [h]
\begin{center}
\scalebox{0.8}{
\begin{tikzpicture}
\foreach \x in {0,...,5} {
\draw (\x, 0) -- (\x, 5);
\draw (0, \x) -- (5, \x);
}
\filldraw[fill = red!20]
		(2,1)
	-- (2,2)
	--(3,2)
	--(3,1)
	-- cycle;
\foreach \x in {3,4,5} {
\foreach \y in {2,3,4,5} {
	\node [red] at (\x,\y) {\textbullet};
}
}
\end{tikzpicture}
}
\end{center}
\caption{A cross-section displaying the membrane operator $\mathcal{M}$ used to construct a single cubic excitation under open boundary conditions. $\sigma^z$ operators are included on the edges at each red dot, and the excitation is shaded red.}
\label{fig:openexcitations}
\end{figure}
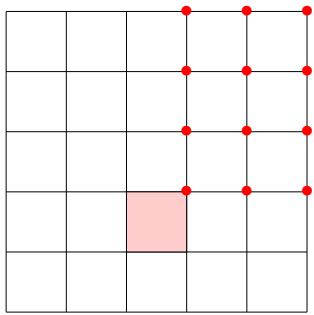

We may construct particular excited states by applying $\sigma_n^z$ to $\ket{\psi_0}$. The operator  $\sigma_n^z$ commutes with all but four $A_c$ operators. Thus, $\sigma_n^z \ket{\psi_0}$ remains an eigenstate of these operators (with eigenvalue $+1$). However, $\sigma^z_n$ will anticommute with the four $A_c$ operators containing $\sigma^x_n$.  
Thus,
\begin{equation}
A_c \left( \sigma^z_n \ket{\psi_0} \right) = - \sigma^z_n A_c \ket{\psi_0} = - \sigma^z_n \ket{\psi_0} \quad \text{for } n \in \partial c .
\end{equation}
It follows that $\sigma^z_n$ flips the eigenvalues of $A_c$ for each cube connected to the $n$th qubit. This can be expanded by instead considering a ``membrane" of $\sigma^z_n$ operators as in figure \ref{membrane}, flipping the eigenvalues of the four cubes at the corners of the membrane. The energy cost of creating these four excitations is $8a$, from flipping the eigenvalue of four $aA_c$ terms in (\ref{hamiltonian}) from $-a$ to $+a$. 

Three of these excitations can be moved ``off to infinity" by increasing the size of the membrane operator at no additional energy cost. A single localized excitation cannot move freely by application of a local operator. However, such an excitation can move while creating two additional excitations as in figure \ref{fig:fractonmovement}, at an energy cost of $4a$. Note that a pair of excitations can move freely in two dimensions, as shown in figure \ref{fig:2dmovement}, but not all three. In addition, a quartet of excitations is fully mobile in all three dimensions via applications of appropriate $\sigma^z$ operators. 

As argued in section \ref{sec:qualitative}, we can construct spectra for the open and cylindrical X-Cube models identical to those of their classical duals using {\it nonlocal} membrane operators. 

First, consider the case of open boundary conditions. The mapping (\ref{openclassical}) identifies each cubic operator $A_c$ with a bond variable $r_m$ of an $L^3+1$ site open Ising chain. The duality suggests that the spectra of $A_c$ operators in the open X-Cube model ought to be identical to that of the open Ising chain, in which each $n$th energy level can be achieved by choosing any arbitrary arrangement of $n$ ``bad bonds" corresponding to $r_m = -1$ (see figure \ref{fig:fractondwduality}). In particular, the duality implies that the excitations of the $A_c$ operators in the open X-Cube model are given not by four-fold excitations at the corners of a membrane operator, but by \textit{any arbitrary arrangement of excitations} corresponding to $A_c \ket{\psi} = -\ket{\psi}$.

\begin{figure} [h]
\begin{center}
\scalebox{0.8}{
\begin{tikzpicture}
\foreach \x in {0,...,5} {
\draw (\x, 0) -- (\x, 5);
\draw (0, \x) -- (5, \x);
}
\filldraw[fill = red!20]
		(1,0)
	-- (1,1)
	--(2,1)
	--(2,0)
	-- cycle;
\filldraw[fill = red!20]
		(3,1)
	-- (3,2)
	--(4,2)
	--(4,1)
	-- cycle;
\filldraw[fill = red!20]
		(3,3)
	-- (3,4)
	--(4,4)
	--(4,3)
	-- cycle;
\filldraw[fill = red!20]
		(0,4)
	-- (0,5)
	--(1,5)
	--(1,4)
	-- cycle;
\foreach \x in {0,1} {
	\node [red] at (\x,0) {\textbullet};
}
\foreach \x in {4,5} {
    \foreach \y in {2,3} {
        \node [red] at (\x,\y) {\textbullet};
    }
}
\node [red] at (0,5) {\textbullet};
\end{tikzpicture}
}
\end{center}
\caption{In the open boundary X-Cube model, with the use of nonlocal membrane operators, we may construct any arbitrary arrangement of cubic excitations in the lattice.}
\label{fig:openarbex}
\end{figure}
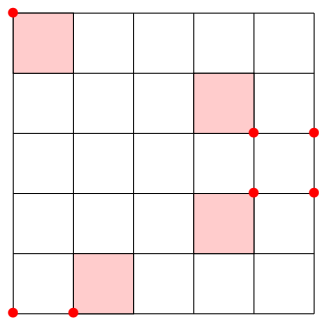

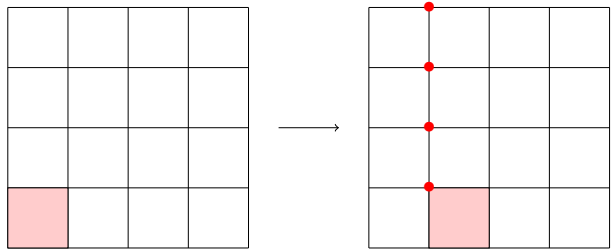
\begin{figure} [h]
    \begin{center}
\scalebox{0.8}{
        \begin{tikzpicture}
        \foreach \x in {0,...,4} {
\draw (\x, 0) -- (\x, 4);
\draw (0, \x) -- (4, \x);
}
\filldraw[fill = red!20]
		(0,0)
	-- (0,1)
	--(1,1)
	--(1,0)
	-- cycle;

\draw[->] (4.5,2) -- (5.5,2);

\foreach \x in {0,...,4} {
\draw (6+\x, 0) -- (6+\x, 4);
\draw (6, \x) -- (10, \x);
}
\filldraw[fill = red!20]
		(7,0)
	-- (7,1)
	--(8,1)
	--(8,0)
	-- cycle;
\foreach \y in {1,2,3,4} {
\node [red] at (7,\y) {\textbullet};
}
        \end{tikzpicture}
        }
    \end{center}
    \caption{In the open boundary X-Cube model, with the use of nonlocal membrane operators, we may move a single excitation in any direction at no additional energy cost.}
    \label{fig:openmovement}
\end{figure}

Indeed, we can explicitly construct each of these excited states using nonlocal membrane operators: to place a lone excitation in any particular cubic location, start with the ground state $\ket{\psi_0}$, and apply the membrane operator $\cal{M}$ described in figure \ref{membrane} with one corner at the desired locaton. Then, expand $\cal{M}$ to move the three extraneous excitations to the boundaries, as in figure \ref{fig:openexcitations}. With $\cal{M}$ extending to the boundaries at all but one corner, only one excitation remains. By overlaying multiple such operators, we may place any number of excitations in any possible arrangement: for instance, we can create four excitations in a non-rectangular arrangement (figure \ref{fig:openarbex}), or we can move a single excitation at no energy cost (figure \ref{fig:openmovement}).

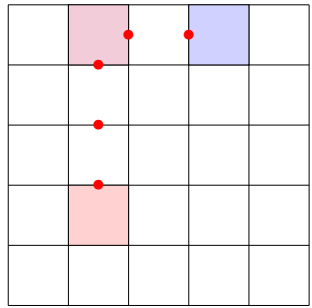
\begin{figure} [h]
\begin{center}
\scalebox{0.8}{
\begin{tikzpicture}
\foreach \x in {0,...,5} {
\draw (\x, 0) -- (\x, 5);
\draw (0, \x) -- (5, \x);
}
\filldraw[fill = red!20, opacity = .9]
		(1,1)
	-- (1,2)
	--(2,2)
	--(2,1)
	-- cycle;
\filldraw[fill = purple!20]
		(1,4)
	-- (1,5)
	--(2,5)
	--(2,4)
	-- cycle;
\filldraw[fill = blue!20, opacity = .9]
		(3,4)
	-- (3,5)
	--(4,5)
	--(4,4)
	-- cycle;
\node [red] at (1.5,2) {\textbullet};
\node [red] at (1.5,3) {\textbullet};
\node [red] at (1.5,4) {\textbullet};
\node [red] at (2,4.5) {\textbullet};
\node [red] at (3,4.5) {\textbullet};
\end{tikzpicture}
}
\end{center}
\caption{Under cylindrical boundary conditions, two excitations may be placed in any desired locations within an $x$-plane (marked in red and blue) by multiplying two membrane operators overlapping at a common corner (marked in purple). Red dots indicate locations of $\sigma^z$ operators within this cross-section of the lattice.}
\label{fig:cylexcitations}
\end{figure}

The case of cylindrical boundary conditions is not as simple. The mapping (\ref{cylindricalmapping}) identifies each plane of cubic operators with an $L^2$ site periodic Ising chain, with each given $A_c$ mapped to a single bond $r^i_m r^i_{m+1}$. Excitations in the bonds of a periodic Ising chain come in multiples of two: any domain wall must be accompanied by another domain wall to return to the original spin direction. The duality implies that the excitations of the cylindrical X-Cube model are also given by arbitrary arrangements of excitations, so long as each $x$-plane has excitations in multiples of two. We may construct such excited states using two nonlocal membrane operators: first, to construct two cubic excitations in the same $x$-plane and lying along a line, simply apply to the ground state a membrane operator $\cal{M}$ perpendicular to the $y$ or $z$ direction, with two corners of the membrane lying in the desired plane and the other two corners extending to the open $x$ boundary. Then, to create two excitations in any location within the same $x$-plane, simply multiply two such perpendicular membrane operators sharing a common corner, as in figure \ref{fig:cylexcitations}. This procedure may be continued to construct arbitrary multiple-of-two arrangements of excitations within each $x$-plane of the lattice.

We have previously shown that the free energy density of the X-Cube model is independent of any choice in boundary conditions in the thermodynamic limit. However, this does not explicitly mean that the spectrum is independent of our choice in boundary conditions: indeed, just as the open and periodic one-dimensional Ising chain have different spectra, the open and cylindrical X-Cube models also do not have the same spectra. For this reason, we do not suggest that the periodic X-Cube spectra is the same as that of the open or cylindrical models --- in fact, under fully periodic boundary conditions, there exists no operator (local or nonlocal) which can create a single cubic excitation in a ground-state wavefunction \cite{PhysRevB.94.235157}. That said, the spectra of the open and cylindrical systems do suggest that we must be careful when discussing the nature of excitations in the X-Cube and other similar models. In particular, while the above depiction of fractons in the model discusses creation and mobility of fractons using local operators, the results of this paper suggest that it may not be sufficient to consider local operators alone. Once nonlocal operators are considered, it's perfectly clear that the open boundary X-Cube model has no special thermodynamics or constrained mobility: all cubic excitations are completely decoupled from one another, and any combination of cubic excitations may be realized from any other via some combination of nonlocal membrane operators. The same is true of the cylindrical X-Cube model: while the additional constraints (\ref{AplanarCylindrical}) introduce some correlation among $A_c$ expectation values within a given $x$-plane, these correlations vanish in the thermodynamic limit, in the same way as the bond variables of the periodic Ising chains to which the cylindrical X-Cube model is dual. 

While we cannot provide a closed form partition function, spectrum, or free energy of the periodic X-Cube model at finite $L$, the results for open and cylindrical boundary conditions at least indicate that we must be careful with how we discuss excited states and their dynamics under periodic boundary conditions as well. The lack of finite temperature phase transitions in all models, as well as the seeming irrelevance of order $L^2$ constraints such as (\ref{AplanarCylindrical}), (\ref{BplanarCylindrical}), (\ref{Aplanar}), and (\ref{Bplanar}) in the thermodynamic limit, suggest that the dynamics and equilibrium thermodynamics of the periodic X-Cube model may be simpler than originally thought. An earlier analysis \cite{PhysRevB.95.155133} found that  ``when a zero temperature type I fracton model is placed in contact with a finite temperature heat bath, the approach to equilibrium is a \textit{logarithmic} function of time over an exponentially wide window of time scales." By contrast, the results of section 8 for the open X-Cube model with Glauber dynamics suggest, instead, an exponential decay at all times.

What's more, if the physical motivation for studying the X-Cube model is as a stabilizer code with a robust quantum memory, then it's likely that a physical realization of the X-Cube model has boundary conditions closer to the open boundary conditions discussed in section \ref{sec:open} than the fully periodic boundary conditions discussed in section \ref{periodic}. In this case, our results suggest that such a physical system may be subject to the same thermal fragility as the Kitaev model \cite{fragility}.

\bibstyle{elsarticle-num}
\bibliography{refs}

\begin{thebibliography}{10}
\bibitem{note}
We clarify that here we allude to operators that may be used to measure for the
  presence of defects. as is well known, domain walls in ferromagnetic ising
  chains (similar to the ones that the x-cube model is dual to) may be created
  by a non-local product of pauli operators that flip all of the spins up the
  to the domain wall location. however, the presence of a domain wall in such
  an ising chain can be measured by taking the product of the spin values on
  two consecutive sites - if the latter (local) two spin product is negative
  then a domain is present.
\end{thebibliography}

\end{document}